\documentclass[aps,prd,twocolumn,superscriptaddress,floatfix,nofootinbib,amsmath,amssymb,preprintnumbers]{revtex4-1}
\usepackage[colorlinks=true,breaklinks=true]{hyperref}

\usepackage{bm}
\usepackage{graphicx}
\usepackage{multirow}
\usepackage{amsmath}
\usepackage{mathrsfs}
\usepackage{amssymb}
\usepackage{cancel}
\usepackage{color}
\usepackage{xspace}
\usepackage{url}
\usepackage{verbatim}
\usepackage{mathtools}
\usepackage{siunitx}
\usepackage{amstext}
\usepackage{booktabs}
\usepackage{tabulary}
\usepackage{tabularx}
\usepackage{etoolbox}
\usepackage{lipsum}
\usepackage{lettrine}
\usepackage{fontawesome}
\usepackage{soul}

\usepackage[utf8]{inputenc}
\usepackage[T1]{fontenc}
\usepackage[dvipsnames,table]{xcolor}
\hypersetup{urlcolor=PineGreen,
	    citecolor=PineGreen,
	    linkcolor=PineGreen}
\definecolor{lightgray}{rgb}{0.9,0.9,0.9}	 
\definecolor{green}{rgb}{0,0.5,0}
\definecolor{red}{rgb}{1,0,0}
\definecolor{blue}{rgb}{0,0,0.5}
\definecolor{grey}{rgb}{0.7, 0.75, 0.71}

\input Zallman.fd

\LettrineTextFont{\itshape}
\newcommand{\DM}{\rm DM}

\newcommand{\aap}{{Astron.~Astrophys.}}
\newcommand{\araa}{{Annu.~Rev.~Astron.~Astrophys.}}

\newcommand{\mnras}{{MNRAS}}
\newcommand{\jcap}{{JCAP}}

\newcommand{\apjs}{{The~Astrophysical~Journal~Supp.~Series}}

\begin{document}
\preprint{SLAC-PUB-17772}
\preprint{LTH-1368}

\title{Dark Matter Halo Parameters from Overheated Exoplanets \\ via Bayesian Hierarchical Inference 
}

\author{Mar\'ia Benito}
\thanks{ \href{mailto:mariabenitocst@gmail.com}{mariabenitocst@gmail.com};  \href{http://orcid.org/0000-0002-4711-6516}{0000-0002-4711-6516}}
\affiliation{Tartu Observatory, University of Tartu, Observatooriumi 1, T\~oravere 61602, Estonia}

\author{Konstantin Karchev}
\thanks{\href{mailto:kkarchev@sissa.it}{kkarchev@sissa.it}; \href{https://orcid.org/0000-0001-9344-736X}{0000-0001-9344-736X}}
\affiliation{Theoretical and Scientific Data Science, SISSA, Trieste, Italy}

\author{Rebecca K. Leane}
\thanks{\href{mailto:rleane@slac.stanford.edu}{rleane@slac.stanford.edu}; \href{http://orcid.org/0000-0002-1287-8780}{0000-0002-1287-8780}}
\affiliation{SLAC National Accelerator Laboratory, Stanford University, Stanford, CA 94039, USA}
\affiliation{Kavli Institute for Particle Astrophysics and Cosmology, Stanford University, Stanford, CA 94039, USA}

\author{Sven P\~oder}
\thanks{\href{mailto:msven.poder@kbfi.ee}{sven.poder@kbfi.ee}; \href{http://orcid.org/0000-0002-0001-7123}{0000-0002-0001-7123}}
\affiliation{National Institute of Chemical Physics and Biophysics, R\"avala 10, Tallinn 10143, Estonia
}

\author{Juri Smirnov}
\thanks{ \href{mailto:juri.smirnov@liverpool.ac.uk}{juri.smirnov@liverpool.ac.uk}; \href{http://orcid.org/0000-0002-3082-0929}{0000-0002-3082-0929}}
\affiliation{Department of Mathematical Sciences, University of Liverpool,
Liverpool, L69 7ZL, United Kingdom}

\author{Roberto Trotta}
\thanks{\href{mailto:rtrotta@sissa.it}{rtrotta@sissa.it}; \href{https://orcid.org/0000-0002-3415-0707}{0000-0002-3415-0707}}
\affiliation{Theoretical and Scientific Data Science, SISSA, Trieste, Italy}
\affiliation{Imperial College London, Physics Department, Blackett Laboratory, Prince Consort Rd, SW7 2AZ London, United Kingdom}
\affiliation{Italian Research Center on High Performance Computing, Big Data and Quantum Computing \& National Institute for Nuclear Physics, Sezione di Trieste, Italy}

\date{\today}
\smallskip
\begin{abstract}
Dark Matter (DM) can become captured, deposit annihilation energy, and hence increase the heat flow in exoplanets and brown dwarfs. Detecting such a DM-induced heating in a population of exoplanets in the inner kpc of the Milky Way thus provides potential sensitivity to the galactic DM halo parameters. We develop a Bayesian Hierarchical Model to investigate the feasibility of DM discovery with exoplanets and examine future prospects to recover the spatial distribution of DM in the Milky Way. We reconstruct from mock exoplanet datasets observable parameters such as exoplanet age, temperature, mass, and location, together with DM halo parameters, for representative choices of measurement uncertainty and the number of exoplanets detected. We find that detection of $\mathcal{O}(100)$ exoplanets in the inner Galaxy can yield quantitative information on the galactic DM density profile, under the assumption of 10\% measurement uncertainty. Even as few as $\mathcal{O}(10)$ exoplanets can deliver meaningful sensitivities if the DM density and inner slope are sufficiently large.~\href{https://github.com/mariabenitocst/exoplanets}{\large\faGithub}
\end{abstract}

\maketitle

\lettrine{T}{he milky way} is estimated to be home to about 100 billion exoplanets and brown dwarfs~\cite{2017MNRAS.471.3699M}. Each of them holds a wealth of data, waiting to reveal their secrets and curiosities. One such secret is whether or not exoplanets and brown dwarfs (hereafter collectively referred to as ``exoplanets'') harbor dark matter (DM) cores and consequently have temperatures much higher than otherwise expected~\cite{Leane:2020wob}.

\begin{figure}[t!]
\includegraphics[width=0.95\columnwidth]{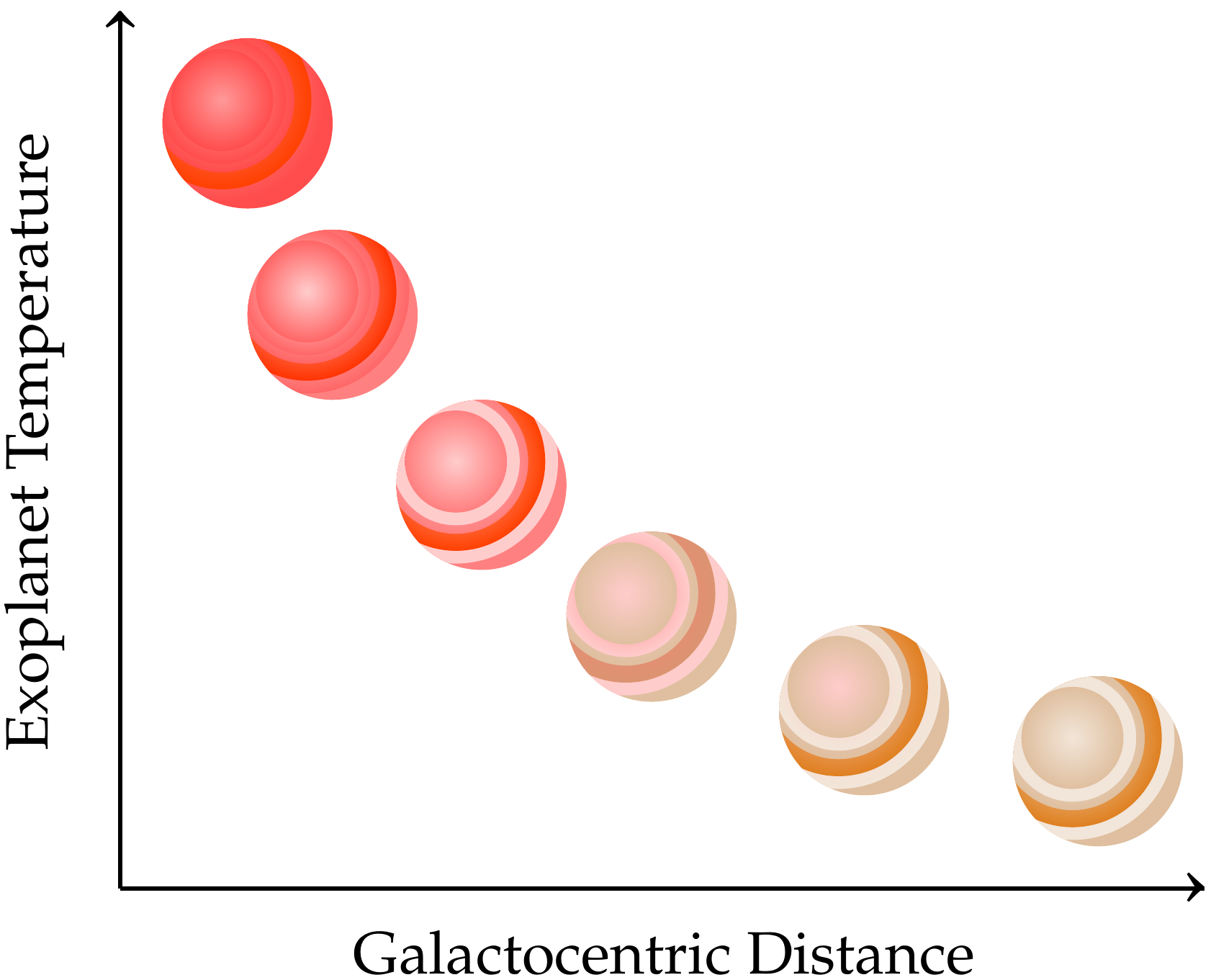}
\caption{Schematic of DM heated exoplanets. In the inner Galaxy, there is more DM, leading to higher levels of exoplanet heating. Therefore, measuring exoplanet temperatures as a function of Galactocentric radius provides a probe of the DM density distribution in our Galaxy.}\vspace{-3mm}
\label{fig:schematic}
\end{figure}

DM heating in exoplanets is expected to potentially arise in the following way. Exoplanets sweep through the Galactic DM halo, collecting DM which has scattered with the exoplanet particles and lost sufficient kinetic energy to become bound to the exoplanet. The accumulated DM can then potentially annihilate, depositing its annihilation energy into the exoplanet and increasing its temperature beyond that expected from models of exoplanet formation and evolution. Once sufficient DM has accumulated, the exoplanet can reach equilibrium between DM scattering and annihilation, such that the measured temperature provides a probe of the DM scattering rate and enables inference of the DM density near the exoplanet, as the "excess" heating scales with the DM density (see Figure~\ref{fig:schematic}). This signal can be potentially detected by upcoming infrared telescopes, such as the James Webb Space Telescope (JWST), or Roman Telescope~\cite{Leane:2020wob}.

DM heating processes have also been considered in other celestial objects, such as nuclear-burning stars, neutron stars and white dwarfs~\cite{Goldman:1989nd,
Gould:1989gw,
Kouvaris:2007ay,
Bertone:2007ae,
Spolyar:2007qv, Iocco:2008xb,Freese:2008hb, Taoso:2008kw, Sivertsson:2010zm, Freese:2015mta, 1989ApJ...338...24S,Fairbairn:2007bn, Scott:2008ns,Lopes:2021jcy,
deLavallaz:2010wp,
Kouvaris:2010vv,
McDermott:2011jp,
Kouvaris:2011fi,
Guver:2012ba,
Bramante:2013hn,
Bell:2013xk,
Bramante:2013nma,
Bertoni:2013bsa,
Kouvaris:2010jy,
McCullough:2010ai,
Perez-Garcia:2014dra,
Bramante:2015cua,
Graham:2015apa,
Cermeno:2016olb,
Krall:2017xij,
Leane:2017vag,
McKeen:2018xwc,
Baryakhtar:2017dbj,
Raj:2017wrv,
Bell:2018pkk,
Chen:2018ohx,
Hamaguchi:2019oev,
Camargo:2019wou,
Bell:2019pyc,
Garani:2019fpa,
Acevedo:2019agu,
Joglekar:2019vzy,
Joglekar:2020liw,
Bell:2020jou,
Garani:2020wge,
Leane:2021ihh,Acevedo:2023xnu,John:2023knt,Croon:2023trk,Acevedo:2024ttq}, the Earth~\cite{Mack:2007xj,Chauhan:2016joa,Bramante:2019fhi}, and other planets and moons~\cite{Adler:2008ky,Kawasaki:1991eu,Mitra:2004fh,Bramante:2019fhi,Garani:2019rcb,Chan:2020vsr,Leane:2021tjj,Blanco:2023qgi,Croon:2023bmu,Linden:2024uph}. Exoplanets can be superior probes to these other celestial objects in several ways~\cite{Leane:2020wob,Leane:2024bvh}. First, they have cooler cores while having relatively high escape velocities, allowing lighter DM particles to not evaporate and be retained in the system, providing a probe of sub-GeV mass DM. Second, they have large radii, allowing them to be detectable far into the inner Galaxy. Third, exoplanets are highly abundant, such that this signature can be probed throughout the Galaxy with potentially high statistics. This allows for a location-dependent probe of the DM density profile.

The exoplanet program is rapidly accelerating. Many new exoplanets are expected to be identified soon, and many new measurements will be performed. The Vera C. Rubin Observatory (Rubin), the Nancy Grace Roman Space Telescope (Roman), and the Gaia Spacecraft have or will have targeted programs to discover as many exoplanets as possible~\cite{Perryman_2014,green2012widefield,Johnson_2020}. While exoplanets have not yet had their spectra measured at the low temperatures and large distances required for this search, the technological breakthroughs of these instruments will allow unprecedented sensitivities to distant exoplanets. There are also many surveys such as the Optical Gravitational Lensing Experiment (OGLE), Two Micron All Sky Survey (2MASS), and the Wide-field Infrared Survey Explorer (WISE), which peer deep into our Galaxy. Detecting the temperature of either free-floating exoplanets, or brown dwarfs, first requires target detection via gravitational microlensing, which can be aided especially with simultaneous use of telescopes. This has already been investigated for example with Roman and Euclid~\cite{Bachelet_2019}. Microlensing has so far led to the identification of candidates of free floating exoplanets and brown dwarfs at large distances, for example OGLE-2015-BLG-1268, a brown dwarf candidate with 50 Jupiter masses and at $5.9 \pm 1.0 \text{ kpc}$~\cite{Zhu2016}.

Overall, the interplay of all these telescopes will only improve the prospects for this search, making it timely to consider the precision required of exoplanet-related measurements to be used in inference of DM distribution and properties. Setting up a first statistical framework to attack this problem, and quantifying the measurement uncertainties required alongside the number of required target exoplanets to test the DM halo, is the goal of this work.

In this work, we develop a Bayesian hierarchical model (BHM) that takes as input measured properties of exoplanets and infers simultaneously: parameters describing their population, such as the shape of their initial mass function and Galactic radial number density profile; the individual properties of all objects; and parameters of the Galactic DM density profile. We use this model because our problem contains many correlated uncertainties, and the model is particularly powerful at accounting for the variability within the exoplanet population and measurement uncertainties. We use this Bayesian hierarchical model to investigate the sample size and measurement precision required for inference of DM halo parameters.

This paper is organized as follows. In Section~\ref{sec:method}, we describe our statistical model and provide details of the luminosities of exoplanets, including DM capture and heating. We discuss our results, including the sensitivity to DM-heating detectability, in Section~\ref{sec:results}. Finally, we conclude and discuss the outlook for detecting DM-heated exoplanets in Section~\ref{sec:conclusion}.

\section{The Bayesian Hierarchical Model}\label{sec:method}

The interplay between the latent ($i.e.$, unobserved) distribution of exoplanet parameters and noise arising from observational error leads to a well-known statistical effect that astronomers call ``Eddington bias'' \cite{1913MNRAS..73..359E} -- the fact that the parameters of the underlying population will be incorrectly estimated from the observed sample (for an illustration of how and why this arises in Gaussian linear models, see Refs.~\cite{Kelly:2007jy,2011MNRAS.418.2308M}). To counter the bias, it is necessary to model the process leading from population parameters to observed (noisy) samples. This can be achieved by means of a hierarchical structure, which models separately the population distribution and the noise. Conditioning on the observed data then leads to posterior distributions for all -- global and latent -- parameters in the model that correctly capture the different properties at each level where uncertainties are introduced (either through intrinsic variability at the population level or through noise). In other words, this requires a Bayesian hierarchical model of the forward generation of the data. 

In this spirit, we present in this section a Bayesian hierarchical model for DM-heated exoplanets, which allows one to infer the normalisation and slope of the Galactic DM halo density profile jointly with parameters describing the population of exoplanets and the a-priori uncertain properties of individual objects from noisy measurements of their temperatures, masses, ages, and positions within the Galaxy. 

\begin{figure}[t]
\includegraphics[scale=0.41]{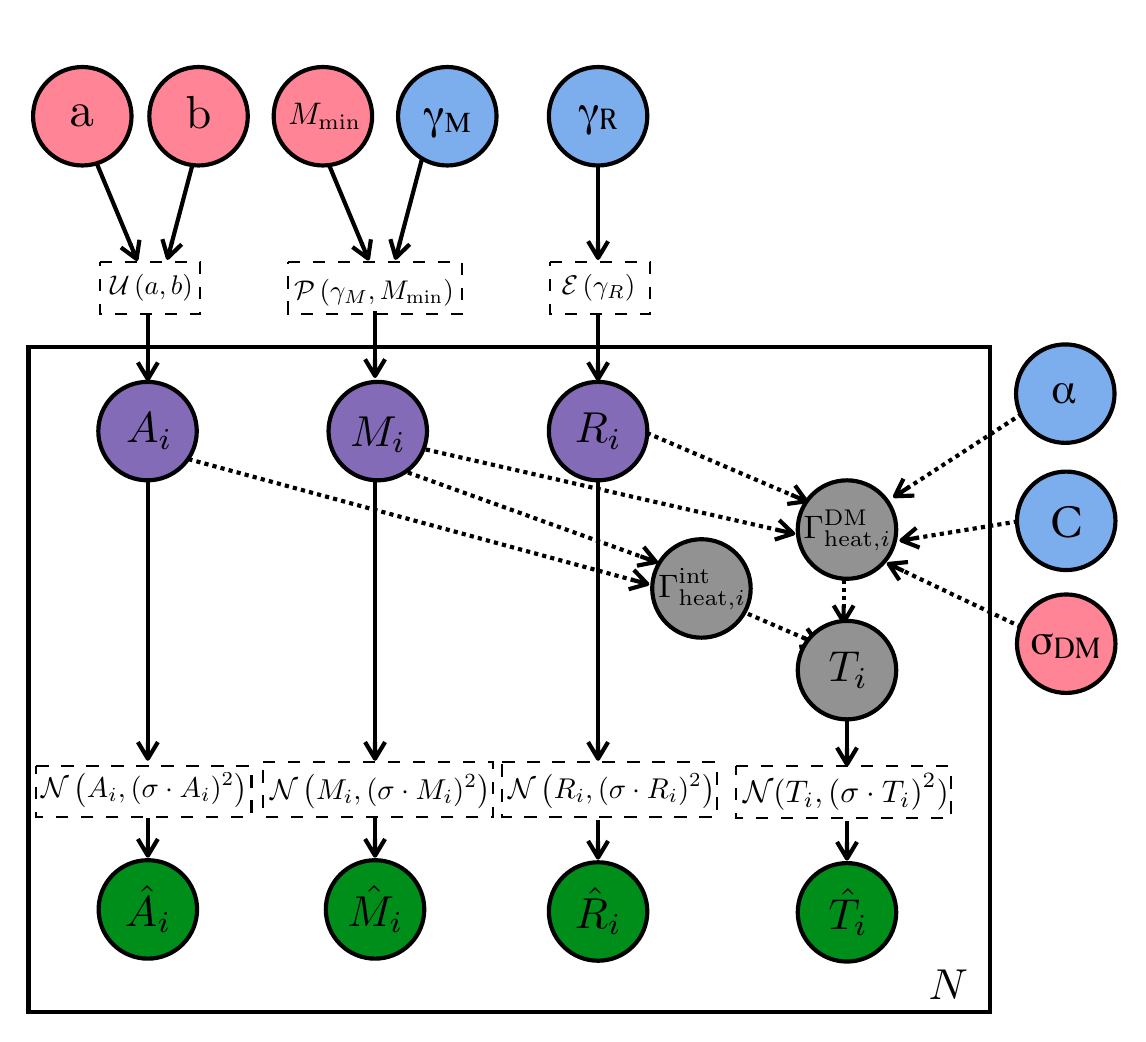}
\caption{Directed acyclic graph depicting our Bayesian hierarchical model used to analyse overheated exoplanets and properties of their population and of the DM density profile. Deterministic and probabilistic connections between variables are indicated by solid and dotted arrows, respectively. Within the $N$ plate, representing enumeration over N objects, data are shown in green, while independent and dependent (unobserved,  {\it latent}) parameters are denoted by purple and grey nodes, respectively. {\it Global} parameters are depicted in blue, with those held fixed in this study shown in pink. 
}
\label{fig:DAG_BHM}
\end{figure}

Figure~\ref{fig:DAG_BHM} shows a graphical network which illustrates the structure of this model, and is summarised in Table~\ref{tab:BHM}. We proceed to elaborate it below.

\renewcommand{\arraystretch}{1.3}
\begin{table}
\centering
\small\addtolength{\tabcolsep}{1pt}
\begin{tabular}{  l  c  c  c }
\hline

Parameter &  &  Prior & Mock value \\
\hline
Measured age, mass, & $\hat{X}$ & $\mathcal{N}(X, (\sigma \cdot X)^2)$ & \\
Gal. distance  \& temp. & & & \\
\hspace{0.02cm} Mean & $X$ & & \\ 
\hspace{0.02cm} Uncertainty & $\sigma \cdot X$ & & $\sigma\in[1, 20]\%$\\ 

Latent age & $A$ & $\mathcal{U}(a, b)$ &  \\ 
\hspace{0.02cm} Min. age & a & & 8 Gyr\\
\hspace{0.02cm} Max. age & b & & 10 Gyr\\ 

Latent mass & $M$ & $\mathcal{P}(\gamma_M, M_{\rm min})$ & $\in[14, 55]\,\rm M_{\rm jup}$ \\ 
\hspace{0.02cm} Slope & $\gamma_M$ & $\mathcal{N}(0.6, 0.1^2)$ & 0.6\\
\hspace{0.02cm} Min. mass & $M_{\rm min}$ & & $14\,\rm M_{\rm jup}$ \\  

Latent Gal. dist. & $R$ & $\mathcal{E}(\gamma_R)$ &  $\in [0.1, 1]\,\rm kpc$ \\ 
\hspace{0.02cm} Slope & $\gamma_R$ & $\mathcal{U}(1, 2)$ & 1.46\\

Latent temp. & $T$ & & \\
\hspace{0.02cm} DM dens. slope & $\alpha$ & $\mathcal{U}(0, 3)$ & $\in[0, 3]$\\
\hspace{0.02cm} DM dens. norm. & $C$ & $\mathcal{U}(0.5, 40)$ & $6, 20 \,\rm GeV/cm^3$\\
\hspace{0.02cm} DM vel. dispersion & $\sigma_{\rm DM}$ &  & $100 \,\rm km/s$\\
\hline
\label{tab:BHM}
\end{tabular}
\caption{Exoplanet parameters, priors and fiducial values used to generate mock observations (see Subsection~\ref{subsec:methods_config}).
}
\end{table}

\subsection{Observables}

The observed data $\boldsymbol{d}$ consists of noisy measurements $\hat{A}_i$, $\hat{M}_i$, $\hat{R}_i$, and $\hat{T}_i$ ($i=1, \dots, N$) of the ages, masses, Galactocentric distances, and temperatures, respectively, of a sample of $N$ exoplanets; the data are shown in green in the Directed Acyclic Graph (DAG) of Fig.~\ref{fig:DAG_BHM}. Under the assumption of normally distributed errors, the sampling distribution of $\boldsymbol{d} \in \mathbb{R}^{4N}$ is Gaussian:
\begin{equation}
    \boldsymbol{d} \sim \mathcal{N}([\boldsymbol{A}, \boldsymbol{M}, \boldsymbol{R}, \boldsymbol{T}], \hat{\boldsymbol{\Sigma}}),
\end{equation}
where $\boldsymbol{A}, \boldsymbol{M}$ and $\boldsymbol{R}$ are $\mathbb{R}^{N}$ vectors containing the latent (unobserved, true) age $A_i$, mass $M_i$, and Galactocentric distance $R_i$, respectively, of individual exoplanets, while $\boldsymbol{T}\in \mathbb{R}^{N}$ is the latent temperature of each exoplanet, which is a deterministic function of the other exoplanet parameters and of properties of the DM halo (see Eq.~\ref{eq:heat_tot}). Latent variables are depicted in the DAG as purple and grey nodes, the latter indicating that they are obtained from other variables via a deterministic relationship (dotted arrows).
The observational covariance matrix $\hat{\boldsymbol{\Sigma}} \in \mathbb{R}^{4N \times 4N}$ accounts for correlations between measurements of individual exoplanets and among different exoplanets. We will assume for simplicity that all measurements are independent, which means that $\hat{\boldsymbol{\Sigma}}$ reduces to a diagonal matrix. In order to explore the effect of measurement uncertainty on our results, we will adopt different noise levels, expressed through the relative uncertainty $\sigma$ in measurements of each quantity $X \in \{A, M, R, T\}$, as detailed in Table~\ref{tab:BHM} and Subsection~\ref{subsec:methods_config}. Thus, the covariance matrix takes the form: 
$$\hat{\boldsymbol{\Sigma}} = \text{diag}(\sigma^2 \cdot A^2_i, \sigma^2 \cdot M^2_i, \sigma^2 \cdot R^2_i, \sigma^2 \cdot T^2_i), \, i=1,\dots, N.$$

\subsection{Heat Model for Exoplanets}

We now discuss the heat contributions to exoplanets, both from DM and internal background processes. Without DM, exoplanets have an effective temperature as a function of their age $A_i$ and mass $M_i$. This heat occurs from $e.g.$ burning processes in the object, as well as heat left over from formation. For the internal heat without DM, $\Gamma^{\rm int}_{{\rm heat}, i}(A_i, M_i)$, we use the theoretical evolution model \textsc{Atmo2020}~\cite{2020A&A...637A..38P}.

The amount of DM heating in an exoplanet depends in part on the flux of DM particles available to be captured. This is characterized by the phase-space distribution of DM particles, which provides a particular DM flux at a given location in the Galaxy.
The most widely adopted generalized Navarro-Frenk-White (gNFW) density profile (see Eq.~\ref{eqn:gNFW}) exhibits a 2D degeneracy between the scale radius $r_s$ and the inner slope $\gamma$~\cite{Karukes:2019jxv}. When the Galactocentric position is smaller than $r_s$, this profile approximates a power-law form (see App.~\ref{sec:appbias} for details). Since we are interested in the inner 1 kpc of the Galaxy and values of $r_s$ less than 10 kpc are disfavoured by observations of the Milky Way's rotation curve~\cite{2019JCAP...03..033B, 2021PDU....3200826B}, we adopt a power-law density profile given by
\begin{equation}
\rho_{\rm DM}(R; \alpha, C) = C\left(\frac{R}{1\,\rm kpc}\right)^{-\alpha},
\end{equation}
where $C$ denotes the DM density at a Galactocentric radius of 1 kpc. 

For the DM velocities, we assume that the DM particles follow an isotropic velocity profile described by a Maxwell-Boltzmann distribution, with an averaged velocity $\bar{v}_{\DM}(R)$ related to the velocity dispersion of DM particles $\sigma_{\DM}(R)$ as $\bar{v}_{\DM}(R)=\sqrt{8/(3\pi)}\,\sigma_{\DM}(R)$. The mean velocity and velocity dispersion of DM particles in the inner part of the Galaxy are not well constrained by observations and must be estimated from simulations (see $e.g.$ Ref.~\cite{2021JCAP...04..070B}). 
Under the assumption of an isothermal sphere, the DM velocity dispersion can be related to the circular velocity curve by $\sigma_{\DM}(r)=\sqrt{3/2}v_c(r)$. However, for simplicity we assume a constant value for the velocity dispersion. Specifically, we adopt $\sigma_{\DM}=100\,\rm km/s$, which is compatible with the circular velocities at a radius of $0.1\,{\rm kpc}-1\,\rm kpc$, obtained by combining the contribution of the stellar bulge, the stellar disk and the DM halo (see \textit{e.g.} Refs.~\cite{2021PDU....3200826B, 2023A&A...676A.134P}). This assumption can be refined in future investigations.

The total DM heat power $\Gamma_{{\rm heat}, i}^{\DM}$ in the exoplanet is given (in natural units) by~\cite{Leane:2020wob}
\begin{equation}
    \Gamma_{{\rm heat}, i}^{\DM} = f\pi R_{E, i}^2 \rho_{\DM}(R_i)\bar{v}_{\DM}\left(1+\frac{3}{2}\frac{v_{esc}^2(R_{E, i}, M_i)}{\sigma_{\DM}^2}\right),
\label{eq:heat_DM}
\end{equation}
where $f$ is the fraction of DM particles captured (and will depend on the particle physics model), $R_{E, i}$ is the radius of the exoplanet, $v_{{\rm esc}, i}$ is the escape velocity at its surface. We neglect the relative motion of the exoplanet, which is only a minor correction, and assume that the object is at rest with respect to the DM halo. For the radius $R_{E, i}$, we adopt $R_{\rm jup}$ for all exoplanets since the radius of exoplanets older than 500 Myr is roughly equal to that of Jupiter according to current evolutionary models (see $e.g.$ \cite{2008ApJ...689.1327S, 2020A&A...637A..38P, 2021arXiv210707434M}) and leave for future work the relaxation of this assumption. 

Under the assumption that there is no nearby host star ($i.e.$ we consider free-floating exoplanets or brown dwarfs), the total heat or luminosity of the exoplanet is then the sum of the contributions from DM heating and internal heat, given by
\begin{equation}
    \Gamma_{{\rm heat}, i}^{\rm int} + \Gamma_{{\rm heat}, i}^{\DM} = 4\pi {R_{E, i}}^2\sigma_{\rm SB}T_i^4 \epsilon_i,
\label{eq:heat_tot}
\end{equation}
where $\sigma_{\rm SB}$ is the Stefan-Boltzmann constant, and $\epsilon_i$ is the dimensionless emissivity of the given exoplanet. The emissivity is a measure of heat retention of the object; $\epsilon_i=1$ corresponds to a blackbody, while values $\epsilon_i<1$ are a greybody and can produce spectra peaked in more favorable wavelength regions for detection (see Ref.~\cite{Leane:2020wob} for an extended discussion of this point). As such, we take the simplifying yet conservative choice of a blackbody, \textit{i.e.} $\epsilon_i=1$. Note that here we have assumed only DM heating from contact interactions. Dark kinetic heating of exoplanets due to long-range forces can increase the signal sensitivity substantially~\cite{Acevedo:2024zkg}, but we do not consider this here.

\begin{figure}[t]
\includegraphics[width=\columnwidth]{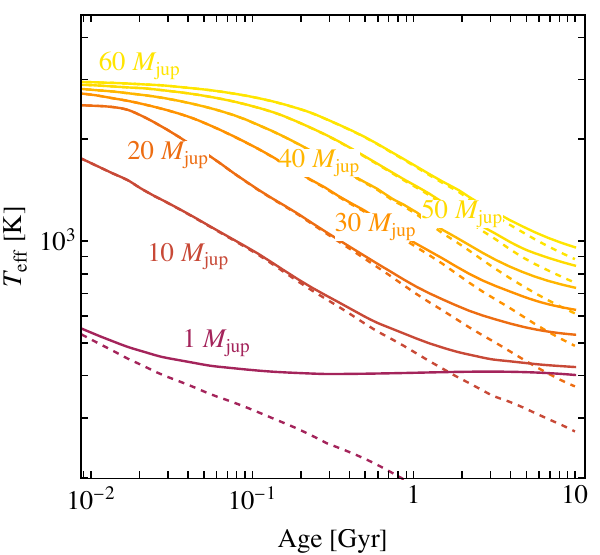}
\caption{Theoretical effective temperature for the evolution of Jupiters, Super-Jupiters, and Brown Dwarfs, as per Ref.~\cite{2020A&A...637A..38P}, with (solid) and without (dashed) DM heating. For the DM-heating lines, a Galactocentric distance of 0.1 kpc and a NFW DM profile with scale radius $r_s = 20\, \text{kpc}$, and DM velocity dispersion $\sigma_{\rm DM} = 100\text{ km/s}$, are assumed for illustration. Larger DM heating rates occur for greater inner slope values.
}
\label{fig:temps}
\end{figure}

Figure~\ref{fig:temps} shows expected exoplanet temperatures as a function of age, with and without DM, for a range of planetary masses. We have used an example NFW DM profile for the DM heating rates; even larger amounts of DM heating occur for greater inner slopes as per gNFW profiles. We see that DM heating is most striking when objects are old, or when objects are smaller such that backgrounds are comparably lower. Note however that this does not take into account telescope detectability, which can be easier to achieve for hotter objects which have higher luminosities. We therefore will focus on candidates with masses greater than 14~$M_{\rm jup}$ (sufficiently hot with DM heating to detect in the inner Galaxy), but not larger than 55~$M_{\rm jup}$ (backgrounds become too high relative to DM heating), which is the optimal detection range~\cite{Leane:2020wob}.

\begin{figure*}[t]
\includegraphics[scale=0.65]{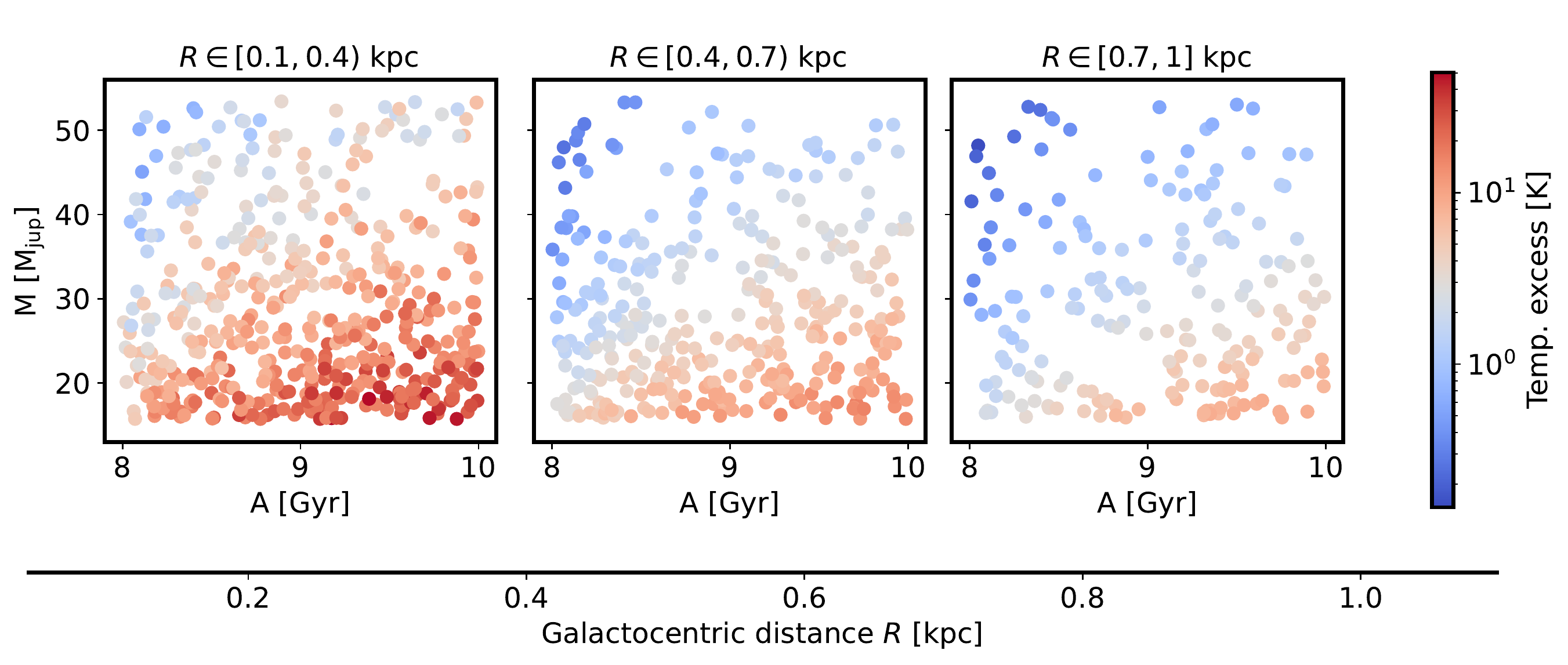}
\caption{Theoretical excess in the exoplanet effective temperature due to DM-heating for exoplanets located between Galactocentric distances of 0.1--0.4 kpc (left panel), 0.4--0.7 kpc (middle panel) and 0.7--1 kpc (right panel), as a function of exoplanet age and mass. We consider a sample of 1000 exoplanets under the assumption of a power-law DM density profile characterized by a normalization of $C=10\,\rm GeV/cm^3$ and an inner slope of $\alpha=1$.
}\label{fig:excess}
\end{figure*}

Figure~\ref{fig:excess} depicts the expected excess in exoplanet effective temperature, as a function of age and mass for exoplanets across three Galactocentric distance ranges: $[0.1, 0.4)\,\rm kpc$, $[0.4, 0.7)\,\rm kpc$ and $[0.7, 1]\,\rm kpc$. The decrease in exoplanet abundance moving away from the Galactic center is attributed to the postulation of an exponential radial number density profile for the exoplanet population (see Subsections~\ref{subsec:method_population} and \ref{subsec:methods_config} for details), and the scatter at a given age and mass is due to the distance mixing.


\subsection{Population Distribution of Exoplanets}\label{subsec:method_population}

For the population of exoplanets we assume independent distributions of $R_i$, $A_i$ and $M_i$ controlled by global parameters 
as explained below.
\begin{itemize}
    \item \textbf{Exoplanet Age:} We consider a uniform ($i.e.$ constant birthrate) age distribution from 8--10 Gyr, 
    \begin{equation}
    A_i \sim \mathcal{U}(8\,\text{Gyr}; 10\,\text{Gyr}),
    \end{equation}
    compatible with observations of an old population of Galactic bulge stars (see Ref.~\cite{2018ARA&A..56..223B, 2023ApJ...946...28J} and references therein).
    \item \textbf{Exoplanet Mass:} We use a Pareto distribution, namely 
    \begin{equation}
    M_i \sim \frac{\gamma_M-1}{M_{\rm min}}\left(\frac{M_i}{M_{\rm min}}\right)^{-\gamma_M},
    \end{equation}
    which describes a power-law initial mass function. As noted above, we set $M_{\rm min}=14\,\rm M_{jup}$ and truncate the distribution at $M_{\rm max}=55\,\rm M_{jup}$ to restrict the simulation within the optimal DM detection range~\cite{Leane:2020wob}. 
    \item \textbf{Exoplanet Spatial Distribution:} We assume that exoplanets follow the distribution of bulge stars, hence we employ an exponential distribution, given by
    \begin{equation}
    R_i \sim \gamma_Re^{-\gamma_R R_i}.
    \end{equation}
    Note that for $\gamma_R=1.46\,\rm kpc^{-1}$ this is equivalent to the exoplanets being spatially distributed according to the E2 bulge profile from Ref.~\cite{1997ApJ...477..163S} (see App.~\ref{app:spatial_distribution} for details). Since the closer to the centre of the Galaxy, the higher the expected temperature excess, we limit the simulation to $R_i\in[0.1, 1]$, with the lower boundary representing JWST's approximate minimum sensitivity threshold for exoplanet temperature measurement~\cite{Leane:2020wob}.
\end{itemize}

\subsection{Priors}

The priors on the population variables are chosen to encompass observational constraints. Specifically, the population parameter $\gamma_M$ is inferred to be $0.6\pm0.1$~\cite{2021ApJS..253....7K} based on a sample of exoplanets located within 20 pc from the Sun. This value is consistent with other estimates in the literature, e.g.~\cite{2017MNRAS.471.3699M}. Conversely, $\gamma_R$ is assigned a uniform distribution that accounts for the uncertain shape of the distribution of stars and sub-stellar objects in the central region of our Galaxy. Mathematically, this is represented as:
\begin{align}
\begin{split}
    \gamma_M &\sim \mathcal{N}(0.6, 0.1^2)\textrm{ and}\\
    \gamma_R &\sim \mathcal{U}(1, 2).
\end{split}
\end{align}
Finally, we adopt uniform priors for the DM halo parameters within the ranges $0<\alpha<3$ and $0.5 < C[{\rm GeV/cm^3}] < 40$. 

\subsection{Simulator Configuration} \label{subsec:methods_config}
In the absence of real data and in order to verify the inference procedure, we analyze mock data. Herein, we justify the selected configuration of global parameters to generate the simulations and outline potential improvements for future investigations. 
\begin{itemize}
\item Regarding exoplanet ages, we opted for an old population with uniform ages spanning from 8 to 10 Gyr. Given the ongoing debate surrounding the age distribution of stars in the Galactic bulge (see Refs.~\cite{2018ARA&A..56..223B, 2023ApJ...946...28J} and references therein), we defer accounting for intermediate-aged stars in the Galactic bulge ($e.g.$, \cite{2017A&A...605A..89B}) and possible contamination from disc stars to future work.
\item Exoplanets are set to have a power-law initial mass function of $\gamma_M=0.6$, consistent with findings in the literature (e.g.~\cite{2021ApJS..253....7K, 2017MNRAS.471.3699M}). 
\item Furthermore, exoplanets are spatially distributed according to the E2 bulge profile from Ref.~\cite{1997ApJ...477..163S} (see App.~\ref{app:spatial_distribution} for details), setting $\gamma_R=1.46\,\rm kpc^{-1}$.
\item For global DM parameters, we simulate datasets from true underlying values of $\alpha$ drawn from its prior range and two settings for the normalization constant, specifically $C=6\,\rm GeV/cm^3$ and $C=20\,\rm GeV/cm^3$. The former value is adopted to reproduce the normalization of the DM density as in the NFW profile with scale radius of $20~\rm kpc$ and local DM density of $\rho_0=0.42\,\rm GeV/cm^3$~\cite{Karukes:2019jxv, 2021PDU....3200826B}, while the latter value could account for an increase of the DM density due to, for example, adiabatic contraction~\cite{1986ApJ...301...27B}. 
\end{itemize}
Furthermore, we adopt relative measurement uncertainties, denoted by $\sigma$, across a wide range to investigate the impact of noise levels. In particular, $\sigma$ is fixed to values within $[1, 20]\,\%$, where the optimistic (but unrealistic in the near future) value of 1\% allows verification of the inference procedure, while a more realistic albeit still optimistic scenario is 20\% uncertainty~\cite{Zhu2016}. Additionally, we vary the number of observed exoplanets $N$ from 10 to 200. In this first study, we do not consider selection effects and focus on the construction and validation of an idealised simulator. Modelling selection effects in a Bayesian hierarchical model framework is computationally demanding, as it usually involves a selection probability correction factor that depends in a non-factorizable way on the parameters of interest, and can only be computed analytically if a number of assumptions about selection probability are met~\cite{Kelly:2007jy}. The next step will be to investigate selection effects with, for example, Simulation-Based Inference (SBI) methods, which offer more flexibility and ease of implementation in modelling selection processes.

\subsection{Inference}

To sample the high-dimensional posterior over all 3N latent and 4 global parameters, we employ the No-U-turn sampler (NUTS)~\cite{2011arXiv1111.4246H}, a self-tuning variant of Hamiltonian Monte Carlo (HMC)~\cite{2011hmcm.book..113N}, as implemented in the Pyro~\cite{2018arXiv181009538B} probabilistic programming language.
After 1400 adaptation and warmup steps, we run a chain of 3200 samples (with near-unity autocorrelation length due to the efficiency of NUTS), which are enough to adequately explore the posterior distribution. Inference took around 4 minutes (25 minutes) per data set with 10 (100) objects on a single CPU. 

\section{Results}
\label{sec:results}

\subsection{Global Parameter Inference}\label{subsec:results_global}
\begin{figure*}[t]
\includegraphics[scale=0.5]{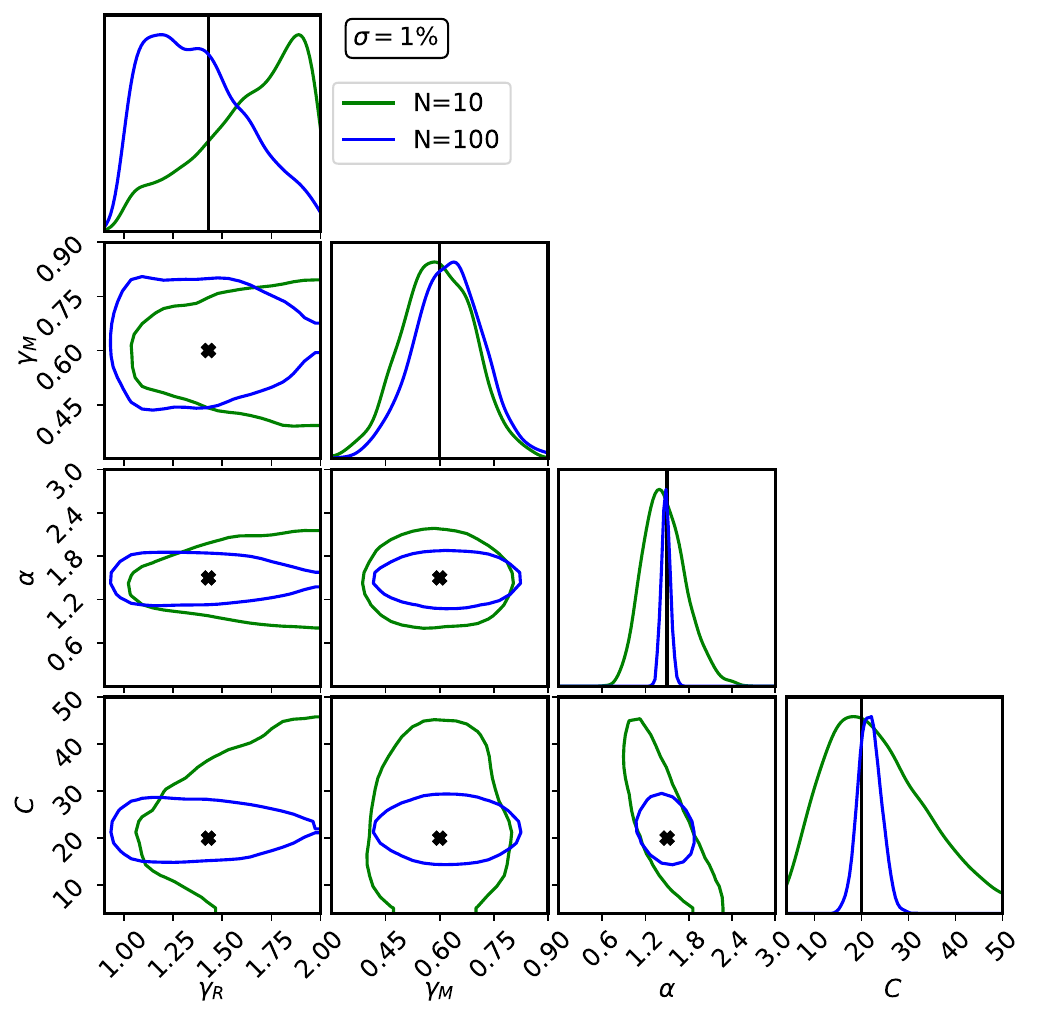}
\includegraphics[scale=0.5]{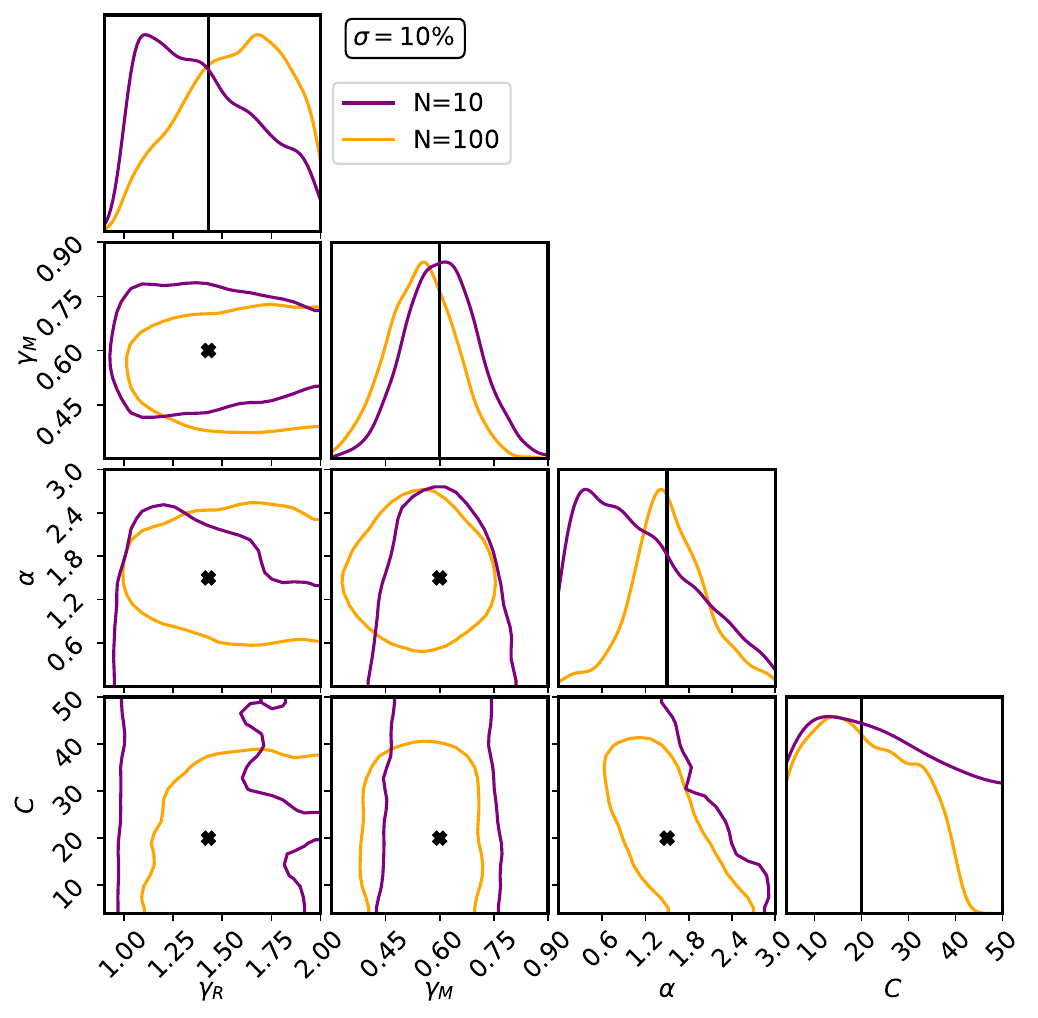}
\caption{Posterior marginalized 2D and 1D distributions for the exoplanet population parameters and DM halo density profile from analyses of mock data (true parameter values indicated by crosses / vertical lines in the 2D / 1D plots) with different noise levels and number of detected objects as indicated in the legends. The 2D contours depict the 2$\sigma$ highest posterior density (HPD) regions (\textit{i.e.} with 86\% credibility).
}\label{fig:results_corner}
\end{figure*}

\begin{figure*}[t]
\includegraphics[scale=0.55]{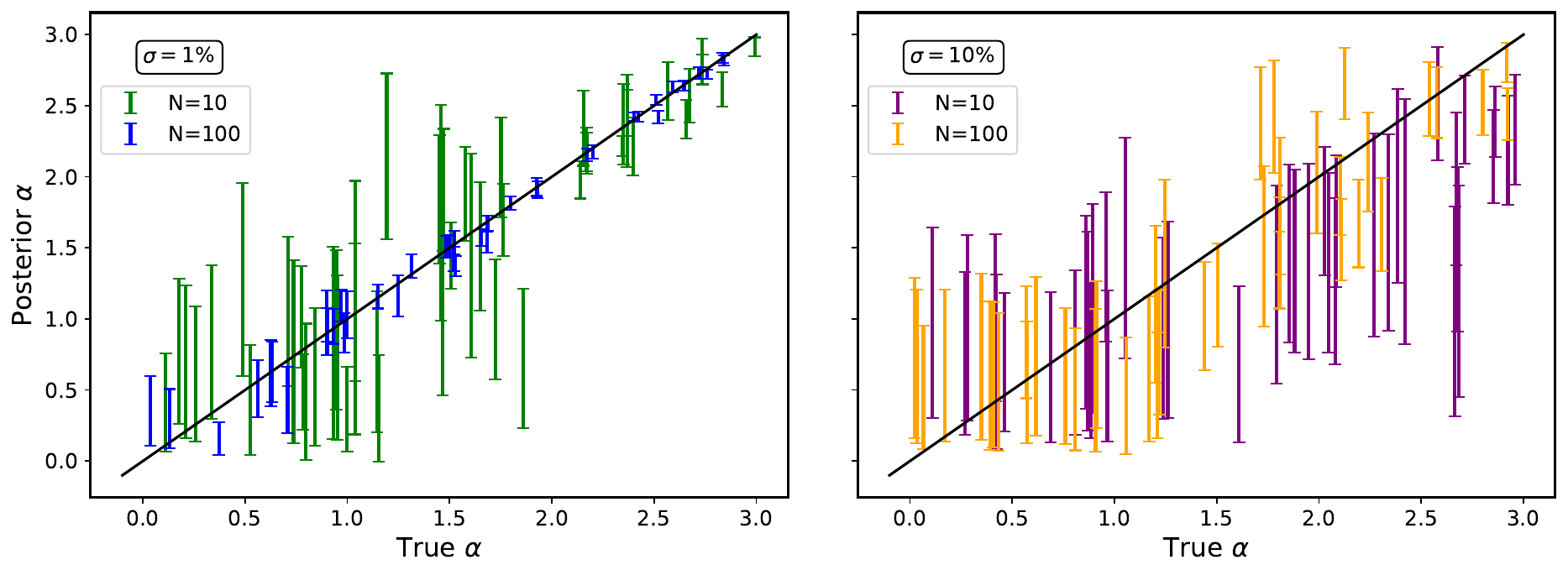}
\caption{Marginal central 68\% credible intervals for the DM density slope $\alpha$, inferred from mock data with true simulated $\alpha$ values drawn from the prior distribution and plotted on the abscissa. The legends indicate the number of exoplanets analysed. As in Fig.~\ref{fig:results_corner}, in the left and right panels the noise levels are fixed to $1\%$ and $10\%$, respectively. In the simulation, $C$, $\gamma_R$ and $\gamma_M$ were fixed to $20\,\rm GeV/cm^3$, $1.43$ and $0.6$ respectively and then inferred jointly with $\alpha$ and the latent variables (exactly as in Fig.~\ref{fig:results_corner}).
}\label{fig:results_posterior}
\end{figure*}

Figure~\ref{fig:results_corner} shows example joint posteriors for the global (DM halo and exoplanet population) parameters obtained from mock data sets comprising 10 or 100 exoplanets, with noise levels set to 1\% or 10\%. This figure illustrates the increase in precision obtained with noise level reduced to 1\% or, similarly, with sample size increased to 100 objects. Regarding the population parameters, the inference is dominated by the initial prior on the slope of the initial mass function (IMF) power law denoted as $\gamma_M$, since it is small compared to the expected constraints of 100 objects or less. On the contrary, for the exponential slope of the number density of exoplanets, $\gamma_R$, the prior distribution is broad compared to the expected constraints of 100 or fewer objects (even with zero uncertainty)\footnote{\label{foot:gamma_R}The trivial estimator $\hat{\gamma}_R = \sum \hat{R}_i / N$ has variance $1/(N\gamma_R^2)$ (if the observational noise is negligible), which amounts to about 10\% of the prior range for $\gamma_R$ and is an indicator for the optimal precision in inferring it. However, the fact that we are restricting $R\in[0.1, 1]\,\rm kpc$ further broadens the $\gamma_R$, and so the $\pm0.1$ precision is not achieved even for 1\% observational uncertainty.}.

Figure~\ref{fig:results_posterior} focuses on the marginal results (under the configurations of $N$ and $\sigma$ indicated in the legend) for the inner DM profile slope $\alpha$. Here we plot central 68\% credible intervals, defined as the 16th and 84th percentiles of the 1D marginal posterior, versus the true\footnote{Here and elsewhere, we use the word `true' to indicate the parameter values adopted in generating the data.} $\alpha$ values used to simulate the analysed data sets. 
From this figure, the increase in precision with increasing number of observed exoplanets and decreasing noise level is also observed. Additionally, we see that larger values of the slope can be recovered with greater precision. This is expected, since for a fixed normalisation $C$, the DM density increases with increasing $\alpha$ value.

\begin{figure*}[t]
\includegraphics[scale=0.6]{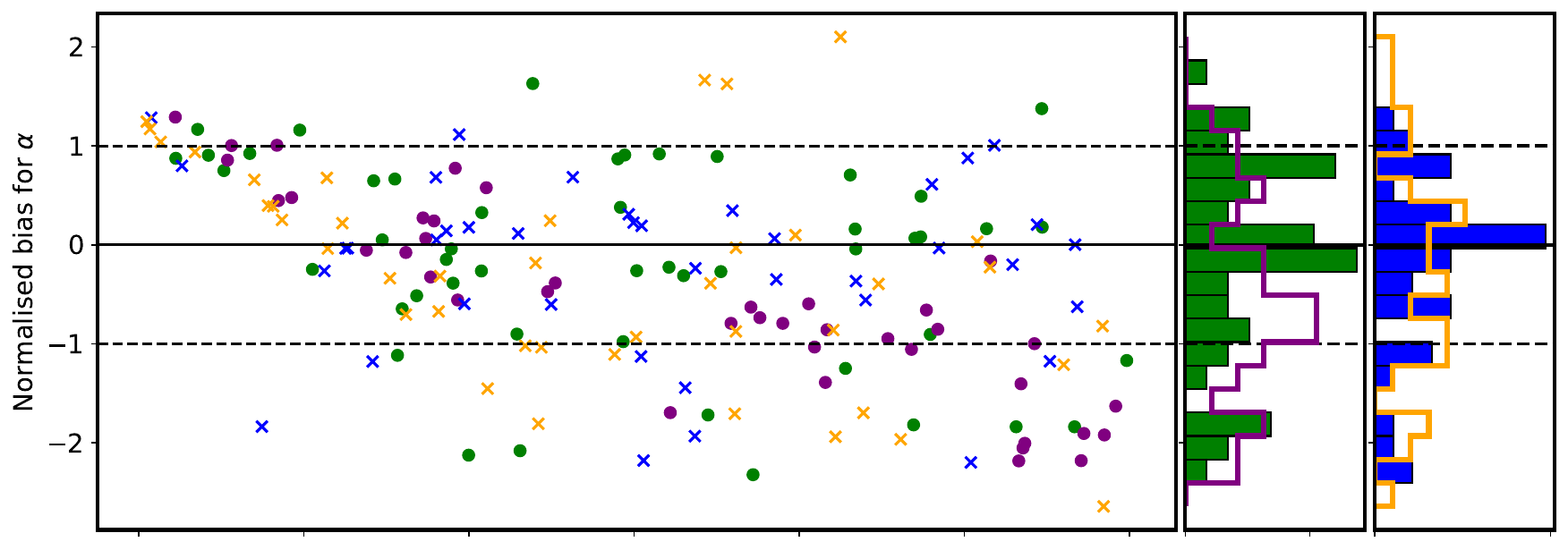}
\includegraphics[scale=0.6]{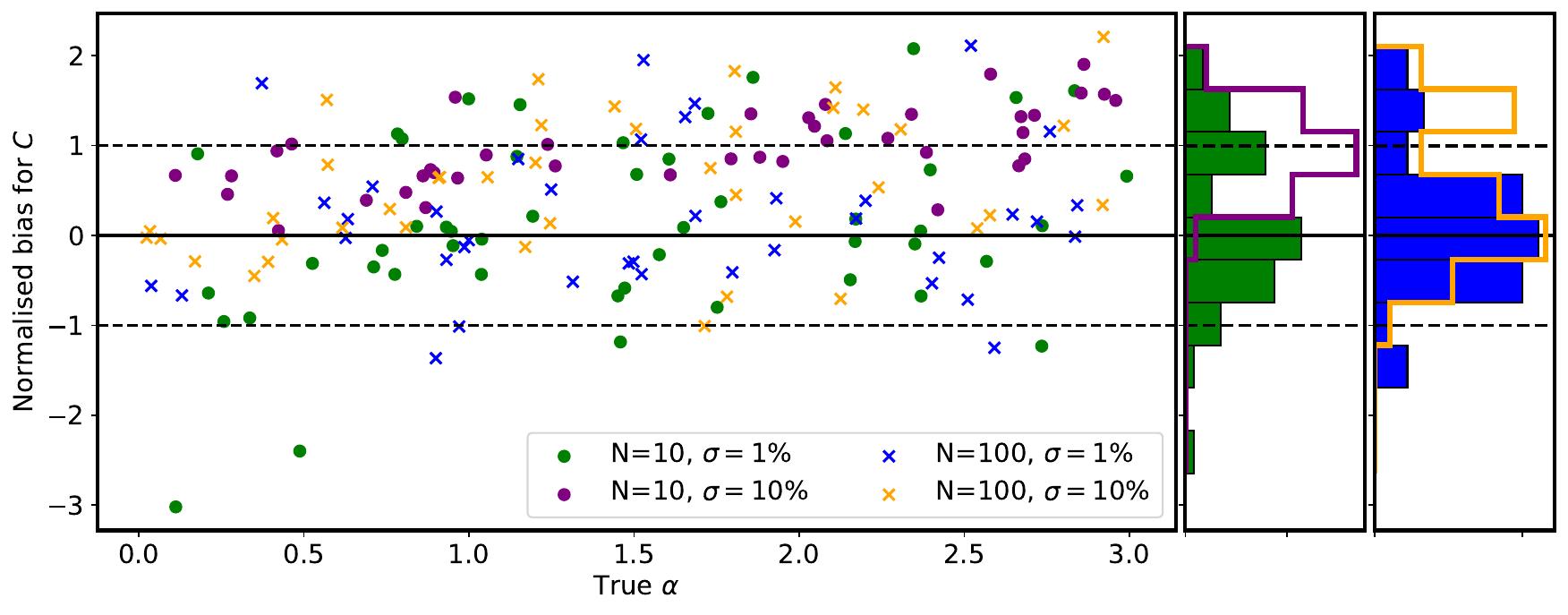}
\caption{Normalized bias for the DM density profile parameters $\alpha$ (top) and $C$ (bottom) as inferred for a given $(N, \sigma)$ configuration using mock data with true values drawn from their priors. The left panels display the true parameter value on the abscissa, while the right panels depict histograms of the bias across the different $(N, \sigma)$ configurations. 
}\label{fig:results_bias}
\end{figure*}

Figure~\ref{fig:results_bias} displays the normalized bias of inferring $\alpha$ and $C$ as measured by the offset of the posterior mean from the true value divided by the standard deviation of the marginal posterior, \textit{i.e.}
\begin{equation} \label{eq:norm_bias}
\textrm{Normalized bias}=\frac{\langle X\rangle -X_{\rm true}}{\sqrt{\langle X^2\rangle - \langle X\rangle^2}},
\end{equation}
where $X \in \{\alpha, C\}$, and inferred from mock data with true parameter values drawn from their respective priors. We observe a systematic shift of the posteriors towards the middle of the prior range when the true values of $\alpha$ are at the extremes of this range, particularly for larger noise level and smaller number of exoplanets. This is a reflection of a statistical effect called `shrinkage', which in the context of the Bayesian hierarchical model employed here `shrinks' posterior estimates towards the prior mean when the data are less constraining.  
These figures also reveal a subtle pattern in which the mean estimate of $\alpha$ tends to be biased low, while that of $C$ tends to be biased high, explained by the anticorrelation between these parameters (see Fig.~\ref{fig:results_corner}). This is, in turn, driven by the  intrinsic scatter in temperatures at a given Galactocentric distance due to the scatter in exoplanet masses and ages which produces an irreducible variance in the distribution of their values. 

\begin{figure}[h!]
\includegraphics[scale=0.45]{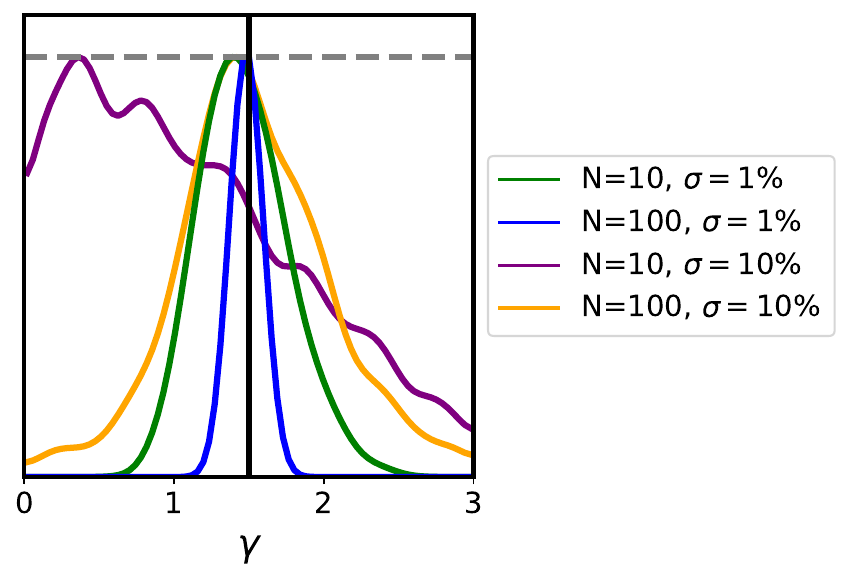}
\caption{1D marginalized posterior for the inner slope of a gNFW density profile, derived from the analyses of the mock data in Figure~\ref{fig:results_corner}. The colour scheme, denoting varying noise levels and number of detected objects, matches that of the reference figure. The grey, dashed line depicts the prior, while the vertical black line is the ground truth.
}\label{fig:posterior_rs}
\end{figure}

Figure~\ref{fig:posterior_rs} illustrates how the inferred posterior distributions, depicted in Fig.~\ref{fig:results_corner}, translate into the marginal posterior for the inner slope $\gamma$ of a gNFW profile (via Eq.~\eqref{eqn:gNFW}). The transformation into a gNFW assumes equal inner steepness in both profiles ($\alpha=\gamma$), a local DM density of $\rho_0=0.43\pm0.02\,\rm GeV/cm^3$~\cite{Karukes:2019jxv, 2021PDU....3200826B}, and equivalence in density at 1 kpc from the Galactic center. As expected, with increasing numbers of objects and smaller observational error the posterior distribution for $\gamma$ becomes increasingly concentrated around the true value. As for the scale radius $r_s$ of the gNFW profile, our choice of priors translates into an informative prior on $r_s$, which cannot anyway be meaningfully constrained using only data in the inner kpc and a single value for the DM density at the solar location. As a consequence of the informative prior and the weak constraining power of the data, $r_s$ cannot be strongly constrained in this setup.

\begin{figure*}[t]
\includegraphics[scale=0.42]{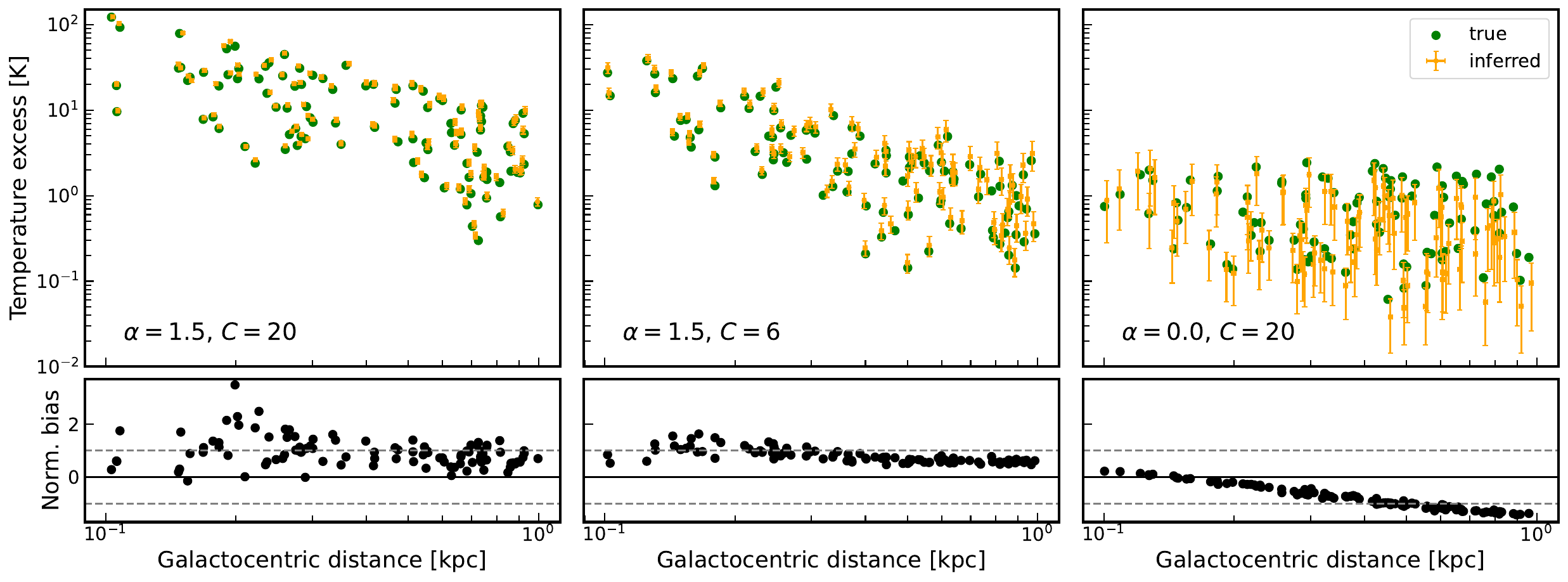}
\caption{Inferred (yellow, with $1\sigma$ posterior ranges) and true (green) temperature excesses as a function of inferred (with $1\sigma$ posterior ranges) and true Galactocentric distance, respectively, for mock observations of $N=100$ exoplanets generated assuming different DM halo shapes and a relative uncertainty in exoplanet parameters of $1\%$. The bottom panels show the normalized bias. Note that errorbars in the inferred Galactocentric distance are tiny due to the low noise level of 1\% and are consequently not discernible.
}
\label{fig:results_N100_excess}
\end{figure*}

\subsection{Latent Parameter Inference}\label{subsec:results_latent}

Our Bayesian hierarchical model allows us to further infer the age, mass, and Galactocentric distance of each exoplanet in a given simulated data set, along with the population and DM halo parameters. We leave for App.~\ref{sec:app_latent} a more thorough examination of the marginal posteriors of the latent parameters. 

Figure \ref{fig:results_N100_excess} compares the true and inferred temperature excesses for simulated datasets with $100$ exoplanets, noise level $\sigma=1\%$ and different shapes of the DM halo as indicated in the text insets. The normalized residuals, given by Eq.~\eqref{eq:norm_bias}, for each DM halo configuration is shown in the bottom panels of this figure. 
The patterns observed in the residuals in the bottom panels of Fig.~\ref{fig:results_N100_excess} can be explained in the context of hierarchical Bayesian modelling by the trends in the inference of global parameters, as described in the previous section, namely the shrinkage of posterior estimates and the 2D degeneracy between $\alpha$ and $C$.

\subsection{Detection}\label{subsec:results_detection}

\newcommand{\ttDM}{\text{\texttt{DM}}}
\newcommand{\ttnoDM}{\text{\texttt{no-DM}}}

Our goal is to detect annihilating DM particles through their impact on the temperature rise of a \emph{population} of exoplanets. A global indication of this signal would be an increase in the temperature of exoplanets with decreasing Galactocentric distance, as the expected excess in effective temperature scales with the DM density in the exoplanet's surrounding. Note however that there is a limitation to the lowest exoplanet temperature that can be measured at large distances. In Ref.~\cite{Leane:2020wob}, it was estimated that JWST could observe exoplanets with Jupiter's radius down to about 650 K. However, a number of simplifying assumptions were made in obtaining this number, including observation time, and that the thermal emission is a blackbody; considering more accurate spectra can produce higher signals in telescope bins which are more detectable~\cite{Leane:2020wob}. Furthermore, other telescopes such as Roman may be more optimal for this search. We therefore do not implement this temperature cut in this work, and while including a cut can reduce signal sensitivity, we defer to future work for a more detailed analysis of this issue.

In the hierarchical Bayesian framework we adopt throughout, the ``detection significance'' can be quantified by the Bayes factor, \textit{i.e.} the ratio of Bayesian evidences\footnote{Adopting equal \emph{model} priors (\textit{i.e.}\ prior beliefs in the existence/non-existence of DM) makes the Bayes factor equal to the ratio of the two models' posterior probabilities.}:
\begin{equation} \label{eq:bayes_ratio}
B=\frac{p(\boldsymbol{d}| \ttDM)}{p(\boldsymbol{d}| \ttnoDM)}.
\end{equation}
Here \ttnoDM{} indicates a model that only considers the standard exoplanet evolution, while \ttDM{} includes the contribution of DM heating. It should be noted that \ttnoDM{} is \emph{nested} within \ttDM: it corresponds to setting $C=0$.
In turn, the Bayesian evidence for each model is the average likelihood of all its global and latent parameters, distributed according to the hierarchical prior. Since the number of parameters scales with $N$, the high-dimensional integral this requires prohibits an exact computation even for as few as 10 exoplanets. 

\begin{figure*}[t]
\includegraphics[scale=0.65]{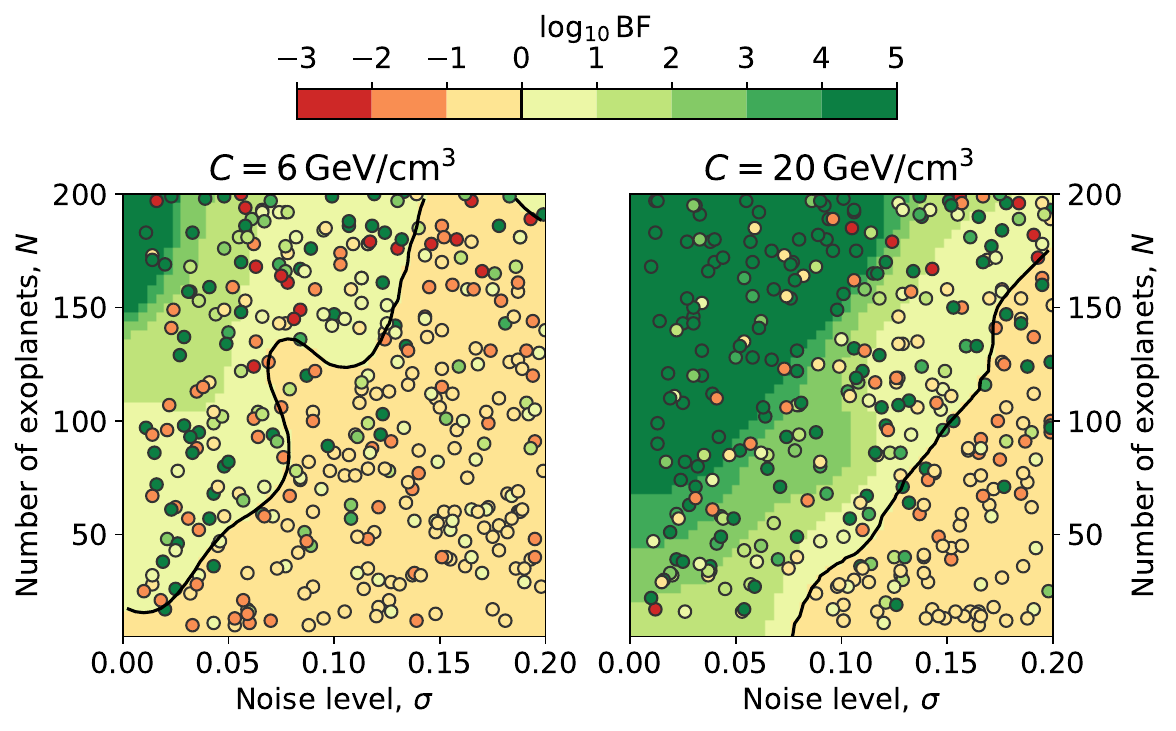}
\caption{Data realizations (dots) are colour-coded according to their approximate Bayes factor, equal to the posterior odds ratio between the \ttDM\ and \ttnoDM\ models,  calculated for simulated observations from the \ttDM{} model with varying numbers of exoplanets and noise levels, $\alpha$ drawn from the prior, and fiducial $\gamma_R=1.43$ and $\gamma_M=0.6$. A log Bayes factor larger than 0 indicates (correctly) a preference for the \ttDM\ model. The left (right) panel assumes a lower (larger) DM density.  The background shading illustrates the median trend of Bayes factors, with the black line indicating the locus of equal posterior odds.}\label{fig:results_bayes} 
\end{figure*}

Instead, we employ a simple approximation scheme for the required evidences, which makes use of the Hamiltonian Monte Carlo posterior inference described above. For a given data set, we first run the Hamiltonian Monte Carlo with each model (labelled $M$). From each set of posterior samples, we then derive a Gaussian posterior \emph{density} estimate centred on the maximum-a-posteriori (MAP) value, $\theta_0$, and with a covariance matrix matching the empirical covariance of the samples. This allows us to approximate $p(\theta_0|\mathbf{d}, M)$; dividing it by the joint probability at $\theta_0$, given model $M$, $p(\theta_0, \mathbf{d} | M) = p(\mathbf{d}|\theta_0, M) p(\theta_0 | M)$ (which is already calculated and recorded in the Hamiltonian Monte Carlo run), gives (an approximation for) the evidence $p(\mathbf{d} | M)$ by virtue of Bayes' theorem.
This approximation allows us to explore via analyses of mock data the expected ``detectability'' of DM under different data set configurations: number of observed objects, $N$, and noise level, $\sigma$, and for different halo parameters: slope, $\alpha$, and normalisation, $C$.

Figure~\ref{fig:results_bayes} illustrates the effect of these four parameters. 
Both panels of the figure show the mean of the Bayes factors within 2D bins of ($N$, $\sigma$), calculated from data set realizations with different $N \in [5, 200]$ and $\sigma \in [1/\%, 20/\%]$, drawn uniformly from their range. Furthermore, the true simulated values of $\alpha$ are drawn from its prior distribution, while $\gamma_R$ and $\gamma_M$ are fixed to $1.43$ and $0.6$, respectively. Finally, the two panels present the results with $C$ fixed at $6\,\rm GeV/cm^3$ (left) and $20\,\rm GeV/cm^3$ (right). As expected, the figure shows that the feasibility of detecting DM-overheated exoplanets -- $i.e.$, values of the log Bayes factor larger than 0 -- increases with increasing number of observed objects and DM density, and decreasing uncertainty, $i.e.$, towards the top left of each panel. 
It is also shown that for small $N$ and large noise level, there is a large scatter of Bayes factors at fixed $(\alpha, C)$ that can be explained by  whether we observe the less massive and older exoplanets close to the Galactic centre. For example, for a sample dataset of about 10 exoplanets, there is the potential to detect the signal when the density slope $\alpha$ exceeds about 1, along with $C=6\,\rm GeV/cm^2$, a normalization constant in agreement with rotation curve constraints.

\section{Conclusions and Outlook}
\label{sec:conclusion}

We have demonstrated the application of Bayesian hierarchical modelling of overheated exoplanets with the goal of detecting and characterising dark matter and applied it to simulated data sets representing realisitic exoplanet populations within the inner kpc of the Galaxy. Considering different configurations of sample size and uncertainty level, our study has demonstrated that it is possible to achieve high accuracy in inferring the shape of the DM halo, both in the inner slope and in the normalisation. In the case of 1\% measurement uncertainties, observations of a sample of about 10 exoplanets can be sufficient for this task, depending on the underlying DM density at a Galactocentric distance of 1 kpc and the inner DM slope. Despite the potential variability, the detection of DM-overheated exoplanets within 1 kpc of our Galaxy remains feasible for data sets containing about 100 exoplanets and a DM density profile in agreement with extrapolations of the inference of the DM density profile in our Galaxy via the stellar disc rotation curve. However, more realistic but still very optimistic measurement uncertainties of about 10\% require around 100 exoplanets for DM parameter inference. To optimise detection strategies, follow-up observations with multiple telescopes should prioritise accurate measurements of the mass, age and Galactocentric distance of these exoplanets

While our model correctly infers both global and per-object latent parameters, an application to real data requires the inclusion of observational selection effects. A viable solution for this is simulation-based inference (SBI), which offers a straightforward approach to handling arbitrarily complex probabilistic models both for parameter inference and model selection. Therefore, integrating the Bayesian hierarchical model developed here into an SBI framework represents a logical progression in advancing this novel and promising approach to indirectly search for DM particles through overheated exoplanets.

\section*{Acknowledgments}
We thank Bruce Macintosh, Rain Kipper, Elmo Tempel and Peeter Tenjes for helpful discussions and comments. RKL was supported in part by the U.S. Department of Energy under Contract DE-AC02-76SF00515.
This work was supported by the Estonian Research Council grants PRG1006, PRG803 and PSG938, and partially supported by the ETAg CoE grant “Foundations of the Universe" (TK202) and University of Tartu ASTRA project 2014-2020.4.01.16-0029 KOMEET. RT acknowledges co-funding from Next Generation EU, in the context of the National Recovery and Resilience Plan, Investment PE1 – Project FAIR ``Future Artificial Intelligence Research''. This resource was co-financed by the Next Generation EU [DM 1555 del 11.10.22]. RT is partially supported by the Fondazione ICSC, Spoke 3 ``Astrophysics and Cosmos Observations'', Piano Nazionale di Ripresa e Resilienza Project ID CN00000013 ``Italian Research Center on High-Performance Computing, Big Data and Quantum Computing'' funded by MUR Missione 4 Componente 2 Investimento 1.4: Potenziamento strutture di ricerca e creazione di ``campioni nazionali di R\&S (M4C2-19 )'' - Next Generation EU (NGEU). This article/publication is based upon work from COST Action COSMIC WISPers CA21106, supported by COST (European Cooperation in Science and Technology)

\appendix

\section{Spatial Distribution}
\label{app:spatial_distribution}
The spatial distribution of exoplanets is assumed to follow the E2 bulge profile from \cite{1997ApJ...477..163S}. Thus, the number density profile, given in arbitrary units, takes the form\footnote{We adopt Galactocentric spherical coordinates with $\phi$ increasing in the direction of Galactic rotation, i.e. clock-wise as seen from the North Galactic Pole (NGP), and where $\theta$ is measured from the NGP.}
\begin{equation}
    n(r, \phi, \theta) = n_0\exp{\left[-\gamma_R(\theta, \phi)r\right]}  
\end{equation}
with 
\begin{equation}
\scalebox{0.8}{$
   \gamma_R(\theta, \phi)=\sqrt{\sin^2{\theta}\left[\left(\frac{\cos{(\phi+\alpha)}}{x_0}\right)^2 + \left(\frac{\sin{(\phi+\alpha)}}{y_0}\right)^2\right] + \left(\frac{\cos{\theta}}{z_0}\right)^2},
$}
\end{equation}
$x_0=\SI{0.899}{kpc}$, $y_0=\SI{0.386}{kpc}$, $z_0=\SI{0.250}{kpc}$, and $\alpha=23.8^{\rm o}$ is the angle between the bulges's major axis and the Sun-Galactic center line. The scale radius and scale length of the disc are $\SI{2.15}{kpc}$ and $\SI{0.40}{kpc}$, respectively\footnote{Scale parameters in the planed defined by $\theta=\pi/2$ (i.e. $xy$-plane), which are taken from observational studies, are re-scaled as $R_0$.}.

We are only interested in the distances of exoplanets to the Galactic center,
rather than in their three-dimensional position. Is it for this reason that we get rid of $\theta$ and $\phi$ by integrating $n(r, \phi, \theta)$ over these coordinates. The integration must be done over the region covered by the observations. However, for simplicity, we have assumed $\phi=0$ and $\theta=\pi/2$, which gives a value of $\gamma_R=1.46\,\rm kpc^{-1}$. 

\section{Power-law Bias}
\label{sec:appbias}

As discussed in the main text for radii that are significantly smaller than the scale radius of the profile $r_s$, we can expand the gNFW equation
\begin{equation}\label{eqn:gNFW}
 \rho_\mathrm{DM}(R;\gamma,r_s,\rho_0)=\rho_0\left(\frac{R_0}{R}\right)^{\gamma}\left(\frac{r_s+R_0}{r_s+R}\right)^{3-\gamma},
\end{equation}
which yields a power law expression 
\begin{equation}\label{eq:power_law}
    \rho_{\rm DM}(R; C, \gamma) \approx C\,R^{-\gamma} \quad \text{for } R \ll r_s.
\end{equation}
As discussed, this helps to simplify the fitting procedure and breaks the degeneracy between the scale radius and the inner steepness parameter $\gamma$. 
However, this procedure introduces a systematic bias to slightly larger $\gamma$ values. A simple shape analysis shows that comparing the shape of the full expression in Eq.~\ref{eqn:gNFW} and the expanded expression in Eq.~\ref{eq:power_law} leads to very similar graphs for slightly different scale parameters $\gamma$, modulo a total normalization correction. 

Table~\ref{tab:bias} shows several values for the $\gamma$ parameters that lead to identical inner slopes for the DM profile. 
We can make two observations. One, is that the bias when using the power law leads to systematically larger values for $\gamma$. The second is,
that the bias can be somewhat substantial for $\gamma$ values smaller than one, and vanishes entirely for steeper profiles, when $\gamma$ approaches two. In the intermediate regime, that corresponds to the standard 
NFW profile, the bias is about $10 \%$. We conclude that the induced bias is well below the experimental accuracy that we can hope to archive in the near and intermediate future, and thus the expansion prescription is a valid method for the $\gamma$ factor reconstruction.  
\begin{table}
  \centering
  \begin{tabular}{|c|c|c|}
    \hline
      $\gamma$ from full gNFW & $\gamma$ from Power Law & Induced bias \\
    \hline
    0.5 & 0.57 & $+14 \%$ \\
    \hline
    1.0 & 1.1 & $+10 \%$ \\
    \hline
     1.2 & 1.3 & $+8 \%$ \\
    \hline
   1.5 & 1.6 & $+6 \%$ \\
    \hline
    1.8 & 1.8 & $0 \%$ \\
    \hline
  \end{tabular}
  \caption{The induced bias for several benchmark values of the DM profile parameter $\gamma$.}
  \label{tab:bias}
\end{table}


\section{Closer Examination of Latent Parameter Inference}
\label{sec:app_latent}

\begin{figure}[t]
\includegraphics[scale=0.5]{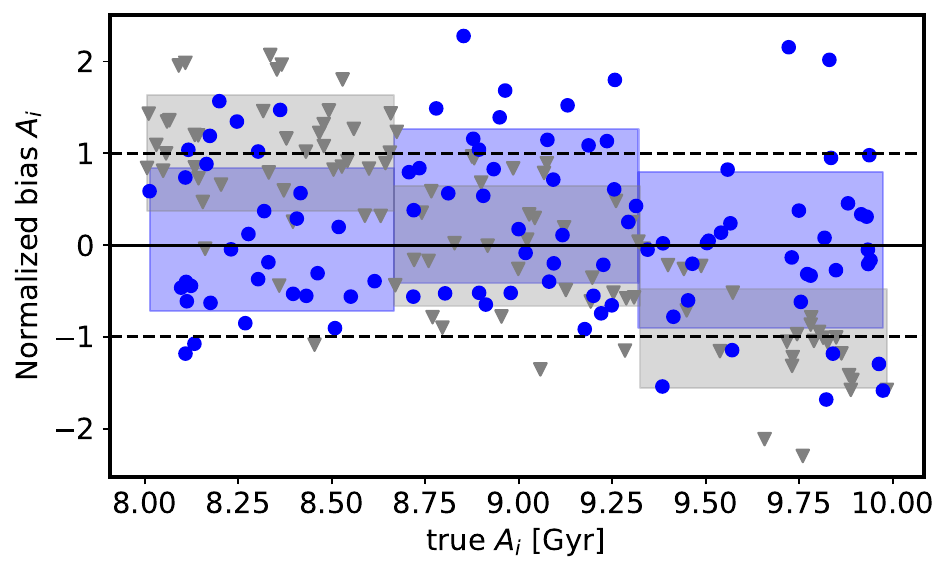}
\includegraphics[scale=0.5]{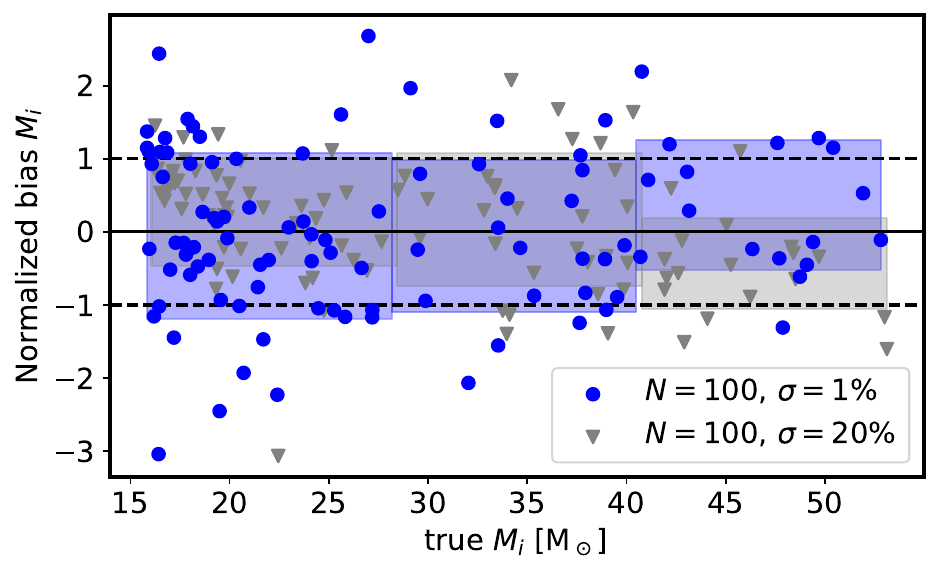}
\includegraphics[scale=0.5]{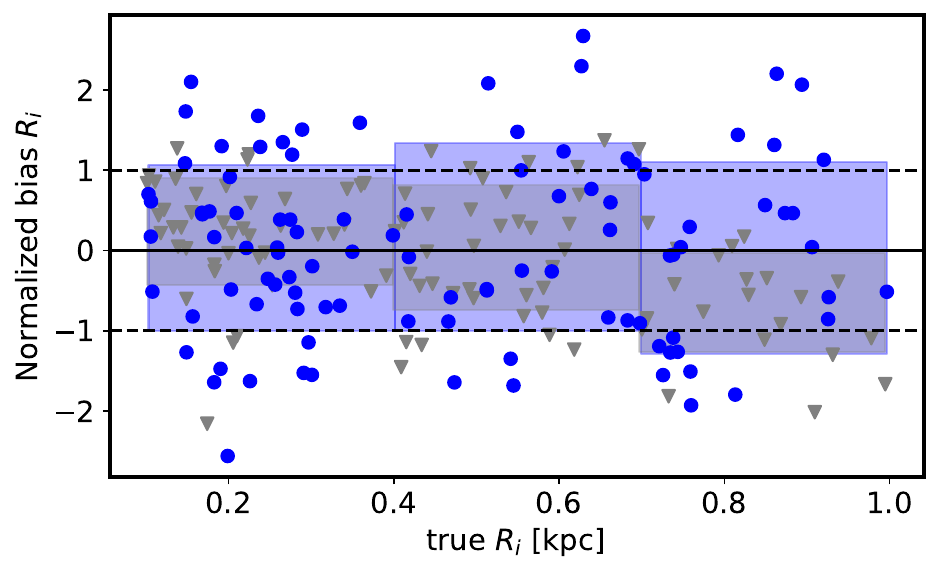}
\caption{Normalized bias for the latent ages (top), masses (middle) and Galactocentric distance (bottom) inferred, along with the global parameters, from a mock observation with 100 exoplanets and $\sigma$ as indicated in the legend. In the simulation, $\alpha$, $C$, $\gamma_R$ and $\gamma_M$ were fixed to $1.5$, $20\,\rm GeV/cm^3$, $1.43$ and $0.6$, respectively. The shaded bands indicate the standard deviation computed within the top-hat bin shown.}\label{fig:bias_latents}
\end{figure}

For a given mock dataset, the accuracy in the inferred age, mass and Galactocentric distance, as measured by the normalized bias, is plotted in the top, middle and bottom panels, respectively, of Figure~\ref{fig:bias_latents}, as a function of the true values of these latent parameters. The true latent parameter space has been divided into three equal-width intervals, with color bands within each interval representing the bias region encompassing 1$\sigma$ of the points is shown to ease visualization. These bands highlight the effect of `shrinkage' discussed in the text, particularly notable at high noise levels, on latent parameter inference (most notably for $A$), demonstrating a tendency for the marginal posteriors to be shrunk, in virtue of the Bayesian hierarchical model, towards the mean of the prior range. 

\newpage 

\bibliography{main.bib}

\begin{thebibliography}{91}%
\makeatletter
\providecommand \@ifxundefined [1]{%
 \@ifx{#1\undefined}
}%
\providecommand \@ifnum [1]{%
 \ifnum #1\expandafter \@firstoftwo
 \else \expandafter \@secondoftwo
 \fi
}%
\providecommand \@ifx [1]{%
 \ifx #1\expandafter \@firstoftwo
 \else \expandafter \@secondoftwo
 \fi
}%
\providecommand \natexlab [1]{#1}%
\providecommand \enquote  [1]{``#1''}%
\providecommand \bibnamefont  [1]{#1}%
\providecommand \bibfnamefont [1]{#1}%
\providecommand \citenamefont [1]{#1}%
\providecommand \href@noop [0]{\@secondoftwo}%
\providecommand \href [0]{\begingroup \@sanitize@url \@href}%
\providecommand \@href[1]{\@@startlink{#1}\@@href}%
\providecommand \@@href[1]{\endgroup#1\@@endlink}%
\providecommand \@sanitize@url [0]{\catcode `\\12\catcode `\$12\catcode
  `\&12\catcode `\#12\catcode `\^12\catcode `\_12\catcode `\%12\relax}%
\providecommand \@@startlink[1]{}%
\providecommand \@@endlink[0]{}%
\providecommand \url  [0]{\begingroup\@sanitize@url \@url }%
\providecommand \@url [1]{\endgroup\@href {#1}{\urlprefix }}%
\providecommand \urlprefix  [0]{URL }%
\providecommand \Eprint [0]{\href }%
\providecommand \doibase [0]{http://dx.doi.org/}%
\providecommand \selectlanguage [0]{\@gobble}%
\providecommand \bibinfo  [0]{\@secondoftwo}%
\providecommand \bibfield  [0]{\@secondoftwo}%
\providecommand \translation [1]{[#1]}%
\providecommand \BibitemOpen [0]{}%
\providecommand \bibitemStop [0]{}%
\providecommand \bibitemNoStop [0]{.\EOS\space}%
\providecommand \EOS [0]{\spacefactor3000\relax}%
\providecommand \BibitemShut  [1]{\csname bibitem#1\endcsname}%
\let\auto@bib@innerbib\@empty
\bibitem [{\citenamefont {{Mu{\v{z}}i{\'c}}}\ \emph {et~al.}(2017)\citenamefont
  {{Mu{\v{z}}i{\'c}}}, \citenamefont {{Sch{\"o}del}}, \citenamefont {{Scholz}},
  \citenamefont {{Geers}}, \citenamefont {{Jayawardhana}}, \citenamefont
  {{Ascenso}},\ and\ \citenamefont {{Cieza}}}]{2017MNRAS.471.3699M}%
  \BibitemOpen
  \bibfield  {author} {\bibinfo {author} {\bibfnamefont {K.}~\bibnamefont
  {{Mu{\v{z}}i{\'c}}}}, \bibinfo {author} {\bibfnamefont {R.}~\bibnamefont
  {{Sch{\"o}del}}}, \bibinfo {author} {\bibfnamefont {A.}~\bibnamefont
  {{Scholz}}}, \bibinfo {author} {\bibfnamefont {V.~C.}\ \bibnamefont
  {{Geers}}}, \bibinfo {author} {\bibfnamefont {R.}~\bibnamefont
  {{Jayawardhana}}}, \bibinfo {author} {\bibfnamefont {J.}~\bibnamefont
  {{Ascenso}}}, \ and\ \bibinfo {author} {\bibfnamefont {L.~A.}\ \bibnamefont
  {{Cieza}}},\ }\href {\doibase 10.1093/mnras/stx1906} {\bibfield  {journal}
  {\bibinfo  {journal} {\mnras}\ }\textbf {\bibinfo {volume} {471}},\ \bibinfo
  {pages} {3699} (\bibinfo {year} {2017})},\ \Eprint
  {http://arxiv.org/abs/1707.00277} {arXiv:1707.00277 [astro-ph.SR]}
  \BibitemShut {NoStop}%
\bibitem [{\citenamefont {{Leane}}\ and\ \citenamefont
  {{Smirnov}}(2021)}]{Leane:2020wob}%
  \BibitemOpen
  \bibfield  {author} {\bibinfo {author} {\bibfnamefont {R.~K.}\ \bibnamefont
  {{Leane}}}\ and\ \bibinfo {author} {\bibfnamefont {J.}~\bibnamefont
  {{Smirnov}}},\ }\href {\doibase 10.1103/PhysRevLett.126.161101} {\bibfield
  {journal} {\bibinfo  {journal} {\prl}\ }\textbf {\bibinfo {volume} {126}},\
  \bibinfo {eid} {161101} (\bibinfo {year} {2021})},\ \Eprint
  {http://arxiv.org/abs/2010.00015} {arXiv:2010.00015 [hep-ph]} \BibitemShut
  {NoStop}%
\bibitem [{\citenamefont {Goldman}\ and\ \citenamefont
  {Nussinov}(1989)}]{Goldman:1989nd}%
  \BibitemOpen
  \bibfield  {author} {\bibinfo {author} {\bibfnamefont {I.}~\bibnamefont
  {Goldman}}\ and\ \bibinfo {author} {\bibfnamefont {S.}~\bibnamefont
  {Nussinov}},\ }\href {\doibase 10.1103/PhysRevD.40.3221} {\bibfield
  {journal} {\bibinfo  {journal} {Phys. Rev.}\ }\textbf {\bibinfo {volume}
  {D40}},\ \bibinfo {pages} {3221} (\bibinfo {year} {1989})}\BibitemShut
  {NoStop}%
\bibitem [{\citenamefont {Gould}\ \emph {et~al.}(1990)\citenamefont {Gould},
  \citenamefont {Draine}, \citenamefont {Romani},\ and\ \citenamefont
  {Nussinov}}]{Gould:1989gw}%
  \BibitemOpen
  \bibfield  {author} {\bibinfo {author} {\bibfnamefont {A.}~\bibnamefont
  {Gould}}, \bibinfo {author} {\bibfnamefont {B.~T.}\ \bibnamefont {Draine}},
  \bibinfo {author} {\bibfnamefont {R.~W.}\ \bibnamefont {Romani}}, \ and\
  \bibinfo {author} {\bibfnamefont {S.}~\bibnamefont {Nussinov}},\ }\href
  {\doibase 10.1016/0370-2693(90)91745-W} {\bibfield  {journal} {\bibinfo
  {journal} {Phys. Lett.}\ }\textbf {\bibinfo {volume} {B238}},\ \bibinfo
  {pages} {337} (\bibinfo {year} {1990})}\BibitemShut {NoStop}%
\bibitem [{\citenamefont {Kouvaris}(2008)}]{Kouvaris:2007ay}%
  \BibitemOpen
  \bibfield  {author} {\bibinfo {author} {\bibfnamefont {C.}~\bibnamefont
  {Kouvaris}},\ }\href {\doibase 10.1103/PhysRevD.77.023006} {\bibfield
  {journal} {\bibinfo  {journal} {Phys. Rev.}\ }\textbf {\bibinfo {volume}
  {D77}},\ \bibinfo {pages} {023006} (\bibinfo {year} {2008})},\ \Eprint
  {http://arxiv.org/abs/0708.2362} {arXiv:0708.2362 [astro-ph]} \BibitemShut
  {NoStop}%
\bibitem [{\citenamefont {Bertone}\ and\ \citenamefont
  {Fairbairn}(2008)}]{Bertone:2007ae}%
  \BibitemOpen
  \bibfield  {author} {\bibinfo {author} {\bibfnamefont {G.}~\bibnamefont
  {Bertone}}\ and\ \bibinfo {author} {\bibfnamefont {M.}~\bibnamefont
  {Fairbairn}},\ }\href {\doibase 10.1103/PhysRevD.77.043515} {\bibfield
  {journal} {\bibinfo  {journal} {Phys. Rev.}\ }\textbf {\bibinfo {volume}
  {D77}},\ \bibinfo {pages} {043515} (\bibinfo {year} {2008})},\ \Eprint
  {http://arxiv.org/abs/0709.1485} {arXiv:0709.1485 [astro-ph]} \BibitemShut
  {NoStop}%
\bibitem [{\citenamefont {Spolyar}\ \emph {et~al.}(2008)\citenamefont
  {Spolyar}, \citenamefont {Freese},\ and\ \citenamefont
  {Gondolo}}]{Spolyar:2007qv}%
  \BibitemOpen
  \bibfield  {author} {\bibinfo {author} {\bibfnamefont {D.}~\bibnamefont
  {Spolyar}}, \bibinfo {author} {\bibfnamefont {K.}~\bibnamefont {Freese}}, \
  and\ \bibinfo {author} {\bibfnamefont {P.}~\bibnamefont {Gondolo}},\ }\href
  {\doibase 10.1103/PhysRevLett.100.051101} {\bibfield  {journal} {\bibinfo
  {journal} {Phys. Rev. Lett.}\ }\textbf {\bibinfo {volume} {100}},\ \bibinfo
  {pages} {051101} (\bibinfo {year} {2008})},\ \Eprint
  {http://arxiv.org/abs/0705.0521} {arXiv:0705.0521 [astro-ph]} \BibitemShut
  {NoStop}%
\bibitem [{\citenamefont {Iocco}(2008)}]{Iocco:2008xb}%
  \BibitemOpen
  \bibfield  {author} {\bibinfo {author} {\bibfnamefont {F.}~\bibnamefont
  {Iocco}},\ }\href {\doibase 10.1086/587959} {\bibfield  {journal} {\bibinfo
  {journal} {Astrophys. J. Lett.}\ }\textbf {\bibinfo {volume} {677}},\
  \bibinfo {pages} {L1} (\bibinfo {year} {2008})},\ \Eprint
  {http://arxiv.org/abs/0802.0941} {arXiv:0802.0941 [astro-ph]} \BibitemShut
  {NoStop}%
\bibitem [{\citenamefont {Freese}\ \emph {et~al.}(2009)\citenamefont {Freese},
  \citenamefont {Gondolo}, \citenamefont {Sellwood},\ and\ \citenamefont
  {Spolyar}}]{Freese:2008hb}%
  \BibitemOpen
  \bibfield  {author} {\bibinfo {author} {\bibfnamefont {K.}~\bibnamefont
  {Freese}}, \bibinfo {author} {\bibfnamefont {P.}~\bibnamefont {Gondolo}},
  \bibinfo {author} {\bibfnamefont {J.}~\bibnamefont {Sellwood}}, \ and\
  \bibinfo {author} {\bibfnamefont {D.}~\bibnamefont {Spolyar}},\ }\href
  {\doibase 10.1088/0004-637X/693/2/1563} {\bibfield  {journal} {\bibinfo
  {journal} {Astrophys. J.}\ }\textbf {\bibinfo {volume} {693}},\ \bibinfo
  {pages} {1563} (\bibinfo {year} {2009})},\ \Eprint
  {http://arxiv.org/abs/0805.3540} {arXiv:0805.3540 [astro-ph]} \BibitemShut
  {NoStop}%
\bibitem [{\citenamefont {Taoso}\ \emph {et~al.}(2008)\citenamefont {Taoso},
  \citenamefont {Bertone}, \citenamefont {Meynet},\ and\ \citenamefont
  {Ekstrom}}]{Taoso:2008kw}%
  \BibitemOpen
  \bibfield  {author} {\bibinfo {author} {\bibfnamefont {M.}~\bibnamefont
  {Taoso}}, \bibinfo {author} {\bibfnamefont {G.}~\bibnamefont {Bertone}},
  \bibinfo {author} {\bibfnamefont {G.}~\bibnamefont {Meynet}}, \ and\ \bibinfo
  {author} {\bibfnamefont {S.}~\bibnamefont {Ekstrom}},\ }\href {\doibase
  10.1103/PhysRevD.78.123510} {\bibfield  {journal} {\bibinfo  {journal} {Phys.
  Rev. D}\ }\textbf {\bibinfo {volume} {78}},\ \bibinfo {pages} {123510}
  (\bibinfo {year} {2008})},\ \Eprint {http://arxiv.org/abs/0806.2681}
  {arXiv:0806.2681 [astro-ph]} \BibitemShut {NoStop}%
\bibitem [{\citenamefont {Sivertsson}\ and\ \citenamefont
  {Gondolo}(2011)}]{Sivertsson:2010zm}%
  \BibitemOpen
  \bibfield  {author} {\bibinfo {author} {\bibfnamefont {S.}~\bibnamefont
  {Sivertsson}}\ and\ \bibinfo {author} {\bibfnamefont {P.}~\bibnamefont
  {Gondolo}},\ }\href {\doibase 10.1088/0004-637X/729/1/51} {\bibfield
  {journal} {\bibinfo  {journal} {Astrophys. J.}\ }\textbf {\bibinfo {volume}
  {729}},\ \bibinfo {pages} {51} (\bibinfo {year} {2011})},\ \Eprint
  {http://arxiv.org/abs/1006.0025} {arXiv:1006.0025 [astro-ph.CO]} \BibitemShut
  {NoStop}%
\bibitem [{\citenamefont {Freese}\ \emph {et~al.}(2016)\citenamefont {Freese},
  \citenamefont {Rindler-Daller}, \citenamefont {Spolyar},\ and\ \citenamefont
  {Valluri}}]{Freese:2015mta}%
  \BibitemOpen
  \bibfield  {author} {\bibinfo {author} {\bibfnamefont {K.}~\bibnamefont
  {Freese}}, \bibinfo {author} {\bibfnamefont {T.}~\bibnamefont
  {Rindler-Daller}}, \bibinfo {author} {\bibfnamefont {D.}~\bibnamefont
  {Spolyar}}, \ and\ \bibinfo {author} {\bibfnamefont {M.}~\bibnamefont
  {Valluri}},\ }\href {\doibase 10.1088/0034-4885/79/6/066902} {\bibfield
  {journal} {\bibinfo  {journal} {Rept. Prog. Phys.}\ }\textbf {\bibinfo
  {volume} {79}},\ \bibinfo {pages} {066902} (\bibinfo {year} {2016})},\
  \Eprint {http://arxiv.org/abs/1501.02394} {arXiv:1501.02394 [astro-ph.CO]}
  \BibitemShut {NoStop}%
\bibitem [{\citenamefont {{Salati}}\ and\ \citenamefont
  {{Silk}}(1989)}]{1989ApJ...338...24S}%
  \BibitemOpen
  \bibfield  {author} {\bibinfo {author} {\bibfnamefont {P.}~\bibnamefont
  {{Salati}}}\ and\ \bibinfo {author} {\bibfnamefont {J.}~\bibnamefont
  {{Silk}}},\ }\href {\doibase 10.1086/167177} {\bibfield  {journal} {\bibinfo
  {journal} {\apj}\ }\textbf {\bibinfo {volume} {338}},\ \bibinfo {pages} {24}
  (\bibinfo {year} {1989})}\BibitemShut {NoStop}%
\bibitem [{\citenamefont {Fairbairn}\ \emph {et~al.}(2008)\citenamefont
  {Fairbairn}, \citenamefont {Scott},\ and\ \citenamefont
  {Edsjo}}]{Fairbairn:2007bn}%
  \BibitemOpen
  \bibfield  {author} {\bibinfo {author} {\bibfnamefont {M.}~\bibnamefont
  {Fairbairn}}, \bibinfo {author} {\bibfnamefont {P.}~\bibnamefont {Scott}}, \
  and\ \bibinfo {author} {\bibfnamefont {J.}~\bibnamefont {Edsjo}},\ }\href
  {\doibase 10.1103/PhysRevD.77.047301} {\bibfield  {journal} {\bibinfo
  {journal} {Phys. Rev. D}\ }\textbf {\bibinfo {volume} {77}},\ \bibinfo
  {pages} {047301} (\bibinfo {year} {2008})},\ \Eprint
  {http://arxiv.org/abs/0710.3396} {arXiv:0710.3396 [astro-ph]} \BibitemShut
  {NoStop}%
\bibitem [{\citenamefont {Scott}\ \emph {et~al.}(2009)\citenamefont {Scott},
  \citenamefont {Fairbairn},\ and\ \citenamefont {Edsjo}}]{Scott:2008ns}%
  \BibitemOpen
  \bibfield  {author} {\bibinfo {author} {\bibfnamefont {P.}~\bibnamefont
  {Scott}}, \bibinfo {author} {\bibfnamefont {M.}~\bibnamefont {Fairbairn}}, \
  and\ \bibinfo {author} {\bibfnamefont {J.}~\bibnamefont {Edsjo}},\ }\href
  {\doibase 10.1111/j.1365-2966.2008.14282.x} {\bibfield  {journal} {\bibinfo
  {journal} {Mon. Not. Roy. Astron. Soc.}\ }\textbf {\bibinfo {volume} {394}},\
  \bibinfo {pages} {82} (\bibinfo {year} {2009})},\ \Eprint
  {http://arxiv.org/abs/0809.1871} {arXiv:0809.1871 [astro-ph]} \BibitemShut
  {NoStop}%
\bibitem [{\citenamefont {Lopes}\ and\ \citenamefont
  {Lopes}(2021)}]{Lopes:2021jcy}%
  \BibitemOpen
  \bibfield  {author} {\bibinfo {author} {\bibfnamefont {J.}~\bibnamefont
  {Lopes}}\ and\ \bibinfo {author} {\bibfnamefont {I.}~\bibnamefont {Lopes}},\
  }\href {\doibase 10.1051/0004-6361/202140750} {\bibfield  {journal} {\bibinfo
   {journal} {Astron. Astrophys.}\ }\textbf {\bibinfo {volume} {651}},\
  \bibinfo {pages} {A101} (\bibinfo {year} {2021})},\ \Eprint
  {http://arxiv.org/abs/2107.13885} {arXiv:2107.13885 [astro-ph.SR]}
  \BibitemShut {NoStop}%
\bibitem [{\citenamefont {de~Lavallaz}\ and\ \citenamefont
  {Fairbairn}(2010)}]{deLavallaz:2010wp}%
  \BibitemOpen
  \bibfield  {author} {\bibinfo {author} {\bibfnamefont {A.}~\bibnamefont
  {de~Lavallaz}}\ and\ \bibinfo {author} {\bibfnamefont {M.}~\bibnamefont
  {Fairbairn}},\ }\href {\doibase 10.1103/PhysRevD.81.123521} {\bibfield
  {journal} {\bibinfo  {journal} {Phys. Rev.}\ }\textbf {\bibinfo {volume}
  {D81}},\ \bibinfo {pages} {123521} (\bibinfo {year} {2010})},\ \Eprint
  {http://arxiv.org/abs/1004.0629} {arXiv:1004.0629 [astro-ph.GA]} \BibitemShut
  {NoStop}%
\bibitem [{\citenamefont {Kouvaris}\ and\ \citenamefont
  {Tinyakov}(2010)}]{Kouvaris:2010vv}%
  \BibitemOpen
  \bibfield  {author} {\bibinfo {author} {\bibfnamefont {C.}~\bibnamefont
  {Kouvaris}}\ and\ \bibinfo {author} {\bibfnamefont {P.}~\bibnamefont
  {Tinyakov}},\ }\href {\doibase 10.1103/PhysRevD.82.063531} {\bibfield
  {journal} {\bibinfo  {journal} {Phys. Rev.}\ }\textbf {\bibinfo {volume}
  {D82}},\ \bibinfo {pages} {063531} (\bibinfo {year} {2010})},\ \Eprint
  {http://arxiv.org/abs/1004.0586} {arXiv:1004.0586 [astro-ph.GA]} \BibitemShut
  {NoStop}%
\bibitem [{\citenamefont {McDermott}\ \emph {et~al.}(2012)\citenamefont
  {McDermott}, \citenamefont {Yu},\ and\ \citenamefont
  {Zurek}}]{McDermott:2011jp}%
  \BibitemOpen
  \bibfield  {author} {\bibinfo {author} {\bibfnamefont {S.~D.}\ \bibnamefont
  {McDermott}}, \bibinfo {author} {\bibfnamefont {H.-B.}\ \bibnamefont {Yu}}, \
  and\ \bibinfo {author} {\bibfnamefont {K.~M.}\ \bibnamefont {Zurek}},\ }\href
  {\doibase 10.1103/PhysRevD.85.023519} {\bibfield  {journal} {\bibinfo
  {journal} {Phys. Rev.}\ }\textbf {\bibinfo {volume} {D85}},\ \bibinfo {pages}
  {023519} (\bibinfo {year} {2012})},\ \Eprint {http://arxiv.org/abs/1103.5472}
  {arXiv:1103.5472 [hep-ph]} \BibitemShut {NoStop}%
\bibitem [{\citenamefont {Kouvaris}\ and\ \citenamefont
  {Tinyakov}(2011{\natexlab{a}})}]{Kouvaris:2011fi}%
  \BibitemOpen
  \bibfield  {author} {\bibinfo {author} {\bibfnamefont {C.}~\bibnamefont
  {Kouvaris}}\ and\ \bibinfo {author} {\bibfnamefont {P.}~\bibnamefont
  {Tinyakov}},\ }\href {\doibase 10.1103/PhysRevLett.107.091301} {\bibfield
  {journal} {\bibinfo  {journal} {Phys. Rev. Lett.}\ }\textbf {\bibinfo
  {volume} {107}},\ \bibinfo {pages} {091301} (\bibinfo {year}
  {2011}{\natexlab{a}})},\ \Eprint {http://arxiv.org/abs/1104.0382}
  {arXiv:1104.0382 [astro-ph.CO]} \BibitemShut {NoStop}%
\bibitem [{\citenamefont {Guver}\ \emph {et~al.}(2014)\citenamefont {Guver},
  \citenamefont {Erkoca}, \citenamefont {Hall~Reno},\ and\ \citenamefont
  {Sarcevic}}]{Guver:2012ba}%
  \BibitemOpen
  \bibfield  {author} {\bibinfo {author} {\bibfnamefont {T.}~\bibnamefont
  {Guver}}, \bibinfo {author} {\bibfnamefont {A.~E.}\ \bibnamefont {Erkoca}},
  \bibinfo {author} {\bibfnamefont {M.}~\bibnamefont {Hall~Reno}}, \ and\
  \bibinfo {author} {\bibfnamefont {I.}~\bibnamefont {Sarcevic}},\ }\href
  {\doibase 10.1088/1475-7516/2014/05/013} {\bibfield  {journal} {\bibinfo
  {journal} {JCAP}\ }\textbf {\bibinfo {volume} {1405}},\ \bibinfo {pages}
  {013} (\bibinfo {year} {2014})},\ \Eprint {http://arxiv.org/abs/1201.2400}
  {arXiv:1201.2400 [hep-ph]} \BibitemShut {NoStop}%
\bibitem [{\citenamefont {Bramante}\ \emph {et~al.}(2013)\citenamefont
  {Bramante}, \citenamefont {Fukushima},\ and\ \citenamefont
  {Kumar}}]{Bramante:2013hn}%
  \BibitemOpen
  \bibfield  {author} {\bibinfo {author} {\bibfnamefont {J.}~\bibnamefont
  {Bramante}}, \bibinfo {author} {\bibfnamefont {K.}~\bibnamefont {Fukushima}},
  \ and\ \bibinfo {author} {\bibfnamefont {J.}~\bibnamefont {Kumar}},\ }\href
  {\doibase 10.1103/PhysRevD.87.055012} {\bibfield  {journal} {\bibinfo
  {journal} {Phys. Rev.}\ }\textbf {\bibinfo {volume} {D87}},\ \bibinfo {pages}
  {055012} (\bibinfo {year} {2013})},\ \Eprint {http://arxiv.org/abs/1301.0036}
  {arXiv:1301.0036 [hep-ph]} \BibitemShut {NoStop}%
\bibitem [{\citenamefont {Bell}\ \emph {et~al.}(2013)\citenamefont {Bell},
  \citenamefont {Melatos},\ and\ \citenamefont {Petraki}}]{Bell:2013xk}%
  \BibitemOpen
  \bibfield  {author} {\bibinfo {author} {\bibfnamefont {N.~F.}\ \bibnamefont
  {Bell}}, \bibinfo {author} {\bibfnamefont {A.}~\bibnamefont {Melatos}}, \
  and\ \bibinfo {author} {\bibfnamefont {K.}~\bibnamefont {Petraki}},\ }\href
  {\doibase 10.1103/PhysRevD.87.123507} {\bibfield  {journal} {\bibinfo
  {journal} {Phys. Rev.}\ }\textbf {\bibinfo {volume} {D87}},\ \bibinfo {pages}
  {123507} (\bibinfo {year} {2013})},\ \Eprint {http://arxiv.org/abs/1301.6811}
  {arXiv:1301.6811 [hep-ph]} \BibitemShut {NoStop}%
\bibitem [{\citenamefont {Bramante}\ \emph {et~al.}(2014)\citenamefont
  {Bramante}, \citenamefont {Fukushima}, \citenamefont {Kumar},\ and\
  \citenamefont {Stopnitzky}}]{Bramante:2013nma}%
  \BibitemOpen
  \bibfield  {author} {\bibinfo {author} {\bibfnamefont {J.}~\bibnamefont
  {Bramante}}, \bibinfo {author} {\bibfnamefont {K.}~\bibnamefont {Fukushima}},
  \bibinfo {author} {\bibfnamefont {J.}~\bibnamefont {Kumar}}, \ and\ \bibinfo
  {author} {\bibfnamefont {E.}~\bibnamefont {Stopnitzky}},\ }\href {\doibase
  10.1103/PhysRevD.89.015010} {\bibfield  {journal} {\bibinfo  {journal} {Phys.
  Rev.}\ }\textbf {\bibinfo {volume} {D89}},\ \bibinfo {pages} {015010}
  (\bibinfo {year} {2014})},\ \Eprint {http://arxiv.org/abs/1310.3509}
  {arXiv:1310.3509 [hep-ph]} \BibitemShut {NoStop}%
\bibitem [{\citenamefont {Bertoni}\ \emph {et~al.}(2013)\citenamefont
  {Bertoni}, \citenamefont {Nelson},\ and\ \citenamefont
  {Reddy}}]{Bertoni:2013bsa}%
  \BibitemOpen
  \bibfield  {author} {\bibinfo {author} {\bibfnamefont {B.}~\bibnamefont
  {Bertoni}}, \bibinfo {author} {\bibfnamefont {A.~E.}\ \bibnamefont {Nelson}},
  \ and\ \bibinfo {author} {\bibfnamefont {S.}~\bibnamefont {Reddy}},\ }\href
  {\doibase 10.1103/PhysRevD.88.123505} {\bibfield  {journal} {\bibinfo
  {journal} {Phys. Rev.}\ }\textbf {\bibinfo {volume} {D88}},\ \bibinfo {pages}
  {123505} (\bibinfo {year} {2013})},\ \Eprint {http://arxiv.org/abs/1309.1721}
  {arXiv:1309.1721 [hep-ph]} \BibitemShut {NoStop}%
\bibitem [{\citenamefont {Kouvaris}\ and\ \citenamefont
  {Tinyakov}(2011{\natexlab{b}})}]{Kouvaris:2010jy}%
  \BibitemOpen
  \bibfield  {author} {\bibinfo {author} {\bibfnamefont {C.}~\bibnamefont
  {Kouvaris}}\ and\ \bibinfo {author} {\bibfnamefont {P.}~\bibnamefont
  {Tinyakov}},\ }\href {\doibase 10.1103/PhysRevD.83.083512} {\bibfield
  {journal} {\bibinfo  {journal} {Phys. Rev.}\ }\textbf {\bibinfo {volume}
  {D83}},\ \bibinfo {pages} {083512} (\bibinfo {year} {2011}{\natexlab{b}})},\
  \Eprint {http://arxiv.org/abs/1012.2039} {arXiv:1012.2039 [astro-ph.HE]}
  \BibitemShut {NoStop}%
\bibitem [{\citenamefont {McCullough}\ and\ \citenamefont
  {Fairbairn}(2010)}]{McCullough:2010ai}%
  \BibitemOpen
  \bibfield  {author} {\bibinfo {author} {\bibfnamefont {M.}~\bibnamefont
  {McCullough}}\ and\ \bibinfo {author} {\bibfnamefont {M.}~\bibnamefont
  {Fairbairn}},\ }\href {\doibase 10.1103/PhysRevD.81.083520} {\bibfield
  {journal} {\bibinfo  {journal} {Phys. Rev.}\ }\textbf {\bibinfo {volume}
  {D81}},\ \bibinfo {pages} {083520} (\bibinfo {year} {2010})},\ \Eprint
  {http://arxiv.org/abs/1001.2737} {arXiv:1001.2737 [hep-ph]} \BibitemShut
  {NoStop}%
\bibitem [{\citenamefont {Angeles Perez-Garcia}\ and\ \citenamefont
  {Silk}(2015)}]{Perez-Garcia:2014dra}%
  \BibitemOpen
  \bibfield  {author} {\bibinfo {author} {\bibfnamefont {M.}~\bibnamefont
  {Angeles Perez-Garcia}}\ and\ \bibinfo {author} {\bibfnamefont
  {J.}~\bibnamefont {Silk}},\ }\href {\doibase 10.1016/j.physletb.2015.03.026}
  {\bibfield  {journal} {\bibinfo  {journal} {Phys. Lett.}\ }\textbf {\bibinfo
  {volume} {B744}},\ \bibinfo {pages} {13} (\bibinfo {year} {2015})},\ \Eprint
  {http://arxiv.org/abs/1403.6111} {arXiv:1403.6111 [astro-ph.SR]} \BibitemShut
  {NoStop}%
\bibitem [{\citenamefont {Bramante}(2015)}]{Bramante:2015cua}%
  \BibitemOpen
  \bibfield  {author} {\bibinfo {author} {\bibfnamefont {J.}~\bibnamefont
  {Bramante}},\ }\href {\doibase 10.1103/PhysRevLett.115.141301} {\bibfield
  {journal} {\bibinfo  {journal} {Phys. Rev. Lett.}\ }\textbf {\bibinfo
  {volume} {115}},\ \bibinfo {pages} {141301} (\bibinfo {year} {2015})},\
  \Eprint {http://arxiv.org/abs/1505.07464} {arXiv:1505.07464 [hep-ph]}
  \BibitemShut {NoStop}%
\bibitem [{\citenamefont {Graham}\ \emph {et~al.}(2015)\citenamefont {Graham},
  \citenamefont {Rajendran},\ and\ \citenamefont {Varela}}]{Graham:2015apa}%
  \BibitemOpen
  \bibfield  {author} {\bibinfo {author} {\bibfnamefont {P.~W.}\ \bibnamefont
  {Graham}}, \bibinfo {author} {\bibfnamefont {S.}~\bibnamefont {Rajendran}}, \
  and\ \bibinfo {author} {\bibfnamefont {J.}~\bibnamefont {Varela}},\ }\href
  {\doibase 10.1103/PhysRevD.92.063007} {\bibfield  {journal} {\bibinfo
  {journal} {Phys. Rev.}\ }\textbf {\bibinfo {volume} {D92}},\ \bibinfo {pages}
  {063007} (\bibinfo {year} {2015})},\ \Eprint
  {http://arxiv.org/abs/1505.04444} {arXiv:1505.04444 [hep-ph]} \BibitemShut
  {NoStop}%
\bibitem [{\citenamefont {Cermeno}\ \emph {et~al.}(2016)\citenamefont
  {Cermeno}, \citenamefont {Perez-Garcia},\ and\ \citenamefont
  {Silk}}]{Cermeno:2016olb}%
  \BibitemOpen
  \bibfield  {author} {\bibinfo {author} {\bibfnamefont {M.}~\bibnamefont
  {Cermeno}}, \bibinfo {author} {\bibfnamefont {M.}~\bibnamefont
  {Perez-Garcia}}, \ and\ \bibinfo {author} {\bibfnamefont {J.}~\bibnamefont
  {Silk}},\ }\href {\doibase 10.1103/PhysRevD.94.063001} {\bibfield  {journal}
  {\bibinfo  {journal} {Phys. Rev.}\ }\textbf {\bibinfo {volume} {D94}},\
  \bibinfo {pages} {063001} (\bibinfo {year} {2016})},\ \Eprint
  {http://arxiv.org/abs/1607.06815} {arXiv:1607.06815 [astro-ph.HE]}
  \BibitemShut {NoStop}%
\bibitem [{\citenamefont {Krall}\ and\ \citenamefont
  {Reece}(2018)}]{Krall:2017xij}%
  \BibitemOpen
  \bibfield  {author} {\bibinfo {author} {\bibfnamefont {R.}~\bibnamefont
  {Krall}}\ and\ \bibinfo {author} {\bibfnamefont {M.}~\bibnamefont {Reece}},\
  }\href {\doibase 10.1088/1674-1137/42/4/043105} {\bibfield  {journal}
  {\bibinfo  {journal} {Chin. Phys.}\ }\textbf {\bibinfo {volume} {C42}},\
  \bibinfo {pages} {043105} (\bibinfo {year} {2018})},\ \Eprint
  {http://arxiv.org/abs/1705.04843} {arXiv:1705.04843 [hep-ph]} \BibitemShut
  {NoStop}%
\bibitem [{\citenamefont {Leane}\ \emph {et~al.}(2017)\citenamefont {Leane},
  \citenamefont {Ng},\ and\ \citenamefont {Beacom}}]{Leane:2017vag}%
  \BibitemOpen
  \bibfield  {author} {\bibinfo {author} {\bibfnamefont {R.~K.}\ \bibnamefont
  {Leane}}, \bibinfo {author} {\bibfnamefont {K.~C.~Y.}\ \bibnamefont {Ng}}, \
  and\ \bibinfo {author} {\bibfnamefont {J.~F.}\ \bibnamefont {Beacom}},\
  }\href {\doibase 10.1103/PhysRevD.95.123016} {\bibfield  {journal} {\bibinfo
  {journal} {Phys. Rev.}\ }\textbf {\bibinfo {volume} {D95}},\ \bibinfo {pages}
  {123016} (\bibinfo {year} {2017})},\ \Eprint
  {http://arxiv.org/abs/1703.04629} {arXiv:1703.04629 [astro-ph.HE]}
  \BibitemShut {NoStop}%
\bibitem [{\citenamefont {McKeen}\ \emph {et~al.}(2018)\citenamefont {McKeen},
  \citenamefont {Nelson}, \citenamefont {Reddy},\ and\ \citenamefont
  {Zhou}}]{McKeen:2018xwc}%
  \BibitemOpen
  \bibfield  {author} {\bibinfo {author} {\bibfnamefont {D.}~\bibnamefont
  {McKeen}}, \bibinfo {author} {\bibfnamefont {A.~E.}\ \bibnamefont {Nelson}},
  \bibinfo {author} {\bibfnamefont {S.}~\bibnamefont {Reddy}}, \ and\ \bibinfo
  {author} {\bibfnamefont {D.}~\bibnamefont {Zhou}},\ }\href {\doibase
  10.1103/PhysRevLett.121.061802} {\bibfield  {journal} {\bibinfo  {journal}
  {Phys. Rev. Lett.}\ }\textbf {\bibinfo {volume} {121}},\ \bibinfo {pages}
  {061802} (\bibinfo {year} {2018})},\ \Eprint
  {http://arxiv.org/abs/1802.08244} {arXiv:1802.08244 [hep-ph]} \BibitemShut
  {NoStop}%
\bibitem [{\citenamefont {Baryakhtar}\ \emph {et~al.}(2017)\citenamefont
  {Baryakhtar}, \citenamefont {Bramante}, \citenamefont {Li}, \citenamefont
  {Linden},\ and\ \citenamefont {Raj}}]{Baryakhtar:2017dbj}%
  \BibitemOpen
  \bibfield  {author} {\bibinfo {author} {\bibfnamefont {M.}~\bibnamefont
  {Baryakhtar}}, \bibinfo {author} {\bibfnamefont {J.}~\bibnamefont
  {Bramante}}, \bibinfo {author} {\bibfnamefont {S.~W.}\ \bibnamefont {Li}},
  \bibinfo {author} {\bibfnamefont {T.}~\bibnamefont {Linden}}, \ and\ \bibinfo
  {author} {\bibfnamefont {N.}~\bibnamefont {Raj}},\ }\href {\doibase
  10.1103/PhysRevLett.119.131801} {\bibfield  {journal} {\bibinfo  {journal}
  {Phys. Rev. Lett.}\ }\textbf {\bibinfo {volume} {119}},\ \bibinfo {pages}
  {131801} (\bibinfo {year} {2017})},\ \Eprint
  {http://arxiv.org/abs/1704.01577} {arXiv:1704.01577 [hep-ph]} \BibitemShut
  {NoStop}%
\bibitem [{\citenamefont {Raj}\ \emph {et~al.}(2018)\citenamefont {Raj},
  \citenamefont {Tanedo},\ and\ \citenamefont {Yu}}]{Raj:2017wrv}%
  \BibitemOpen
  \bibfield  {author} {\bibinfo {author} {\bibfnamefont {N.}~\bibnamefont
  {Raj}}, \bibinfo {author} {\bibfnamefont {P.}~\bibnamefont {Tanedo}}, \ and\
  \bibinfo {author} {\bibfnamefont {H.-B.}\ \bibnamefont {Yu}},\ }\href
  {\doibase 10.1103/PhysRevD.97.043006} {\bibfield  {journal} {\bibinfo
  {journal} {Phys. Rev.}\ }\textbf {\bibinfo {volume} {D97}},\ \bibinfo {pages}
  {043006} (\bibinfo {year} {2018})},\ \Eprint
  {http://arxiv.org/abs/1707.09442} {arXiv:1707.09442 [hep-ph]} \BibitemShut
  {NoStop}%
\bibitem [{\citenamefont {Bell}\ \emph {et~al.}(2018)\citenamefont {Bell},
  \citenamefont {Busoni},\ and\ \citenamefont {Robles}}]{Bell:2018pkk}%
  \BibitemOpen
  \bibfield  {author} {\bibinfo {author} {\bibfnamefont {N.~F.}\ \bibnamefont
  {Bell}}, \bibinfo {author} {\bibfnamefont {G.}~\bibnamefont {Busoni}}, \ and\
  \bibinfo {author} {\bibfnamefont {S.}~\bibnamefont {Robles}},\ }\href
  {\doibase 10.1088/1475-7516/2018/09/018} {\bibfield  {journal} {\bibinfo
  {journal} {JCAP}\ }\textbf {\bibinfo {volume} {1809}},\ \bibinfo {pages}
  {018} (\bibinfo {year} {2018})},\ \Eprint {http://arxiv.org/abs/1807.02840}
  {arXiv:1807.02840 [hep-ph]} \BibitemShut {NoStop}%
\bibitem [{\citenamefont {Chen}\ and\ \citenamefont
  {Lin}(2018)}]{Chen:2018ohx}%
  \BibitemOpen
  \bibfield  {author} {\bibinfo {author} {\bibfnamefont {C.-S.}\ \bibnamefont
  {Chen}}\ and\ \bibinfo {author} {\bibfnamefont {Y.-H.}\ \bibnamefont {Lin}},\
  }\href {\doibase 10.1007/JHEP08(2018)069} {\bibfield  {journal} {\bibinfo
  {journal} {JHEP}\ }\textbf {\bibinfo {volume} {08}},\ \bibinfo {pages} {069}
  (\bibinfo {year} {2018})},\ \Eprint {http://arxiv.org/abs/1804.03409}
  {arXiv:1804.03409 [hep-ph]} \BibitemShut {NoStop}%
\bibitem [{\citenamefont {Hamaguchi}\ \emph {et~al.}(2019)\citenamefont
  {Hamaguchi}, \citenamefont {Nagata},\ and\ \citenamefont
  {Yanagi}}]{Hamaguchi:2019oev}%
  \BibitemOpen
  \bibfield  {author} {\bibinfo {author} {\bibfnamefont {K.}~\bibnamefont
  {Hamaguchi}}, \bibinfo {author} {\bibfnamefont {N.}~\bibnamefont {Nagata}}, \
  and\ \bibinfo {author} {\bibfnamefont {K.}~\bibnamefont {Yanagi}},\ }\href
  {\doibase 10.1016/j.physletb.2019.06.060} {\bibfield  {journal} {\bibinfo
  {journal} {Phys. Lett.}\ }\textbf {\bibinfo {volume} {B795}},\ \bibinfo
  {pages} {484} (\bibinfo {year} {2019})},\ \Eprint
  {http://arxiv.org/abs/1905.02991} {arXiv:1905.02991 [hep-ph]} \BibitemShut
  {NoStop}%
\bibitem [{\citenamefont {Camargo}\ \emph {et~al.}(2019)\citenamefont
  {Camargo}, \citenamefont {Queiroz},\ and\ \citenamefont
  {Sturani}}]{Camargo:2019wou}%
  \BibitemOpen
  \bibfield  {author} {\bibinfo {author} {\bibfnamefont {D.~A.}\ \bibnamefont
  {Camargo}}, \bibinfo {author} {\bibfnamefont {F.~S.}\ \bibnamefont
  {Queiroz}}, \ and\ \bibinfo {author} {\bibfnamefont {R.}~\bibnamefont
  {Sturani}},\ }\href {\doibase 10.1088/1475-7516/2019/09/051} {\bibfield
  {journal} {\bibinfo  {journal} {JCAP}\ }\textbf {\bibinfo {volume} {1909}},\
  \bibinfo {pages} {051} (\bibinfo {year} {2019})},\ \Eprint
  {http://arxiv.org/abs/1901.05474} {arXiv:1901.05474 [hep-ph]} \BibitemShut
  {NoStop}%
\bibitem [{\citenamefont {Bell}\ \emph {et~al.}(2019)\citenamefont {Bell},
  \citenamefont {Busoni},\ and\ \citenamefont {Robles}}]{Bell:2019pyc}%
  \BibitemOpen
  \bibfield  {author} {\bibinfo {author} {\bibfnamefont {N.~F.}\ \bibnamefont
  {Bell}}, \bibinfo {author} {\bibfnamefont {G.}~\bibnamefont {Busoni}}, \ and\
  \bibinfo {author} {\bibfnamefont {S.}~\bibnamefont {Robles}},\ }\href
  {\doibase 10.1088/1475-7516/2019/06/054} {\bibfield  {journal} {\bibinfo
  {journal} {JCAP}\ }\textbf {\bibinfo {volume} {1906}},\ \bibinfo {pages}
  {054} (\bibinfo {year} {2019})},\ \Eprint {http://arxiv.org/abs/1904.09803}
  {arXiv:1904.09803 [hep-ph]} \BibitemShut {NoStop}%
\bibitem [{\citenamefont {Garani}\ and\ \citenamefont
  {Heeck}(2019)}]{Garani:2019fpa}%
  \BibitemOpen
  \bibfield  {author} {\bibinfo {author} {\bibfnamefont {R.}~\bibnamefont
  {Garani}}\ and\ \bibinfo {author} {\bibfnamefont {J.}~\bibnamefont {Heeck}},\
  }\href {\doibase 10.1103/PhysRevD.100.035039} {\bibfield  {journal} {\bibinfo
   {journal} {Phys. Rev.}\ }\textbf {\bibinfo {volume} {D100}},\ \bibinfo
  {pages} {035039} (\bibinfo {year} {2019})},\ \Eprint
  {http://arxiv.org/abs/1906.10145} {arXiv:1906.10145 [hep-ph]} \BibitemShut
  {NoStop}%
\bibitem [{\citenamefont {Acevedo}\ \emph {et~al.}(2020)\citenamefont
  {Acevedo}, \citenamefont {Bramante}, \citenamefont {Leane},\ and\
  \citenamefont {Raj}}]{Acevedo:2019agu}%
  \BibitemOpen
  \bibfield  {author} {\bibinfo {author} {\bibfnamefont {J.~F.}\ \bibnamefont
  {Acevedo}}, \bibinfo {author} {\bibfnamefont {J.}~\bibnamefont {Bramante}},
  \bibinfo {author} {\bibfnamefont {R.~K.}\ \bibnamefont {Leane}}, \ and\
  \bibinfo {author} {\bibfnamefont {N.}~\bibnamefont {Raj}},\ }\href {\doibase
  10.1088/1475-7516/2020/03/038} {\bibfield  {journal} {\bibinfo  {journal}
  {JCAP}\ }\textbf {\bibinfo {volume} {03}},\ \bibinfo {pages} {038} (\bibinfo
  {year} {2020})},\ \Eprint {http://arxiv.org/abs/1911.06334} {arXiv:1911.06334
  [hep-ph]} \BibitemShut {NoStop}%
\bibitem [{\citenamefont {{Joglekar}}\ \emph
  {et~al.}(2020{\natexlab{a}})\citenamefont {{Joglekar}}, \citenamefont
  {{Raj}}, \citenamefont {{Tanedo}},\ and\ \citenamefont
  {{Yu}}}]{Joglekar:2019vzy}%
  \BibitemOpen
  \bibfield  {author} {\bibinfo {author} {\bibfnamefont {A.}~\bibnamefont
  {{Joglekar}}}, \bibinfo {author} {\bibfnamefont {N.}~\bibnamefont {{Raj}}},
  \bibinfo {author} {\bibfnamefont {P.}~\bibnamefont {{Tanedo}}}, \ and\
  \bibinfo {author} {\bibfnamefont {H.-B.}\ \bibnamefont {{Yu}}},\ }\href
  {\doibase 10.1016/j.physletb.2020.135767} {\bibfield  {journal} {\bibinfo
  {journal} {Physics Letters B}\ }\textbf {\bibinfo {volume} {809}},\ \bibinfo
  {eid} {135767} (\bibinfo {year} {2020}{\natexlab{a}})},\ \Eprint
  {http://arxiv.org/abs/1911.13293} {arXiv:1911.13293 [hep-ph]} \BibitemShut
  {NoStop}%
\bibitem [{\citenamefont {{Joglekar}}\ \emph
  {et~al.}(2020{\natexlab{b}})\citenamefont {{Joglekar}}, \citenamefont
  {{Raj}}, \citenamefont {{Tanedo}},\ and\ \citenamefont
  {{Yu}}}]{Joglekar:2020liw}%
  \BibitemOpen
  \bibfield  {author} {\bibinfo {author} {\bibfnamefont {A.}~\bibnamefont
  {{Joglekar}}}, \bibinfo {author} {\bibfnamefont {N.}~\bibnamefont {{Raj}}},
  \bibinfo {author} {\bibfnamefont {P.}~\bibnamefont {{Tanedo}}}, \ and\
  \bibinfo {author} {\bibfnamefont {H.-B.}\ \bibnamefont {{Yu}}},\ }\href
  {\doibase 10.1103/PhysRevD.102.123002} {\bibfield  {journal} {\bibinfo
  {journal} {\prd}\ }\textbf {\bibinfo {volume} {102}},\ \bibinfo {eid}
  {123002} (\bibinfo {year} {2020}{\natexlab{b}})},\ \Eprint
  {http://arxiv.org/abs/2004.09539} {arXiv:2004.09539 [hep-ph]} \BibitemShut
  {NoStop}%
\bibitem [{\citenamefont {{Bell}}\ \emph {et~al.}(2020)\citenamefont {{Bell}},
  \citenamefont {{Busoni}}, \citenamefont {{Robles}},\ and\ \citenamefont
  {{Virgato}}}]{Bell:2020jou}%
  \BibitemOpen
  \bibfield  {author} {\bibinfo {author} {\bibfnamefont {N.~F.}\ \bibnamefont
  {{Bell}}}, \bibinfo {author} {\bibfnamefont {G.}~\bibnamefont {{Busoni}}},
  \bibinfo {author} {\bibfnamefont {S.}~\bibnamefont {{Robles}}}, \ and\
  \bibinfo {author} {\bibfnamefont {M.}~\bibnamefont {{Virgato}}},\ }\href
  {\doibase 10.1088/1475-7516/2020/09/028} {\bibfield  {journal} {\bibinfo
  {journal} {\jcap}\ }\textbf {\bibinfo {volume} {2020}},\ \bibinfo {eid} {028}
  (\bibinfo {year} {2020})},\ \Eprint {http://arxiv.org/abs/2004.14888}
  {arXiv:2004.14888 [hep-ph]} \BibitemShut {NoStop}%
\bibitem [{\citenamefont {{Garani}}\ \emph {et~al.}(2021)\citenamefont
  {{Garani}}, \citenamefont {{Gupta}},\ and\ \citenamefont
  {{Raj}}}]{Garani:2020wge}%
  \BibitemOpen
  \bibfield  {author} {\bibinfo {author} {\bibfnamefont {R.}~\bibnamefont
  {{Garani}}}, \bibinfo {author} {\bibfnamefont {A.}~\bibnamefont {{Gupta}}}, \
  and\ \bibinfo {author} {\bibfnamefont {N.}~\bibnamefont {{Raj}}},\ }\href
  {\doibase 10.1103/PhysRevD.103.043019} {\bibfield  {journal} {\bibinfo
  {journal} {\prd}\ }\textbf {\bibinfo {volume} {103}},\ \bibinfo {eid}
  {043019} (\bibinfo {year} {2021})},\ \Eprint
  {http://arxiv.org/abs/2009.10728} {arXiv:2009.10728 [hep-ph]} \BibitemShut
  {NoStop}%
\bibitem [{\citenamefont {Leane}\ \emph {et~al.}(2021)\citenamefont {Leane},
  \citenamefont {Linden}, \citenamefont {Mukhopadhyay},\ and\ \citenamefont
  {Toro}}]{Leane:2021ihh}%
  \BibitemOpen
  \bibfield  {author} {\bibinfo {author} {\bibfnamefont {R.~K.}\ \bibnamefont
  {Leane}}, \bibinfo {author} {\bibfnamefont {T.}~\bibnamefont {Linden}},
  \bibinfo {author} {\bibfnamefont {P.}~\bibnamefont {Mukhopadhyay}}, \ and\
  \bibinfo {author} {\bibfnamefont {N.}~\bibnamefont {Toro}},\ }\href@noop {}
  {\  (\bibinfo {year} {2021})},\ \Eprint {http://arxiv.org/abs/2101.12213}
  {arXiv:2101.12213 [astro-ph.HE]} \BibitemShut {NoStop}%
\bibitem [{\citenamefont {Acevedo}\ \emph
  {et~al.}(2024{\natexlab{a}})\citenamefont {Acevedo}, \citenamefont {Leane},\
  and\ \citenamefont {Santos-Olmsted}}]{Acevedo:2023xnu}%
  \BibitemOpen
  \bibfield  {author} {\bibinfo {author} {\bibfnamefont {J.~F.}\ \bibnamefont
  {Acevedo}}, \bibinfo {author} {\bibfnamefont {R.~K.}\ \bibnamefont {Leane}},
  \ and\ \bibinfo {author} {\bibfnamefont {L.}~\bibnamefont {Santos-Olmsted}},\
  }\href {\doibase 10.1088/1475-7516/2024/03/042} {\bibfield  {journal}
  {\bibinfo  {journal} {JCAP}\ }\textbf {\bibinfo {volume} {03}},\ \bibinfo
  {pages} {042} (\bibinfo {year} {2024}{\natexlab{a}})},\ \Eprint
  {http://arxiv.org/abs/2309.10843} {arXiv:2309.10843 [hep-ph]} \BibitemShut
  {NoStop}%
\bibitem [{\citenamefont {John}\ \emph {et~al.}(2023)\citenamefont {John},
  \citenamefont {Leane},\ and\ \citenamefont {Linden}}]{John:2023knt}%
  \BibitemOpen
  \bibfield  {author} {\bibinfo {author} {\bibfnamefont {I.}~\bibnamefont
  {John}}, \bibinfo {author} {\bibfnamefont {R.~K.}\ \bibnamefont {Leane}}, \
  and\ \bibinfo {author} {\bibfnamefont {T.}~\bibnamefont {Linden}},\
  }\href@noop {} {\  (\bibinfo {year} {2023})},\ \Eprint
  {http://arxiv.org/abs/2311.16228} {arXiv:2311.16228 [astro-ph.HE]}
  \BibitemShut {NoStop}%
\bibitem [{\citenamefont {Croon}\ and\ \citenamefont
  {Sakstein}(2023)}]{Croon:2023trk}%
  \BibitemOpen
  \bibfield  {author} {\bibinfo {author} {\bibfnamefont {D.}~\bibnamefont
  {Croon}}\ and\ \bibinfo {author} {\bibfnamefont {J.}~\bibnamefont
  {Sakstein}},\ }\href@noop {} {\  (\bibinfo {year} {2023})},\ \Eprint
  {http://arxiv.org/abs/2310.20044} {arXiv:2310.20044 [astro-ph.HE]}
  \BibitemShut {NoStop}%
\bibitem [{\citenamefont {Acevedo}\ \emph
  {et~al.}(2024{\natexlab{b}})\citenamefont {Acevedo}, \citenamefont
  {Bramante}, \citenamefont {Liu},\ and\ \citenamefont
  {Tyagi}}]{Acevedo:2024ttq}%
  \BibitemOpen
  \bibfield  {author} {\bibinfo {author} {\bibfnamefont {J.~F.}\ \bibnamefont
  {Acevedo}}, \bibinfo {author} {\bibfnamefont {J.}~\bibnamefont {Bramante}},
  \bibinfo {author} {\bibfnamefont {Q.}~\bibnamefont {Liu}}, \ and\ \bibinfo
  {author} {\bibfnamefont {N.}~\bibnamefont {Tyagi}},\ }\href@noop {} {\
  (\bibinfo {year} {2024}{\natexlab{b}})},\ \Eprint
  {http://arxiv.org/abs/2404.10039} {arXiv:2404.10039 [hep-ph]} \BibitemShut
  {NoStop}%
\bibitem [{\citenamefont {Mack}\ \emph {et~al.}(2007)\citenamefont {Mack},
  \citenamefont {Beacom},\ and\ \citenamefont {Bertone}}]{Mack:2007xj}%
  \BibitemOpen
  \bibfield  {author} {\bibinfo {author} {\bibfnamefont {G.~D.}\ \bibnamefont
  {Mack}}, \bibinfo {author} {\bibfnamefont {J.~F.}\ \bibnamefont {Beacom}}, \
  and\ \bibinfo {author} {\bibfnamefont {G.}~\bibnamefont {Bertone}},\ }\href
  {\doibase 10.1103/PhysRevD.76.043523} {\bibfield  {journal} {\bibinfo
  {journal} {Phys. Rev. D}\ }\textbf {\bibinfo {volume} {76}},\ \bibinfo
  {pages} {043523} (\bibinfo {year} {2007})},\ \Eprint
  {http://arxiv.org/abs/0705.4298} {arXiv:0705.4298 [astro-ph]} \BibitemShut
  {NoStop}%
\bibitem [{\citenamefont {Chauhan}\ and\ \citenamefont
  {Mohanty}(2016)}]{Chauhan:2016joa}%
  \BibitemOpen
  \bibfield  {author} {\bibinfo {author} {\bibfnamefont {B.}~\bibnamefont
  {Chauhan}}\ and\ \bibinfo {author} {\bibfnamefont {S.}~\bibnamefont
  {Mohanty}},\ }\href {\doibase 10.1103/PhysRevD.94.035024} {\bibfield
  {journal} {\bibinfo  {journal} {Phys. Rev. D}\ }\textbf {\bibinfo {volume}
  {94}},\ \bibinfo {pages} {035024} (\bibinfo {year} {2016})},\ \Eprint
  {http://arxiv.org/abs/1603.06350} {arXiv:1603.06350 [hep-ph]} \BibitemShut
  {NoStop}%
\bibitem [{\citenamefont {Bramante}\ \emph {et~al.}(2020)\citenamefont
  {Bramante}, \citenamefont {Buchanan}, \citenamefont {Goodman},\ and\
  \citenamefont {Lodhi}}]{Bramante:2019fhi}%
  \BibitemOpen
  \bibfield  {author} {\bibinfo {author} {\bibfnamefont {J.}~\bibnamefont
  {Bramante}}, \bibinfo {author} {\bibfnamefont {A.}~\bibnamefont {Buchanan}},
  \bibinfo {author} {\bibfnamefont {A.}~\bibnamefont {Goodman}}, \ and\
  \bibinfo {author} {\bibfnamefont {E.}~\bibnamefont {Lodhi}},\ }\href
  {\doibase 10.1103/PhysRevD.101.043001} {\bibfield  {journal} {\bibinfo
  {journal} {Phys. Rev. D}\ }\textbf {\bibinfo {volume} {101}},\ \bibinfo
  {pages} {043001} (\bibinfo {year} {2020})},\ \Eprint
  {http://arxiv.org/abs/1909.11683} {arXiv:1909.11683 [hep-ph]} \BibitemShut
  {NoStop}%
\bibitem [{\citenamefont {Adler}(2009)}]{Adler:2008ky}%
  \BibitemOpen
  \bibfield  {author} {\bibinfo {author} {\bibfnamefont {S.~L.}\ \bibnamefont
  {Adler}},\ }\href {\doibase 10.1016/j.physletb.2008.12.023} {\bibfield
  {journal} {\bibinfo  {journal} {Phys. Lett. B}\ }\textbf {\bibinfo {volume}
  {671}},\ \bibinfo {pages} {203} (\bibinfo {year} {2009})},\ \Eprint
  {http://arxiv.org/abs/0808.2823} {arXiv:0808.2823 [astro-ph]} \BibitemShut
  {NoStop}%
\bibitem [{\citenamefont {Kawasaki}\ \emph {et~al.}(1992)\citenamefont
  {Kawasaki}, \citenamefont {Murayama},\ and\ \citenamefont
  {Yanagida}}]{Kawasaki:1991eu}%
  \BibitemOpen
  \bibfield  {author} {\bibinfo {author} {\bibfnamefont {M.}~\bibnamefont
  {Kawasaki}}, \bibinfo {author} {\bibfnamefont {H.}~\bibnamefont {Murayama}},
  \ and\ \bibinfo {author} {\bibfnamefont {T.}~\bibnamefont {Yanagida}},\
  }\href {\doibase 10.1143/PTP.87.685} {\bibfield  {journal} {\bibinfo
  {journal} {Prog. Theor. Phys.}\ }\textbf {\bibinfo {volume} {87}},\ \bibinfo
  {pages} {685} (\bibinfo {year} {1992})}\BibitemShut {NoStop}%
\bibitem [{\citenamefont {Mitra}(2004)}]{Mitra:2004fh}%
  \BibitemOpen
  \bibfield  {author} {\bibinfo {author} {\bibfnamefont {S.}~\bibnamefont
  {Mitra}},\ }\href {\doibase 10.1103/PhysRevD.70.103517} {\bibfield  {journal}
  {\bibinfo  {journal} {Phys. Rev. D}\ }\textbf {\bibinfo {volume} {70}},\
  \bibinfo {pages} {103517} (\bibinfo {year} {2004})},\ \Eprint
  {http://arxiv.org/abs/astro-ph/0408341} {arXiv:astro-ph/0408341} \BibitemShut
  {NoStop}%
\bibitem [{\citenamefont {Garani}\ and\ \citenamefont
  {Tinyakov}(2020)}]{Garani:2019rcb}%
  \BibitemOpen
  \bibfield  {author} {\bibinfo {author} {\bibfnamefont {R.}~\bibnamefont
  {Garani}}\ and\ \bibinfo {author} {\bibfnamefont {P.}~\bibnamefont
  {Tinyakov}},\ }\href {\doibase 10.1016/j.physletb.2020.135403} {\bibfield
  {journal} {\bibinfo  {journal} {Phys. Lett. B}\ }\textbf {\bibinfo {volume}
  {804}},\ \bibinfo {pages} {135403} (\bibinfo {year} {2020})},\ \Eprint
  {http://arxiv.org/abs/1912.00443} {arXiv:1912.00443 [hep-ph]} \BibitemShut
  {NoStop}%
\bibitem [{\citenamefont {Chan}\ and\ \citenamefont
  {Lee}(2020)}]{Chan:2020vsr}%
  \BibitemOpen
  \bibfield  {author} {\bibinfo {author} {\bibfnamefont {M.~H.}\ \bibnamefont
  {Chan}}\ and\ \bibinfo {author} {\bibfnamefont {C.~M.}\ \bibnamefont {Lee}},\
  }\href {\doibase 10.1103/PhysRevD.102.023024} {\bibfield  {journal} {\bibinfo
   {journal} {Phys. Rev. D}\ }\textbf {\bibinfo {volume} {102}},\ \bibinfo
  {pages} {023024} (\bibinfo {year} {2020})},\ \Eprint
  {http://arxiv.org/abs/2007.01589} {arXiv:2007.01589 [astro-ph.HE]}
  \BibitemShut {NoStop}%
\bibitem [{\citenamefont {Leane}\ and\ \citenamefont
  {Linden}(2023)}]{Leane:2021tjj}%
  \BibitemOpen
  \bibfield  {author} {\bibinfo {author} {\bibfnamefont {R.~K.}\ \bibnamefont
  {Leane}}\ and\ \bibinfo {author} {\bibfnamefont {T.}~\bibnamefont {Linden}},\
  }\href {\doibase 10.1103/PhysRevLett.131.071001} {\bibfield  {journal}
  {\bibinfo  {journal} {Phys. Rev. Lett.}\ }\textbf {\bibinfo {volume} {131}},\
  \bibinfo {pages} {071001} (\bibinfo {year} {2023})},\ \Eprint
  {http://arxiv.org/abs/2104.02068} {arXiv:2104.02068 [astro-ph.HE]}
  \BibitemShut {NoStop}%
\bibitem [{\citenamefont {Blanco}\ and\ \citenamefont
  {Leane}(2023)}]{Blanco:2023qgi}%
  \BibitemOpen
  \bibfield  {author} {\bibinfo {author} {\bibfnamefont {C.}~\bibnamefont
  {Blanco}}\ and\ \bibinfo {author} {\bibfnamefont {R.~K.}\ \bibnamefont
  {Leane}},\ }\href@noop {} {\  (\bibinfo {year} {2023})},\ \Eprint
  {http://arxiv.org/abs/2312.06758} {arXiv:2312.06758 [hep-ph]} \BibitemShut
  {NoStop}%
\bibitem [{\citenamefont {Croon}\ and\ \citenamefont
  {Smirnov}(2023)}]{Croon:2023bmu}%
  \BibitemOpen
  \bibfield  {author} {\bibinfo {author} {\bibfnamefont {D.}~\bibnamefont
  {Croon}}\ and\ \bibinfo {author} {\bibfnamefont {J.}~\bibnamefont
  {Smirnov}},\ }\href@noop {} {\  (\bibinfo {year} {2023})},\ \Eprint
  {http://arxiv.org/abs/2309.02495} {arXiv:2309.02495 [hep-ph]} \BibitemShut
  {NoStop}%
\bibitem [{\citenamefont {Linden}\ \emph {et~al.}(2024)\citenamefont {Linden},
  \citenamefont {Nguyen},\ and\ \citenamefont {Tait}}]{Linden:2024uph}%
  \BibitemOpen
  \bibfield  {author} {\bibinfo {author} {\bibfnamefont {T.}~\bibnamefont
  {Linden}}, \bibinfo {author} {\bibfnamefont {T.~T.~Q.}\ \bibnamefont
  {Nguyen}}, \ and\ \bibinfo {author} {\bibfnamefont {T.~M.~P.}\ \bibnamefont
  {Tait}},\ }\href@noop {} {\  (\bibinfo {year} {2024})},\ \Eprint
  {http://arxiv.org/abs/2402.01839} {arXiv:2402.01839 [hep-ph]} \BibitemShut
  {NoStop}%
\bibitem [{\citenamefont {Leane}\ and\ \citenamefont
  {Tong}(2024)}]{Leane:2024bvh}%
  \BibitemOpen
  \bibfield  {author} {\bibinfo {author} {\bibfnamefont {R.~K.}\ \bibnamefont
  {Leane}}\ and\ \bibinfo {author} {\bibfnamefont {J.}~\bibnamefont {Tong}},\
  }\href@noop {} {\  (\bibinfo {year} {2024})},\ \Eprint
  {http://arxiv.org/abs/2405.05312} {arXiv:2405.05312 [hep-ph]} \BibitemShut
  {NoStop}%
\bibitem [{\citenamefont {{Perryman}}\ \emph {et~al.}(2014)\citenamefont
  {{Perryman}}, \citenamefont {{Hartman}}, \citenamefont {{Bakos}},\ and\
  \citenamefont {{Lindegren}}}]{Perryman_2014}%
  \BibitemOpen
  \bibfield  {author} {\bibinfo {author} {\bibfnamefont {M.}~\bibnamefont
  {{Perryman}}}, \bibinfo {author} {\bibfnamefont {J.}~\bibnamefont
  {{Hartman}}}, \bibinfo {author} {\bibfnamefont {G.~{\'A}.}\ \bibnamefont
  {{Bakos}}}, \ and\ \bibinfo {author} {\bibfnamefont {L.}~\bibnamefont
  {{Lindegren}}},\ }\href {\doibase 10.1088/0004-637X/797/1/14} {\bibfield
  {journal} {\bibinfo  {journal} {\apj}\ }\textbf {\bibinfo {volume} {797}},\
  \bibinfo {eid} {14} (\bibinfo {year} {2014})},\ \Eprint
  {http://arxiv.org/abs/1411.1173} {arXiv:1411.1173 [astro-ph.EP]} \BibitemShut
  {NoStop}%
\bibitem [{\citenamefont {Green}\ \emph {et~al.}(2012)\citenamefont {Green}
  \emph {et~al.}}]{green2012widefield}%
  \BibitemOpen
  \bibfield  {author} {\bibinfo {author} {\bibfnamefont {J.}~\bibnamefont
  {Green}} \emph {et~al.},\ }\href@noop {} {\enquote {\bibinfo {title}
  {Wide-field infrared survey telescope (wfirst) final report},}\ } (\bibinfo
  {year} {2012}),\ \Eprint {http://arxiv.org/abs/1208.4012} {arXiv:1208.4012
  [astro-ph.IM]} \BibitemShut {NoStop}%
\bibitem [{\citenamefont {Johnson}\ \emph {et~al.}(2020)\citenamefont
  {Johnson}, \citenamefont {Penny}, \citenamefont {Gaudi}, \citenamefont
  {Kerins}, \citenamefont {Rattenbury}, \citenamefont {Robin}, \citenamefont
  {Novati},\ and\ \citenamefont {Henderson}}]{Johnson_2020}%
  \BibitemOpen
  \bibfield  {author} {\bibinfo {author} {\bibfnamefont {S.~A.}\ \bibnamefont
  {Johnson}}, \bibinfo {author} {\bibfnamefont {M.}~\bibnamefont {Penny}},
  \bibinfo {author} {\bibfnamefont {B.~S.}\ \bibnamefont {Gaudi}}, \bibinfo
  {author} {\bibfnamefont {E.}~\bibnamefont {Kerins}}, \bibinfo {author}
  {\bibfnamefont {N.~J.}\ \bibnamefont {Rattenbury}}, \bibinfo {author}
  {\bibfnamefont {A.~C.}\ \bibnamefont {Robin}}, \bibinfo {author}
  {\bibfnamefont {S.~C.}\ \bibnamefont {Novati}}, \ and\ \bibinfo {author}
  {\bibfnamefont {C.~B.}\ \bibnamefont {Henderson}},\ }\href {\doibase
  10.3847/1538-3881/aba75b} {\bibfield  {journal} {\bibinfo  {journal} {The
  Astronomical Journal}\ }\textbf {\bibinfo {volume} {160}},\ \bibinfo {pages}
  {123} (\bibinfo {year} {2020})}\BibitemShut {NoStop}%
\bibitem [{\citenamefont {Bachelet}\ and\ \citenamefont
  {Penny}(2019)}]{Bachelet_2019}%
  \BibitemOpen
  \bibfield  {author} {\bibinfo {author} {\bibfnamefont {E.}~\bibnamefont
  {Bachelet}}\ and\ \bibinfo {author} {\bibfnamefont {M.}~\bibnamefont
  {Penny}},\ }\href {\doibase 10.3847/2041-8213/ab2da5} {\bibfield  {journal}
  {\bibinfo  {journal} {The Astrophysical Journal}\ }\textbf {\bibinfo {volume}
  {880}},\ \bibinfo {pages} {L32} (\bibinfo {year} {2019})}\BibitemShut
  {NoStop}%
\bibitem [{\citenamefont {{Zhu}}\ \emph {et~al.}(2016)\citenamefont {{Zhu}},
  \citenamefont {{Calchi Novati}}, \citenamefont {{Gould}}, \citenamefont
  {{Udalski}}, \citenamefont {{Han}}, \citenamefont {{Shvartzvald}},
  \citenamefont {{Ranc}}, \citenamefont {{J{\o}rgensen}}, \citenamefont
  {{Poleski}}, \citenamefont {{Bozza}}, \citenamefont {{Beichman}},
  \citenamefont {{Bryden}}, \citenamefont {{Carey}}, \citenamefont {{Gaudi}},
  \citenamefont {{Henderson}}, \citenamefont {{Pogge}}, \citenamefont
  {{Porritt}}, \citenamefont {{Wibking}}, \citenamefont {{Yee}}, \citenamefont
  {{SPITZER Team}}, \citenamefont {{Pawlak}}, \citenamefont {{Szyma{\'n}ski}},
  \citenamefont {{Skowron}}, \citenamefont {{Mr{\'o}z}}, \citenamefont
  {{Koz{\l}owski}}, \citenamefont {{Wyrzykowski}}, \citenamefont
  {{Pietrukowicz}}, \citenamefont {{Pietrzy{\'n}ski}}, \citenamefont
  {{Soszy{\'n}ski}}, \citenamefont {{Ulaczyk}}, \citenamefont {{OGLE Group}},
  \citenamefont {{Choi}}, \citenamefont {{Park}}, \citenamefont {{Jung}},
  \citenamefont {{Shin}}, \citenamefont {{Albrow}}, \citenamefont {{Park}},
  \citenamefont {{Kim}}, \citenamefont {{Lee}}, \citenamefont {{Cha}},
  \citenamefont {{Kim}}, \citenamefont {{Lee}}, \citenamefont {{KMTNET Group}},
  \citenamefont {{Friedmann}}, \citenamefont {{Kaspi}}, \citenamefont {{Maoz}},
  \citenamefont {{WISE Group}}, \citenamefont {{Hundertmark}}, \citenamefont
  {{Street}}, \citenamefont {{Tsapras}}, \citenamefont {{Bramich}},
  \citenamefont {{Cassan}}, \citenamefont {{Dominik}}, \citenamefont
  {{Bachelet}}, \citenamefont {{Dong}}, \citenamefont {{Figuera Jaimes}},
  \citenamefont {{Horne}}, \citenamefont {{Mao}}, \citenamefont {{Menzies}},
  \citenamefont {{Schmidt}}, \citenamefont {{Snodgrass}}, \citenamefont
  {{Steele}}, \citenamefont {{Wambsganss}}, \citenamefont {{RoboNeT Team}},
  \citenamefont {{Skottfelt}}, \citenamefont {{Andersen}}, \citenamefont
  {{Burgdorf}}, \citenamefont {{Ciceri}}, \citenamefont {{D'Ago}},
  \citenamefont {{Evans}}, \citenamefont {{Gu}}, \citenamefont {{Hinse}},
  \citenamefont {{Kerins}}, \citenamefont {{Korhonen}}, \citenamefont
  {{Kuffmeier}}, \citenamefont {{Mancini}}, \citenamefont {{Peixinho}},
  \citenamefont {{Popovas}}, \citenamefont {{Rabus}}, \citenamefont {{Rahvar}},
  \citenamefont {{Tronsgaard}}, \citenamefont {{Scarpetta}}, \citenamefont
  {{Southworth}}, \citenamefont {{Surdej}}, \citenamefont {{von Essen}},
  \citenamefont {{Wang}}, \citenamefont {{Wertz}},\ and\ \citenamefont
  {{MiNDSTEP Group}}}]{Zhu2016}%
  \BibitemOpen
  \bibfield  {author} {\bibinfo {author} {\bibfnamefont {W.}~\bibnamefont
  {{Zhu}}}, \bibinfo {author} {\bibfnamefont {S.}~\bibnamefont {{Calchi
  Novati}}}, \bibinfo {author} {\bibfnamefont {A.}~\bibnamefont {{Gould}}},
  \bibinfo {author} {\bibfnamefont {A.}~\bibnamefont {{Udalski}}}, \bibinfo
  {author} {\bibfnamefont {C.}~\bibnamefont {{Han}}}, \bibinfo {author}
  {\bibfnamefont {Y.}~\bibnamefont {{Shvartzvald}}}, \bibinfo {author}
  {\bibfnamefont {C.}~\bibnamefont {{Ranc}}}, \bibinfo {author} {\bibfnamefont
  {U.~G.}\ \bibnamefont {{J{\o}rgensen}}}, \bibinfo {author} {\bibfnamefont
  {R.}~\bibnamefont {{Poleski}}}, \bibinfo {author} {\bibfnamefont
  {V.}~\bibnamefont {{Bozza}}}, \bibinfo {author} {\bibfnamefont
  {C.}~\bibnamefont {{Beichman}}}, \bibinfo {author} {\bibfnamefont
  {G.}~\bibnamefont {{Bryden}}}, \bibinfo {author} {\bibfnamefont
  {S.}~\bibnamefont {{Carey}}}, \bibinfo {author} {\bibfnamefont {B.~S.}\
  \bibnamefont {{Gaudi}}}, \bibinfo {author} {\bibfnamefont {C.~B.}\
  \bibnamefont {{Henderson}}}, \bibinfo {author} {\bibfnamefont {R.~W.}\
  \bibnamefont {{Pogge}}}, \bibinfo {author} {\bibfnamefont {I.}~\bibnamefont
  {{Porritt}}}, \bibinfo {author} {\bibfnamefont {B.}~\bibnamefont
  {{Wibking}}}, \bibinfo {author} {\bibfnamefont {J.~C.}\ \bibnamefont
  {{Yee}}}, \bibinfo {author} {\bibnamefont {{SPITZER Team}}}, \bibinfo
  {author} {\bibfnamefont {M.}~\bibnamefont {{Pawlak}}}, \bibinfo {author}
  {\bibfnamefont {M.~K.}\ \bibnamefont {{Szyma{\'n}ski}}}, \bibinfo {author}
  {\bibfnamefont {J.}~\bibnamefont {{Skowron}}}, \bibinfo {author}
  {\bibfnamefont {P.}~\bibnamefont {{Mr{\'o}z}}}, \bibinfo {author}
  {\bibfnamefont {S.}~\bibnamefont {{Koz{\l}owski}}}, \bibinfo {author}
  {\bibfnamefont {{\L}.}~\bibnamefont {{Wyrzykowski}}}, \bibinfo {author}
  {\bibfnamefont {P.}~\bibnamefont {{Pietrukowicz}}}, \bibinfo {author}
  {\bibfnamefont {G.}~\bibnamefont {{Pietrzy{\'n}ski}}}, \bibinfo {author}
  {\bibfnamefont {I.}~\bibnamefont {{Soszy{\'n}ski}}}, \bibinfo {author}
  {\bibfnamefont {K.}~\bibnamefont {{Ulaczyk}}}, \bibinfo {author}
  {\bibnamefont {{OGLE Group}}}, \bibinfo {author} {\bibfnamefont {J.~Y.}\
  \bibnamefont {{Choi}}}, \bibinfo {author} {\bibfnamefont {H.}~\bibnamefont
  {{Park}}}, \bibinfo {author} {\bibfnamefont {Y.~K.}\ \bibnamefont {{Jung}}},
  \bibinfo {author} {\bibfnamefont {I.~G.}\ \bibnamefont {{Shin}}}, \bibinfo
  {author} {\bibfnamefont {M.~D.}\ \bibnamefont {{Albrow}}}, \bibinfo {author}
  {\bibfnamefont {B.~G.}\ \bibnamefont {{Park}}}, \bibinfo {author}
  {\bibfnamefont {S.~L.}\ \bibnamefont {{Kim}}}, \bibinfo {author}
  {\bibfnamefont {C.~U.}\ \bibnamefont {{Lee}}}, \bibinfo {author}
  {\bibfnamefont {S.~M.}\ \bibnamefont {{Cha}}}, \bibinfo {author}
  {\bibfnamefont {D.~J.}\ \bibnamefont {{Kim}}}, \bibinfo {author}
  {\bibfnamefont {Y.}~\bibnamefont {{Lee}}}, \bibinfo {author} {\bibnamefont
  {{KMTNET Group}}}, \bibinfo {author} {\bibfnamefont {M.}~\bibnamefont
  {{Friedmann}}}, \bibinfo {author} {\bibfnamefont {S.}~\bibnamefont
  {{Kaspi}}}, \bibinfo {author} {\bibfnamefont {D.}~\bibnamefont {{Maoz}}},
  \bibinfo {author} {\bibnamefont {{WISE Group}}}, \bibinfo {author}
  {\bibfnamefont {M.}~\bibnamefont {{Hundertmark}}}, \bibinfo {author}
  {\bibfnamefont {R.~A.}\ \bibnamefont {{Street}}}, \bibinfo {author}
  {\bibfnamefont {Y.}~\bibnamefont {{Tsapras}}}, \bibinfo {author}
  {\bibfnamefont {D.~M.}\ \bibnamefont {{Bramich}}}, \bibinfo {author}
  {\bibfnamefont {A.}~\bibnamefont {{Cassan}}}, \bibinfo {author}
  {\bibfnamefont {M.}~\bibnamefont {{Dominik}}}, \bibinfo {author}
  {\bibfnamefont {E.}~\bibnamefont {{Bachelet}}}, \bibinfo {author}
  {\bibfnamefont {S.}~\bibnamefont {{Dong}}}, \bibinfo {author} {\bibfnamefont
  {R.}~\bibnamefont {{Figuera Jaimes}}}, \bibinfo {author} {\bibfnamefont
  {K.}~\bibnamefont {{Horne}}}, \bibinfo {author} {\bibfnamefont
  {S.}~\bibnamefont {{Mao}}}, \bibinfo {author} {\bibfnamefont
  {J.}~\bibnamefont {{Menzies}}}, \bibinfo {author} {\bibfnamefont
  {R.}~\bibnamefont {{Schmidt}}}, \bibinfo {author} {\bibfnamefont
  {C.}~\bibnamefont {{Snodgrass}}}, \bibinfo {author} {\bibfnamefont {I.~A.}\
  \bibnamefont {{Steele}}}, \bibinfo {author} {\bibfnamefont {J.}~\bibnamefont
  {{Wambsganss}}}, \bibinfo {author} {\bibnamefont {{RoboNeT Team}}}, \bibinfo
  {author} {\bibfnamefont {J.}~\bibnamefont {{Skottfelt}}}, \bibinfo {author}
  {\bibfnamefont {M.~I.}\ \bibnamefont {{Andersen}}}, \bibinfo {author}
  {\bibfnamefont {M.~J.}\ \bibnamefont {{Burgdorf}}}, \bibinfo {author}
  {\bibfnamefont {S.}~\bibnamefont {{Ciceri}}}, \bibinfo {author}
  {\bibfnamefont {G.}~\bibnamefont {{D'Ago}}}, \bibinfo {author} {\bibfnamefont
  {D.~F.}\ \bibnamefont {{Evans}}}, \bibinfo {author} {\bibfnamefont {S.~H.}\
  \bibnamefont {{Gu}}}, \bibinfo {author} {\bibfnamefont {T.~C.}\ \bibnamefont
  {{Hinse}}}, \bibinfo {author} {\bibfnamefont {E.}~\bibnamefont {{Kerins}}},
  \bibinfo {author} {\bibfnamefont {H.}~\bibnamefont {{Korhonen}}}, \bibinfo
  {author} {\bibfnamefont {M.}~\bibnamefont {{Kuffmeier}}}, \bibinfo {author}
  {\bibfnamefont {L.}~\bibnamefont {{Mancini}}}, \bibinfo {author}
  {\bibfnamefont {N.}~\bibnamefont {{Peixinho}}}, \bibinfo {author}
  {\bibfnamefont {A.}~\bibnamefont {{Popovas}}}, \bibinfo {author}
  {\bibfnamefont {M.}~\bibnamefont {{Rabus}}}, \bibinfo {author} {\bibfnamefont
  {S.}~\bibnamefont {{Rahvar}}}, \bibinfo {author} {\bibfnamefont
  {R.}~\bibnamefont {{Tronsgaard}}}, \bibinfo {author} {\bibfnamefont
  {G.}~\bibnamefont {{Scarpetta}}}, \bibinfo {author} {\bibfnamefont
  {J.}~\bibnamefont {{Southworth}}}, \bibinfo {author} {\bibfnamefont
  {J.}~\bibnamefont {{Surdej}}}, \bibinfo {author} {\bibfnamefont
  {C.}~\bibnamefont {{von Essen}}}, \bibinfo {author} {\bibfnamefont {Y.~B.}\
  \bibnamefont {{Wang}}}, \bibinfo {author} {\bibfnamefont {O.}~\bibnamefont
  {{Wertz}}}, \ and\ \bibinfo {author} {\bibnamefont {{MiNDSTEP Group}}},\
  }\href {\doibase 10.3847/0004-637X/825/1/60} {\bibfield  {journal} {\bibinfo
  {journal} {\apj}\ }\textbf {\bibinfo {volume} {825}},\ \bibinfo {eid} {60}
  (\bibinfo {year} {2016})},\ \Eprint {http://arxiv.org/abs/1510.02097}
  {arXiv:1510.02097 [astro-ph.SR]} \BibitemShut {NoStop}%
\bibitem [{\citenamefont {{Eddington}}(1913)}]{1913MNRAS..73..359E}%
  \BibitemOpen
  \bibfield  {author} {\bibinfo {author} {\bibfnamefont {A.~S.}\ \bibnamefont
  {{Eddington}}},\ }\href {\doibase 10.1093/mnras/73.5.359} {\bibfield
  {journal} {\bibinfo  {journal} {\mnras}\ }\textbf {\bibinfo {volume} {73}},\
  \bibinfo {pages} {359} (\bibinfo {year} {1913})}\BibitemShut {NoStop}%
\bibitem [{\citenamefont {Kelly}(2007)}]{Kelly:2007jy}%
  \BibitemOpen
  \bibfield  {author} {\bibinfo {author} {\bibfnamefont {B.~C.}\ \bibnamefont
  {Kelly}},\ }\href {\doibase 10.1086/519947} {\bibfield  {journal} {\bibinfo
  {journal} {Astrophys. J.}\ }\textbf {\bibinfo {volume} {665}},\ \bibinfo
  {pages} {1489} (\bibinfo {year} {2007})},\ \Eprint
  {http://arxiv.org/abs/0705.2774} {arXiv:0705.2774 [astro-ph]} \BibitemShut
  {NoStop}%
\bibitem [{\citenamefont {{March}}\ \emph {et~al.}(2011)\citenamefont
  {{March}}, \citenamefont {{Trotta}}, \citenamefont {{Berkes}}, \citenamefont
  {{Starkman}},\ and\ \citenamefont {{Vaudrevange}}}]{2011MNRAS.418.2308M}%
  \BibitemOpen
  \bibfield  {author} {\bibinfo {author} {\bibfnamefont {M.~C.}\ \bibnamefont
  {{March}}}, \bibinfo {author} {\bibfnamefont {R.}~\bibnamefont {{Trotta}}},
  \bibinfo {author} {\bibfnamefont {P.}~\bibnamefont {{Berkes}}}, \bibinfo
  {author} {\bibfnamefont {G.~D.}\ \bibnamefont {{Starkman}}}, \ and\ \bibinfo
  {author} {\bibfnamefont {P.~M.}\ \bibnamefont {{Vaudrevange}}},\ }\href
  {\doibase 10.1111/j.1365-2966.2011.19584.x} {\bibfield  {journal} {\bibinfo
  {journal} {\mnras}\ }\textbf {\bibinfo {volume} {418}},\ \bibinfo {pages}
  {2308} (\bibinfo {year} {2011})},\ \Eprint {http://arxiv.org/abs/1102.3237}
  {arXiv:1102.3237 [astro-ph.CO]} \BibitemShut {NoStop}%
\bibitem [{\citenamefont {{Phillips}}\ \emph {et~al.}(2020)\citenamefont
  {{Phillips}}, \citenamefont {{Tremblin}}, \citenamefont {{Baraffe}},
  \citenamefont {{Chabrier}}, \citenamefont {{Allard}}, \citenamefont
  {{Spiegelman}}, \citenamefont {{Goyal}}, \citenamefont {{Drummond}},\ and\
  \citenamefont {{H{\'e}brard}}}]{2020A&A...637A..38P}%
  \BibitemOpen
  \bibfield  {author} {\bibinfo {author} {\bibfnamefont {M.~W.}\ \bibnamefont
  {{Phillips}}}, \bibinfo {author} {\bibfnamefont {P.}~\bibnamefont
  {{Tremblin}}}, \bibinfo {author} {\bibfnamefont {I.}~\bibnamefont
  {{Baraffe}}}, \bibinfo {author} {\bibfnamefont {G.}~\bibnamefont
  {{Chabrier}}}, \bibinfo {author} {\bibfnamefont {N.~F.}\ \bibnamefont
  {{Allard}}}, \bibinfo {author} {\bibfnamefont {F.}~\bibnamefont
  {{Spiegelman}}}, \bibinfo {author} {\bibfnamefont {J.~M.}\ \bibnamefont
  {{Goyal}}}, \bibinfo {author} {\bibfnamefont {B.}~\bibnamefont {{Drummond}}},
  \ and\ \bibinfo {author} {\bibfnamefont {E.}~\bibnamefont {{H{\'e}brard}}},\
  }\href {\doibase 10.1051/0004-6361/201937381} {\bibfield  {journal} {\bibinfo
   {journal} {\aap}\ }\textbf {\bibinfo {volume} {637}},\ \bibinfo {eid} {A38}
  (\bibinfo {year} {2020})},\ \Eprint {http://arxiv.org/abs/2003.13717}
  {arXiv:2003.13717 [astro-ph.SR]} \BibitemShut {NoStop}%
\bibitem [{\citenamefont {Karukes}\ \emph {et~al.}(2019)\citenamefont
  {Karukes}, \citenamefont {Benito}, \citenamefont {Iocco}, \citenamefont
  {Trotta},\ and\ \citenamefont {Geringer-Sameth}}]{Karukes:2019jxv}%
  \BibitemOpen
  \bibfield  {author} {\bibinfo {author} {\bibfnamefont {E.~V.}\ \bibnamefont
  {Karukes}}, \bibinfo {author} {\bibfnamefont {M.}~\bibnamefont {Benito}},
  \bibinfo {author} {\bibfnamefont {F.}~\bibnamefont {Iocco}}, \bibinfo
  {author} {\bibfnamefont {R.}~\bibnamefont {Trotta}}, \ and\ \bibinfo {author}
  {\bibfnamefont {A.}~\bibnamefont {Geringer-Sameth}},\ }\href {\doibase
  10.1088/1475-7516/2019/09/046} {\bibfield  {journal} {\bibinfo  {journal}
  {JCAP}\ }\textbf {\bibinfo {volume} {09}},\ \bibinfo {pages} {046} (\bibinfo
  {year} {2019})},\ \Eprint {http://arxiv.org/abs/1901.02463} {arXiv:1901.02463
  [astro-ph.GA]} \BibitemShut {NoStop}%
\bibitem [{\citenamefont {{Benito}}\ \emph {et~al.}(2019)\citenamefont
  {{Benito}}, \citenamefont {{Cuoco}},\ and\ \citenamefont
  {{Iocco}}}]{2019JCAP...03..033B}%
  \BibitemOpen
  \bibfield  {author} {\bibinfo {author} {\bibfnamefont {M.}~\bibnamefont
  {{Benito}}}, \bibinfo {author} {\bibfnamefont {A.}~\bibnamefont {{Cuoco}}}, \
  and\ \bibinfo {author} {\bibfnamefont {F.}~\bibnamefont {{Iocco}}},\ }\href
  {\doibase 10.1088/1475-7516/2019/03/033} {\bibfield  {journal} {\bibinfo
  {journal} {\jcap}\ }\textbf {\bibinfo {volume} {2019}},\ \bibinfo {eid} {033}
  (\bibinfo {year} {2019})},\ \Eprint {http://arxiv.org/abs/1901.02460}
  {arXiv:1901.02460 [astro-ph.GA]} \BibitemShut {NoStop}%
\bibitem [{\citenamefont {{Benito}}\ \emph {et~al.}(2021)\citenamefont
  {{Benito}}, \citenamefont {{Iocco}},\ and\ \citenamefont
  {{Cuoco}}}]{2021PDU....3200826B}%
  \BibitemOpen
  \bibfield  {author} {\bibinfo {author} {\bibfnamefont {M.}~\bibnamefont
  {{Benito}}}, \bibinfo {author} {\bibfnamefont {F.}~\bibnamefont {{Iocco}}}, \
  and\ \bibinfo {author} {\bibfnamefont {A.}~\bibnamefont {{Cuoco}}},\ }\href
  {\doibase 10.1016/j.dark.2021.100826} {\bibfield  {journal} {\bibinfo
  {journal} {Physics of the Dark Universe}\ }\textbf {\bibinfo {volume} {32}},\
  \bibinfo {eid} {100826} (\bibinfo {year} {2021})},\ \Eprint
  {http://arxiv.org/abs/2009.13523} {arXiv:2009.13523 [astro-ph.GA]}
  \BibitemShut {NoStop}%
\bibitem [{\citenamefont {{Board}}\ \emph {et~al.}(2021)\citenamefont
  {{Board}}, \citenamefont {{Bozorgnia}}, \citenamefont {{Strigari}},
  \citenamefont {{Grand}}, \citenamefont {{Fattahi}}, \citenamefont {{Frenk}},
  \citenamefont {{Marinacci}}, \citenamefont {{Navarro}},\ and\ \citenamefont
  {{Oman}}}]{2021JCAP...04..070B}%
  \BibitemOpen
  \bibfield  {author} {\bibinfo {author} {\bibfnamefont {E.}~\bibnamefont
  {{Board}}}, \bibinfo {author} {\bibfnamefont {N.}~\bibnamefont
  {{Bozorgnia}}}, \bibinfo {author} {\bibfnamefont {L.~E.}\ \bibnamefont
  {{Strigari}}}, \bibinfo {author} {\bibfnamefont {R.~J.~J.}\ \bibnamefont
  {{Grand}}}, \bibinfo {author} {\bibfnamefont {A.}~\bibnamefont {{Fattahi}}},
  \bibinfo {author} {\bibfnamefont {C.~S.}\ \bibnamefont {{Frenk}}}, \bibinfo
  {author} {\bibfnamefont {F.}~\bibnamefont {{Marinacci}}}, \bibinfo {author}
  {\bibfnamefont {J.~F.}\ \bibnamefont {{Navarro}}}, \ and\ \bibinfo {author}
  {\bibfnamefont {K.~A.}\ \bibnamefont {{Oman}}},\ }\href {\doibase
  10.1088/1475-7516/2021/04/070} {\bibfield  {journal} {\bibinfo  {journal}
  {\jcap}\ }\textbf {\bibinfo {volume} {2021}},\ \bibinfo {eid} {070} (\bibinfo
  {year} {2021})},\ \Eprint {http://arxiv.org/abs/2101.06284} {arXiv:2101.06284
  [astro-ph.CO]} \BibitemShut {NoStop}%
\bibitem [{\citenamefont {{P{\~o}der}}\ \emph {et~al.}(2023)\citenamefont
  {{P{\~o}der}}, \citenamefont {{Benito}}, \citenamefont {{Pata}},
  \citenamefont {{Kipper}}, \citenamefont {{Ramler}}, \citenamefont
  {{H{\"u}tsi}}, \citenamefont {{Kolka}},\ and\ \citenamefont
  {{Thomas}}}]{2023A&A...676A.134P}%
  \BibitemOpen
  \bibfield  {author} {\bibinfo {author} {\bibfnamefont {S.}~\bibnamefont
  {{P{\~o}der}}}, \bibinfo {author} {\bibfnamefont {M.}~\bibnamefont
  {{Benito}}}, \bibinfo {author} {\bibfnamefont {J.}~\bibnamefont {{Pata}}},
  \bibinfo {author} {\bibfnamefont {R.}~\bibnamefont {{Kipper}}}, \bibinfo
  {author} {\bibfnamefont {H.}~\bibnamefont {{Ramler}}}, \bibinfo {author}
  {\bibfnamefont {G.}~\bibnamefont {{H{\"u}tsi}}}, \bibinfo {author}
  {\bibfnamefont {I.}~\bibnamefont {{Kolka}}}, \ and\ \bibinfo {author}
  {\bibfnamefont {G.~F.}\ \bibnamefont {{Thomas}}},\ }\href {\doibase
  10.1051/0004-6361/202346474} {\bibfield  {journal} {\bibinfo  {journal}
  {\aap}\ }\textbf {\bibinfo {volume} {676}},\ \bibinfo {eid} {A134} (\bibinfo
  {year} {2023})},\ \Eprint {http://arxiv.org/abs/2309.02895} {arXiv:2309.02895
  [astro-ph.GA]} \BibitemShut {NoStop}%
\bibitem [{\citenamefont {{Saumon}}\ and\ \citenamefont
  {{Marley}}(2008)}]{2008ApJ...689.1327S}%
  \BibitemOpen
  \bibfield  {author} {\bibinfo {author} {\bibfnamefont {D.}~\bibnamefont
  {{Saumon}}}\ and\ \bibinfo {author} {\bibfnamefont {M.~S.}\ \bibnamefont
  {{Marley}}},\ }\href {\doibase 10.1086/592734} {\bibfield  {journal}
  {\bibinfo  {journal} {\apj}\ }\textbf {\bibinfo {volume} {689}},\ \bibinfo
  {pages} {1327} (\bibinfo {year} {2008})},\ \Eprint
  {http://arxiv.org/abs/0808.2611} {arXiv:0808.2611 [astro-ph]} \BibitemShut
  {NoStop}%
\bibitem [{\citenamefont {{Marley}}\ \emph {et~al.}(2021)\citenamefont
  {{Marley}}, \citenamefont {{Saumon}}, \citenamefont {{Visscher}},
  \citenamefont {{Lupu}}, \citenamefont {{Freedman}}, \citenamefont {{Morley}},
  \citenamefont {{Fortney}}, \citenamefont {{Seay}}, \citenamefont {{Smith}},
  \citenamefont {{Teal}},\ and\ \citenamefont {{Wang}}}]{2021arXiv210707434M}%
  \BibitemOpen
  \bibfield  {author} {\bibinfo {author} {\bibfnamefont {M.~S.}\ \bibnamefont
  {{Marley}}}, \bibinfo {author} {\bibfnamefont {D.}~\bibnamefont {{Saumon}}},
  \bibinfo {author} {\bibfnamefont {C.}~\bibnamefont {{Visscher}}}, \bibinfo
  {author} {\bibfnamefont {R.}~\bibnamefont {{Lupu}}}, \bibinfo {author}
  {\bibfnamefont {R.}~\bibnamefont {{Freedman}}}, \bibinfo {author}
  {\bibfnamefont {C.}~\bibnamefont {{Morley}}}, \bibinfo {author}
  {\bibfnamefont {J.~J.}\ \bibnamefont {{Fortney}}}, \bibinfo {author}
  {\bibfnamefont {C.}~\bibnamefont {{Seay}}}, \bibinfo {author} {\bibfnamefont
  {A.~J.~R.~W.}\ \bibnamefont {{Smith}}}, \bibinfo {author} {\bibfnamefont
  {D.~J.}\ \bibnamefont {{Teal}}}, \ and\ \bibinfo {author} {\bibfnamefont
  {R.}~\bibnamefont {{Wang}}},\ }\href@noop {} {\bibfield  {journal} {\bibinfo
  {journal} {arXiv e-prints}\ ,\ \bibinfo {eid} {arXiv:2107.07434}} (\bibinfo
  {year} {2021})},\ \Eprint {http://arxiv.org/abs/2107.07434} {arXiv:2107.07434
  [astro-ph.SR]} \BibitemShut {NoStop}%
\bibitem [{\citenamefont {Acevedo}\ \emph
  {et~al.}(2024{\natexlab{c}})\citenamefont {Acevedo}, \citenamefont {Leane},\
  and\ \citenamefont {Reilly}}]{Acevedo:2024zkg}%
  \BibitemOpen
  \bibfield  {author} {\bibinfo {author} {\bibfnamefont {J.~F.}\ \bibnamefont
  {Acevedo}}, \bibinfo {author} {\bibfnamefont {R.~K.}\ \bibnamefont {Leane}},
  \ and\ \bibinfo {author} {\bibfnamefont {A.~J.}\ \bibnamefont {Reilly}},\
  }\href@noop {} {\  (\bibinfo {year} {2024}{\natexlab{c}})},\ \Eprint
  {http://arxiv.org/abs/2405.02393} {arXiv:2405.02393 [astro-ph.EP]}
  \BibitemShut {NoStop}%
\bibitem [{\citenamefont {{Barbuy}}\ \emph {et~al.}(2018)\citenamefont
  {{Barbuy}}, \citenamefont {{Chiappini}},\ and\ \citenamefont
  {{Gerhard}}}]{2018ARA&A..56..223B}%
  \BibitemOpen
  \bibfield  {author} {\bibinfo {author} {\bibfnamefont {B.}~\bibnamefont
  {{Barbuy}}}, \bibinfo {author} {\bibfnamefont {C.}~\bibnamefont
  {{Chiappini}}}, \ and\ \bibinfo {author} {\bibfnamefont {O.}~\bibnamefont
  {{Gerhard}}},\ }\href {\doibase 10.1146/annurev-astro-081817-051826}
  {\bibfield  {journal} {\bibinfo  {journal} {\araa}\ }\textbf {\bibinfo
  {volume} {56}},\ \bibinfo {pages} {223} (\bibinfo {year} {2018})},\ \Eprint
  {http://arxiv.org/abs/1805.01142} {arXiv:1805.01142 [astro-ph.GA]}
  \BibitemShut {NoStop}%
\bibitem [{\citenamefont {{Joyce}}\ \emph {et~al.}(2023)\citenamefont
  {{Joyce}}, \citenamefont {{Johnson}}, \citenamefont {{Marchetti}},
  \citenamefont {{Rich}}, \citenamefont {{Simion}},\ and\ \citenamefont
  {{Bourke}}}]{2023ApJ...946...28J}%
  \BibitemOpen
  \bibfield  {author} {\bibinfo {author} {\bibfnamefont {M.}~\bibnamefont
  {{Joyce}}}, \bibinfo {author} {\bibfnamefont {C.~I.}\ \bibnamefont
  {{Johnson}}}, \bibinfo {author} {\bibfnamefont {T.}~\bibnamefont
  {{Marchetti}}}, \bibinfo {author} {\bibfnamefont {R.~M.}\ \bibnamefont
  {{Rich}}}, \bibinfo {author} {\bibfnamefont {I.}~\bibnamefont {{Simion}}}, \
  and\ \bibinfo {author} {\bibfnamefont {J.}~\bibnamefont {{Bourke}}},\ }\href
  {\doibase 10.3847/1538-4357/acb692} {\bibfield  {journal} {\bibinfo
  {journal} {\apj}\ }\textbf {\bibinfo {volume} {946}},\ \bibinfo {eid} {28}
  (\bibinfo {year} {2023})},\ \Eprint {http://arxiv.org/abs/2205.07964}
  {arXiv:2205.07964 [astro-ph.SR]} \BibitemShut {NoStop}%
\bibitem [{\citenamefont {{Stanek}}\ \emph {et~al.}(1997)\citenamefont
  {{Stanek}}, \citenamefont {{Udalski}}, \citenamefont {{Szyma{\'N}ski}},
  \citenamefont {{Ka{\L}u{\.Z}ny}}, \citenamefont {{Kubiak}}, \citenamefont
  {{Mateo}},\ and\ \citenamefont {{Krzemi{\'N}ski}}}]{1997ApJ...477..163S}%
  \BibitemOpen
  \bibfield  {author} {\bibinfo {author} {\bibfnamefont {K.~Z.}\ \bibnamefont
  {{Stanek}}}, \bibinfo {author} {\bibfnamefont {A.}~\bibnamefont {{Udalski}}},
  \bibinfo {author} {\bibfnamefont {M.}~\bibnamefont {{Szyma{\'N}ski}}},
  \bibinfo {author} {\bibfnamefont {J.}~\bibnamefont {{Ka{\L}u{\.Z}ny}}},
  \bibinfo {author} {\bibfnamefont {Z.~M.}\ \bibnamefont {{Kubiak}}}, \bibinfo
  {author} {\bibfnamefont {M.}~\bibnamefont {{Mateo}}}, \ and\ \bibinfo
  {author} {\bibfnamefont {W.}~\bibnamefont {{Krzemi{\'N}ski}}},\ }\href
  {\doibase 10.1086/303702} {\bibfield  {journal} {\bibinfo  {journal} {\apj}\
  }\textbf {\bibinfo {volume} {477}},\ \bibinfo {pages} {163} (\bibinfo {year}
  {1997})},\ \Eprint {http://arxiv.org/abs/astro-ph/9605162}
  {arXiv:astro-ph/9605162 [astro-ph]} \BibitemShut {NoStop}%
\bibitem [{\citenamefont {{Kirkpatrick}}\ \emph {et~al.}(2021)\citenamefont
  {{Kirkpatrick}}, \citenamefont {{Gelino}}, \citenamefont {{Faherty}},
  \citenamefont {{Meisner}}, \citenamefont {{Caselden}}, \citenamefont
  {{Schneider}}, \citenamefont {{Marocco}}, \citenamefont {{Cayago}},
  \citenamefont {{Smart}}, \citenamefont {{Eisenhardt}}, \citenamefont
  {{Kuchner}}, \citenamefont {{Wright}}, \citenamefont {{Cushing}},
  \citenamefont {{Allers}}, \citenamefont {{Bardalez Gagliuffi}}, \citenamefont
  {{Burgasser}}, \citenamefont {{Gagn{\'e}}}, \citenamefont {{Logsdon}},
  \citenamefont {{Martin}}, \citenamefont {{Ingalls}}, \citenamefont
  {{Lowrance}}, \citenamefont {{Abrahams}}, \citenamefont {{Aganze}},
  \citenamefont {{Gerasimov}}, \citenamefont {{Gonzales}}, \citenamefont
  {{Hsu}}, \citenamefont {{Kamraj}}, \citenamefont {{Kiman}}, \citenamefont
  {{Rees}}, \citenamefont {{Theissen}}, \citenamefont {{Ammar}}, \citenamefont
  {{Andersen}}, \citenamefont {{Beaulieu}}, \citenamefont {{Colin}},
  \citenamefont {{Elachi}}, \citenamefont {{Goodman}}, \citenamefont
  {{Gramaize}}, \citenamefont {{Hamlet}}, \citenamefont {{Hong}}, \citenamefont
  {{Jonkeren}}, \citenamefont {{Khalil}}, \citenamefont {{Martin}},
  \citenamefont {{Pendrill}}, \citenamefont {{Pumphrey}}, \citenamefont
  {{Rothermich}}, \citenamefont {{Sainio}}, \citenamefont {{Stenner}},
  \citenamefont {{Tanner}}, \citenamefont {{Th{\'e}venot}}, \citenamefont
  {{Voloshin}}, \citenamefont {{Walla}}, \citenamefont {{W{\k{e}}dracki}},\
  and\ \citenamefont {{Backyard Worlds: Planet 9
  Collaboration}}}]{2021ApJS..253....7K}%
  \BibitemOpen
  \bibfield  {author} {\bibinfo {author} {\bibfnamefont {J.~D.}\ \bibnamefont
  {{Kirkpatrick}}}, \bibinfo {author} {\bibfnamefont {C.~R.}\ \bibnamefont
  {{Gelino}}}, \bibinfo {author} {\bibfnamefont {J.~K.}\ \bibnamefont
  {{Faherty}}}, \bibinfo {author} {\bibfnamefont {A.~M.}\ \bibnamefont
  {{Meisner}}}, \bibinfo {author} {\bibfnamefont {D.}~\bibnamefont
  {{Caselden}}}, \bibinfo {author} {\bibfnamefont {A.~C.}\ \bibnamefont
  {{Schneider}}}, \bibinfo {author} {\bibfnamefont {F.}~\bibnamefont
  {{Marocco}}}, \bibinfo {author} {\bibfnamefont {A.~J.}\ \bibnamefont
  {{Cayago}}}, \bibinfo {author} {\bibfnamefont {R.~L.}\ \bibnamefont
  {{Smart}}}, \bibinfo {author} {\bibfnamefont {P.~R.}\ \bibnamefont
  {{Eisenhardt}}}, \bibinfo {author} {\bibfnamefont {M.~J.}\ \bibnamefont
  {{Kuchner}}}, \bibinfo {author} {\bibfnamefont {E.~L.}\ \bibnamefont
  {{Wright}}}, \bibinfo {author} {\bibfnamefont {M.~C.}\ \bibnamefont
  {{Cushing}}}, \bibinfo {author} {\bibfnamefont {K.~N.}\ \bibnamefont
  {{Allers}}}, \bibinfo {author} {\bibfnamefont {D.~C.}\ \bibnamefont
  {{Bardalez Gagliuffi}}}, \bibinfo {author} {\bibfnamefont {A.~J.}\
  \bibnamefont {{Burgasser}}}, \bibinfo {author} {\bibfnamefont
  {J.}~\bibnamefont {{Gagn{\'e}}}}, \bibinfo {author} {\bibfnamefont {S.~E.}\
  \bibnamefont {{Logsdon}}}, \bibinfo {author} {\bibfnamefont {E.~C.}\
  \bibnamefont {{Martin}}}, \bibinfo {author} {\bibfnamefont {J.~G.}\
  \bibnamefont {{Ingalls}}}, \bibinfo {author} {\bibfnamefont {P.~J.}\
  \bibnamefont {{Lowrance}}}, \bibinfo {author} {\bibfnamefont {E.~S.}\
  \bibnamefont {{Abrahams}}}, \bibinfo {author} {\bibfnamefont
  {C.}~\bibnamefont {{Aganze}}}, \bibinfo {author} {\bibfnamefont
  {R.}~\bibnamefont {{Gerasimov}}}, \bibinfo {author} {\bibfnamefont {E.~C.}\
  \bibnamefont {{Gonzales}}}, \bibinfo {author} {\bibfnamefont {C.-C.}\
  \bibnamefont {{Hsu}}}, \bibinfo {author} {\bibfnamefont {N.}~\bibnamefont
  {{Kamraj}}}, \bibinfo {author} {\bibfnamefont {R.}~\bibnamefont {{Kiman}}},
  \bibinfo {author} {\bibfnamefont {J.}~\bibnamefont {{Rees}}}, \bibinfo
  {author} {\bibfnamefont {C.}~\bibnamefont {{Theissen}}}, \bibinfo {author}
  {\bibfnamefont {K.}~\bibnamefont {{Ammar}}}, \bibinfo {author} {\bibfnamefont
  {N.~S.}\ \bibnamefont {{Andersen}}}, \bibinfo {author} {\bibfnamefont
  {P.}~\bibnamefont {{Beaulieu}}}, \bibinfo {author} {\bibfnamefont
  {G.}~\bibnamefont {{Colin}}}, \bibinfo {author} {\bibfnamefont {C.~A.}\
  \bibnamefont {{Elachi}}}, \bibinfo {author} {\bibfnamefont {S.~J.}\
  \bibnamefont {{Goodman}}}, \bibinfo {author} {\bibfnamefont {L.}~\bibnamefont
  {{Gramaize}}}, \bibinfo {author} {\bibfnamefont {L.~K.}\ \bibnamefont
  {{Hamlet}}}, \bibinfo {author} {\bibfnamefont {J.}~\bibnamefont {{Hong}}},
  \bibinfo {author} {\bibfnamefont {A.}~\bibnamefont {{Jonkeren}}}, \bibinfo
  {author} {\bibfnamefont {M.}~\bibnamefont {{Khalil}}}, \bibinfo {author}
  {\bibfnamefont {D.~W.}\ \bibnamefont {{Martin}}}, \bibinfo {author}
  {\bibfnamefont {W.}~\bibnamefont {{Pendrill}}}, \bibinfo {author}
  {\bibfnamefont {B.}~\bibnamefont {{Pumphrey}}}, \bibinfo {author}
  {\bibfnamefont {A.}~\bibnamefont {{Rothermich}}}, \bibinfo {author}
  {\bibfnamefont {A.}~\bibnamefont {{Sainio}}}, \bibinfo {author}
  {\bibfnamefont {A.}~\bibnamefont {{Stenner}}}, \bibinfo {author}
  {\bibfnamefont {C.}~\bibnamefont {{Tanner}}}, \bibinfo {author}
  {\bibfnamefont {M.}~\bibnamefont {{Th{\'e}venot}}}, \bibinfo {author}
  {\bibfnamefont {N.~V.}\ \bibnamefont {{Voloshin}}}, \bibinfo {author}
  {\bibfnamefont {J.}~\bibnamefont {{Walla}}}, \bibinfo {author} {\bibfnamefont
  {Z.}~\bibnamefont {{W{\k{e}}dracki}}}, \ and\ \bibinfo {author} {\bibnamefont
  {{Backyard Worlds: Planet 9 Collaboration}}},\ }\href {\doibase
  10.3847/1538-4365/abd107} {\bibfield  {journal} {\bibinfo  {journal} {\apjs}\
  }\textbf {\bibinfo {volume} {253}},\ \bibinfo {eid} {7} (\bibinfo {year}
  {2021})},\ \Eprint {http://arxiv.org/abs/2011.11616} {arXiv:2011.11616
  [astro-ph.SR]} \BibitemShut {NoStop}%
\bibitem [{\citenamefont {{Bensby}}\ \emph {et~al.}(2017)\citenamefont
  {{Bensby}}, \citenamefont {{Feltzing}}, \citenamefont {{Gould}},
  \citenamefont {{Yee}}, \citenamefont {{Johnson}}, \citenamefont {{Asplund}},
  \citenamefont {{Mel{\'e}ndez}}, \citenamefont {{Lucatello}}, \citenamefont
  {{Howes}}, \citenamefont {{McWilliam}}, \citenamefont {{Udalski}},
  \citenamefont {{Szyma{\'n}ski}}, \citenamefont {{Soszy{\'n}ski}},
  \citenamefont {{Poleski}}, \citenamefont {{Wyrzykowski}}, \citenamefont
  {{Ulaczyk}}, \citenamefont {{Koz{\l}owski}}, \citenamefont {{Pietrukowicz}},
  \citenamefont {{Skowron}}, \citenamefont {{Mr{\'o}z}}, \citenamefont
  {{Pawlak}}, \citenamefont {{Abe}}, \citenamefont {{Asakura}}, \citenamefont
  {{Bhattacharya}}, \citenamefont {{Bond}}, \citenamefont {{Bennett}},
  \citenamefont {{Hirao}}, \citenamefont {{Nagakane}}, \citenamefont
  {{Koshimoto}}, \citenamefont {{Sumi}}, \citenamefont {{Suzuki}},\ and\
  \citenamefont {{Tristram}}}]{2017A&A...605A..89B}%
  \BibitemOpen
  \bibfield  {author} {\bibinfo {author} {\bibfnamefont {T.}~\bibnamefont
  {{Bensby}}}, \bibinfo {author} {\bibfnamefont {S.}~\bibnamefont
  {{Feltzing}}}, \bibinfo {author} {\bibfnamefont {A.}~\bibnamefont {{Gould}}},
  \bibinfo {author} {\bibfnamefont {J.~C.}\ \bibnamefont {{Yee}}}, \bibinfo
  {author} {\bibfnamefont {J.~A.}\ \bibnamefont {{Johnson}}}, \bibinfo {author}
  {\bibfnamefont {M.}~\bibnamefont {{Asplund}}}, \bibinfo {author}
  {\bibfnamefont {J.}~\bibnamefont {{Mel{\'e}ndez}}}, \bibinfo {author}
  {\bibfnamefont {S.}~\bibnamefont {{Lucatello}}}, \bibinfo {author}
  {\bibfnamefont {L.~M.}\ \bibnamefont {{Howes}}}, \bibinfo {author}
  {\bibfnamefont {A.}~\bibnamefont {{McWilliam}}}, \bibinfo {author}
  {\bibfnamefont {A.}~\bibnamefont {{Udalski}}}, \bibinfo {author}
  {\bibfnamefont {M.~K.}\ \bibnamefont {{Szyma{\'n}ski}}}, \bibinfo {author}
  {\bibfnamefont {I.}~\bibnamefont {{Soszy{\'n}ski}}}, \bibinfo {author}
  {\bibfnamefont {R.}~\bibnamefont {{Poleski}}}, \bibinfo {author}
  {\bibfnamefont {{\L}.}~\bibnamefont {{Wyrzykowski}}}, \bibinfo {author}
  {\bibfnamefont {K.}~\bibnamefont {{Ulaczyk}}}, \bibinfo {author}
  {\bibfnamefont {S.}~\bibnamefont {{Koz{\l}owski}}}, \bibinfo {author}
  {\bibfnamefont {P.}~\bibnamefont {{Pietrukowicz}}}, \bibinfo {author}
  {\bibfnamefont {J.}~\bibnamefont {{Skowron}}}, \bibinfo {author}
  {\bibfnamefont {P.}~\bibnamefont {{Mr{\'o}z}}}, \bibinfo {author}
  {\bibfnamefont {M.}~\bibnamefont {{Pawlak}}}, \bibinfo {author}
  {\bibfnamefont {F.}~\bibnamefont {{Abe}}}, \bibinfo {author} {\bibfnamefont
  {Y.}~\bibnamefont {{Asakura}}}, \bibinfo {author} {\bibfnamefont
  {A.}~\bibnamefont {{Bhattacharya}}}, \bibinfo {author} {\bibfnamefont
  {I.~A.}\ \bibnamefont {{Bond}}}, \bibinfo {author} {\bibfnamefont {D.~P.}\
  \bibnamefont {{Bennett}}}, \bibinfo {author} {\bibfnamefont {Y.}~\bibnamefont
  {{Hirao}}}, \bibinfo {author} {\bibfnamefont {M.}~\bibnamefont {{Nagakane}}},
  \bibinfo {author} {\bibfnamefont {N.}~\bibnamefont {{Koshimoto}}}, \bibinfo
  {author} {\bibfnamefont {T.}~\bibnamefont {{Sumi}}}, \bibinfo {author}
  {\bibfnamefont {D.}~\bibnamefont {{Suzuki}}}, \ and\ \bibinfo {author}
  {\bibfnamefont {P.~J.}\ \bibnamefont {{Tristram}}},\ }\href {\doibase
  10.1051/0004-6361/201730560} {\bibfield  {journal} {\bibinfo  {journal}
  {\aap}\ }\textbf {\bibinfo {volume} {605}},\ \bibinfo {eid} {A89} (\bibinfo
  {year} {2017})},\ \Eprint {http://arxiv.org/abs/1702.02971} {arXiv:1702.02971
  [astro-ph.GA]} \BibitemShut {NoStop}%
\bibitem [{\citenamefont {{Blumenthal}}\ \emph {et~al.}(1986)\citenamefont
  {{Blumenthal}}, \citenamefont {{Faber}}, \citenamefont {{Flores}},\ and\
  \citenamefont {{Primack}}}]{1986ApJ...301...27B}%
  \BibitemOpen
  \bibfield  {author} {\bibinfo {author} {\bibfnamefont {G.~R.}\ \bibnamefont
  {{Blumenthal}}}, \bibinfo {author} {\bibfnamefont {S.~M.}\ \bibnamefont
  {{Faber}}}, \bibinfo {author} {\bibfnamefont {R.}~\bibnamefont {{Flores}}}, \
  and\ \bibinfo {author} {\bibfnamefont {J.~R.}\ \bibnamefont {{Primack}}},\
  }\href {\doibase 10.1086/163867} {\bibfield  {journal} {\bibinfo  {journal}
  {\apj}\ }\textbf {\bibinfo {volume} {301}},\ \bibinfo {pages} {27} (\bibinfo
  {year} {1986})}\BibitemShut {NoStop}%
\bibitem [{\citenamefont {{Hoffman}}\ and\ \citenamefont
  {{Gelman}}(2011)}]{2011arXiv1111.4246H}%
  \BibitemOpen
  \bibfield  {author} {\bibinfo {author} {\bibfnamefont {M.~D.}\ \bibnamefont
  {{Hoffman}}}\ and\ \bibinfo {author} {\bibfnamefont {A.}~\bibnamefont
  {{Gelman}}},\ }\href {\doibase 10.48550/arXiv.1111.4246} {\bibfield
  {journal} {\bibinfo  {journal} {arXiv e-prints}\ ,\ \bibinfo {eid}
  {arXiv:1111.4246}} (\bibinfo {year} {2011})},\ \Eprint
  {http://arxiv.org/abs/1111.4246} {arXiv:1111.4246 [stat.CO]} \BibitemShut
  {NoStop}%
\bibitem [{\citenamefont {{Neal}}(2011)}]{2011hmcm.book..113N}%
  \BibitemOpen
  \bibfield  {author} {\bibinfo {author} {\bibfnamefont {R.}~\bibnamefont
  {{Neal}}},\ }in\ \href {\doibase 10.1201/b10905} {\emph {\bibinfo {booktitle}
  {Handbook of Markov Chain Monte Carlo}}}\ (\bibinfo {year} {2011})\ pp.\
  \bibinfo {pages} {113--162}\BibitemShut {NoStop}%
\bibitem [{\citenamefont {{Bingham}}\ \emph {et~al.}(2018)\citenamefont
  {{Bingham}}, \citenamefont {{Chen}}, \citenamefont {{Jankowiak}},
  \citenamefont {{Obermeyer}}, \citenamefont {{Pradhan}}, \citenamefont
  {{Karaletsos}}, \citenamefont {{Singh}}, \citenamefont {{Szerlip}},
  \citenamefont {{Horsfall}},\ and\ \citenamefont
  {{Goodman}}}]{2018arXiv181009538B}%
  \BibitemOpen
  \bibfield  {author} {\bibinfo {author} {\bibfnamefont {E.}~\bibnamefont
  {{Bingham}}}, \bibinfo {author} {\bibfnamefont {J.~P.}\ \bibnamefont
  {{Chen}}}, \bibinfo {author} {\bibfnamefont {M.}~\bibnamefont {{Jankowiak}}},
  \bibinfo {author} {\bibfnamefont {F.}~\bibnamefont {{Obermeyer}}}, \bibinfo
  {author} {\bibfnamefont {N.}~\bibnamefont {{Pradhan}}}, \bibinfo {author}
  {\bibfnamefont {T.}~\bibnamefont {{Karaletsos}}}, \bibinfo {author}
  {\bibfnamefont {R.}~\bibnamefont {{Singh}}}, \bibinfo {author} {\bibfnamefont
  {P.}~\bibnamefont {{Szerlip}}}, \bibinfo {author} {\bibfnamefont
  {P.}~\bibnamefont {{Horsfall}}}, \ and\ \bibinfo {author} {\bibfnamefont
  {N.~D.}\ \bibnamefont {{Goodman}}},\ }\href {\doibase
  10.48550/arXiv.1810.09538} {\bibfield  {journal} {\bibinfo  {journal} {arXiv
  e-prints}\ ,\ \bibinfo {eid} {arXiv:1810.09538}} (\bibinfo {year} {2018})},\
  \Eprint {http://arxiv.org/abs/1810.09538} {arXiv:1810.09538 [cs.LG]}
  \BibitemShut {NoStop}%
\end{thebibliography}%

\end{document}